\documentstyle[psfig]{mn}
\title[New subdwarfs from Hipparcos]
{A search for previously unrecognised metal-poor subdwarfs in the Hipparcos astrometric
catalogue}
\author[Reid et al ]
{I. Neill Reid$^{1}$,  F. van Wyk$^2$, F. Marang $^2$, G. Roberts$^2$,
D. Kilkenny$^2$ and S. Mahoney$^3$\\
$^1$Dept. of Physics \& Astronomy, University of Pennsylvania, 209 S. 33rd
Street, Philadelphia, PA 19104-6396; \\
e-mail: inr@morales.physics.upenn.ed \\
$^2$South African Astronomical Observatory, PO Box 9, Observatory 7935, South Africa \\
$^3$Hinman Box 1709, Dartmouth College, Hanover, NH 03755 \\}
\begin {document}

\maketitle

\begin {abstract}

We have identified 317 stars included in the Hipparcos astrometric catalogue
which have parallaxes measured to a precision of better than
15\%, and whose location in the (M$_V$, (B-V)$_T$) diagram implies
a metallicity comparable to or less than that of the intermediate-abundance
globular cluster, M5. We have undertaken an extensive literature search to locate Str\"omgren, 
Johnson/Cousins and Walraven photometry for over 120 stars. 
In addition, we present new UBV(RI)$_C$ photometry of 201 of these candidate halo stars, together
with similar data for a further 14 known metal-poor subdwarfs. Those observations provide the
first extensive dataset of R$_C$I$_C$ photometry of metal-poor, main-sequence stars with 
well-determined trigonometric
parallaxes. Finally, we have obtained intermediate-resolution optical spectroscopy of 175 stars.

Forty-seven stars still lack sufficient supplementary observations for population classification; 
however, we are able to estimate 
abundances for 270 stars, or over 80\% of the sample. The overwhelming majority have
near-solar abundance, with their inclusion in the present sample 
stemming from errors in the colours listed in the Hipparcos catalogue. 
Only forty-four stars show consistent evidence of abundances below [Fe/H]= -1.0.
Nine are additions to the small sample of metal-poor subdwarfs with
accurate photometry. We consider briefly the
implication of these results for cluster main-sequence fitting.

\end{abstract}


\section {Introduction}

Main-sequence fitting remains one of the principal methods of determining distances,
and hence turnoff luminosities and age estimates, for Galactic globular clusters.
While recent investigations suggest that cluster ages may no longer set stringent
constraints on  cosmological
models (Perlmutter {\sl et al.}, 1998; Schmidt {\sl et al.},
1998; Riess  {\sl et al.}, 2000), these measurements remain an important probe
of the formation and evolution of the Milky Way. Empirical main-sequence fitting demands
calibrators with chemical abundances well matched to the target cluster, and spanning a
sufficient range in colour on the unevolved main-sequence to provide a reliable
distance modulus estimate. Unfortunately, those conditions are met for scarcely
any Galactic globulars, even with the addition of milliarcsecond-accuracy astrometric
data from the Hipparcos satellite (ESA, 1997). The local space density of the halo
is sufficiently low that distance analyses 
(Reid, 1997, 1998; Gratton {\sl et al.}, 1997b; Pont {\sl et al.}, 1997; Carretta {\sl et al.},
2000) rest on data for barely two dozen subdwarfs which have both accurate
parallax and abundance determinations, and, to the best of present knowledge, are single stars. 

The situation is particularly acute for metal poor globular clusters, with abundances
[m/H]$< -1.6$. While several halo stars with extreme abundances have reliable 
Hipparcos parallax measurements, most of those stars have $M_V < 5$, placing them near 
the main-sequence turnoff. Evolutionary effects lead to a steepening of
the main sequence at those luminosities in globular clusters, and a small mismatch
in age, colour or abundance between calibrator and cluster can lead to substantial systematic
errors in inferred distance moduli in main-sequence fitting. Of the few lower main-sequence subdwarfs
with [m/H]$< -1.5$, the only star widely used as a distance calibrator
is BD+66 268. Unfortunately, that star is a binary of uncertain mass ratio, and
hence uncertain true luminosity. 

The scarcity of G and K subdwarfs with accurate parallaxes is not surprising. Unevolved subdwarfs
have $M_V > 5.5$; the Hipparcos sample is complete to a magnitude limit of
$V = 7.9 + 1.1\sin{b}$. Adopting $\langle V_{lim} \rangle = 8.5$, this gives a sampling volume of
$\sim1.3 \times 10^5$ pc$^3$ at M$_V = 6$. The space density of subdwarfs at that 
absolute magnitude is only
$\sim 0.2\% \rho_{disk}$, or $\sim 6 \times 10^{-6}$ stars pc$^{-1}$ mag$^{-1}$, so we expect only
one such halo star (spanning the full abundance range) in the Hipparcos catalogue. HD 103095 meets
these criteria, with $M_V = 6.61$, [m/H] = -1.22 and V = 6.42. 

Clementini {\sl et al.} (1998) have addressed this issue, the scarcity of reliable subdwarfs
calibrators, by undertaking improved abundance determinations of known subdwarfs; we adopt
an alternative strategy, searching for previously unrecognised subdwarfs.  More distant
subdwarfs have lower precision parallax measurements, which can lead to systematic biases,
such as the Lutz-Kelker effect, if the relative uncertainties, $\sigma_\pi \over \pi$, exceed $\sim15\%$.
Since Hipparcos parallaxes have typical uncertainties, $\epsilon_\pi$, of 1 to 1.5 milliarcseconds (mas),
this effectively limits the survey volume to $r < 100$ parsecs. Nonetheless, given
the local subdwarf space density estimated above, we might expect 15 to 20 subdwarfs with 
$M_V > 5.5$ and $V < 12$ to lie within range of Hipparcos measurement.  

The Hipparcos catalogue is far from complete at magnitudes fainter than V=9, but does include a
significant number of proper motion stars, mainly from the
Lowell survey (Giclas {\sl et al.}, 1958). Proper motion surveys are biased toward high velocity stars,
and are therefore good hunting grounds for halo subdwarfs. Despite extensive work by Carney and collaborators
(Carney {\sl et al.}, 1994 and refs within), only a subset of the Lowell stars have accurate photometry and/or
spectroscopy. Thus, it remains possible that subdwarfs lie unrecognised amongst the fainter stars
observed by Hipparcos.

Besides proper motions and parallaxes, the Hipparcos catalogue provides photometric data.
We can therefore place each star on the HR diagram. Since metal-poor stars are known to
be subluminous with respect to the disk main sequence at a given colour (or, more correctly,
hotter at a given mass), those photometric data offer the possibility of identifying
additional subdwarfs. This paper marks the first phase in such an analysis. We identify 317 stars with
colours consistent with abundances [m/H]$\le -1.1$. Using both previously published data and
our own observations, we have collected accurate photometry and spectroscopy for 80\%
of the sample. Almost all are eliminated as likely subdwarfs, and most of the  survivors
are well known halo stars. Nonetheless, our observations identify a handful of new
calibrators, and provide the first extensive catalogue of RI photometry
of confirmed metal-poor subdwarfs. 

\section {The Sample}

\subsection {Selection criteria}

The primary source of photometry for Hipparcos stars is the Tycho experiment, which used grids of
star mappers to measure stellar fluxes in two passbands approximating the Johnson BV system. As
discussed in vol. 3 of the Hipparcos catalogue, 
there are systematic differences between Tycho photometry and the standard system (see also
Bessell, 2000). 
The Hipparcos catalogue lists BV photometry culled from the literature for many sources, and Figures
1a and 1b use the latter measurements to illustrate the colour terms present in both V$_T$ and
B$_T$. Figure 1c shows that those terms cancel to a large extent for stars bluer than (B-V)$_T$ = 0.6,
i.e. for F and G dwarfs $(B-V)_T \approx (B-V)$\footnote{ This accounts for the apparent agreement
between ground-based and Tycho data noted by Clementini {\sl et al.} in their limited photometric
comparison.} At later spectral types, (B-V)$_T$ is redder than the Johnson (B-V) colour. 
The Hipparcos catalogue also lists (V-I) colours, but these are drawn from such a wide variety of
sources that, as noted by Clementini {\sl et al.} (1998), the measurements are of little practical benefit
for our purposes. 

\begin{figure}
\psfig{figure=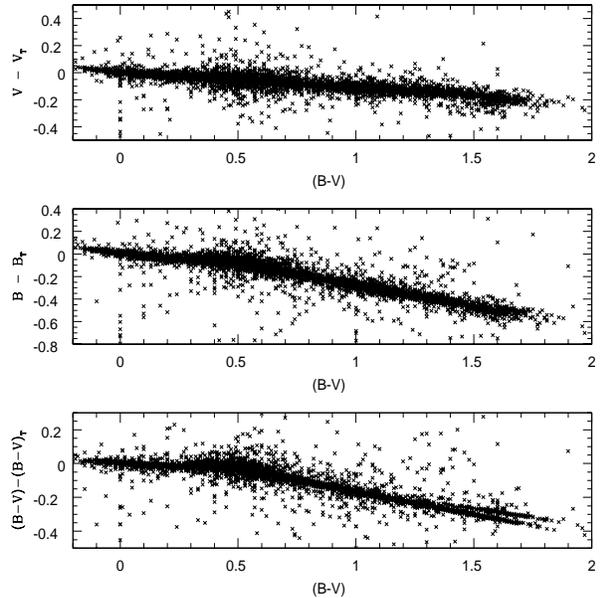,height=8cm,width=8cm
,bbllx=8mm,bblly=57mm,bburx=205mm,bbury=245mm,rheight=9cm}
\caption{Colour terms in the Tycho B$_T$, V$_T$ system. The comparison is based on data for 7000 stars
from the Hipparcos catalogue, comparing the Tycho photometry against the literature BV data listed. }
\end{figure}

Our aim is to identify candidate metal-poor subdwarfs with reliable parallax measurements. Thus,
as a first step, we consider only stars in the Hipparcos catalogue with parallaxes
measured to a formal precision, $\sigma_\pi \over \pi$, better than 15\%. This corresponds to a systematic 
Lutz-Kelker correction of $\Delta M_V = 0.3$ magnitudes for a sample with a uniform space distribution
($N(\pi) \propto \pi^{-4}$); in reality, subdwarf samples are likely to have a flatter
parallax distribution ($N(\pi) \propto \pi^{-n}$, $n < 4$), with correspondingly smaller Lutz-Kelker
corrections. 

Candidate metal-poor stars have been selected based on their location in the (M$_V$, (B-V))
colour-magnitude diagram. Figure 2 plots the (M$_V$, (B-V)$_T$) diagram for Hipparcos
stars with ${\sigma_\pi \over \pi} < 0.15$ and $\pi > 32$ mas - 
a sample dominated by the disk population.
Superimposed on that diagram are the mean colour-magnitude relations for three globular clusters:
the metal-rich cluster, 47 Tucanae, [Fe/H] = -0.7, at (m-M)=13.55; the intermediate-abundance
cluster, M5, [Fe/H]=-1.11, at (m-M)=14.5; and the metal-poor cluster, NGC 6397, [Fe/H]=-1.82,
at (m-M)=12.14 (abundances from Carretta \& Gratton, 1997; see Reid, 1999, for a summary of the
distance modulus determinations\footnote{Anthony-Twarog \& Twarog (2000) have recently derived a 
distance modulus of (m-M)$_0$=12.15 for NGC 6397 based on main-sequence fitting using
Str\"omgren photometry.} ). 
Those clusters provide a reference grid for the identification of
candidate F- and G-type subdwarfs. 

 \setcounter{table}{0}
 \begin {table*}
  \centering
\caption{Candidate Subdwarf Sample} 
 \begin{tabular}{rrrrrcrrrcccl} 
  \hline \hline
 HIP & V$_T$ & (B-V)$_T$ & $\pi$ &$ \sigma_\pi \over \pi$ & M$_V$ & V$_l$ & (B-V)$_l$ & ID & Phot & Spec & sd?&Comments \\  
  \hline
     435&   9.36&   0.46&   10.3&  0.129&   4.42&   9.30&   0.54&HD 34          &C s     & Y   &I& uvby, UBV      \\  
     895&   8.61&   0.44&   12.6&  0.094&   4.11&   8.55&   0.44&HD 664         &        & Y   &I&  LCO        \\  
    1051&   8.80&   0.41&   11.1&  0.100&   4.03&   8.75&   0.40&HD 864         &C       & Y   &I& UBV          \\  
    1300&   9.80&   0.46&    7.7&  0.123&   4.22&   9.74&   0.45&BD+71 9       &        &     &          \\  
    1719&   9.79&   0.79&   23.2&  0.044&   6.62&   9.65&   0.83&CPD -67 23     &C       & Y   &D& UBV, Dbl.         \\  
    3022&  10.01&   0.78&   19.9&  0.094&   6.49&   9.96&   0.79&BD+42 126     &        &     &          \\  
    3139&   9.38&   0.41&    8.6&  0.142&   4.07&   9.27&   0.47&HD 3734        &C  s     & Y   &D& UBV          \\  
    3531&  11.43&   0.70&   18.8&  0.123&   7.80&  10.91&   1.21&G 270-59       &C       & Y   &D& M dwarf          \\  
    3855&  10.94&   0.59&   19.5&  0.102&   7.39&  10.57&   0.60&LP 406-78      &C O     & Y   &D& M dwarf          \\  
    4450&   9.67&   0.49&    9.9&  0.120&   4.65&   9.65&   0.50&HD 5592        &C       & Y   &D& UBV          \\  
    4576&   9.54&   0.44&    9.2&  0.132&   4.36&   9.52&   0.44&HD 5770        &C       & Y   &D&UBV          \\  
    4750&  10.20&   0.57&   10.9&  0.148&   5.39&  10.08&   0.57&G 269-75       &C O     & Y   &H&UBV          \\  
    4981&   9.22&   0.57&   17.4&  0.073&   5.41&   9.11&   0.58&HD 6323        &C s O     & Y   &I&UBV, uvby     \\  
    5004&  10.37&   0.76&   16.3&  0.108&   6.43&  10.25&   0.76&G 269-87       &C s O   & Y   &H& RN, uvby, UBV  \\  
    5097&   9.36&   0.46&   10.0&  0.127&   4.37&   9.30&   0.45&HD 6507        &C       & Y   &D&UBV          \\  
    5896&   8.44&   0.48&   48.9&  0.011&   6.89&   4.25&   0.48&HD 7788 B      & s O      & Y   &D& uvby, Dbl. C A         \\  
    6251&   8.50&   0.51&   18.9$^F$&  0.059&   4.88&   8.43&   0.52&HD 8068        &C       & Y   &I&UBV          \\  
    6758&   9.74&   0.58&   14.1&  0.116&   5.49&   9.67&   0.64&               &C       & Y   &D&UBV          \\  
    7303&   9.89&   0.57&   13.0&  0.131&   5.46&   9.84&   0.60&G 271-94       &C        & Y   &I&UBV, Dbl. G          \\  
    7459&  10.18&   0.50&   11.6&  0.102&   5.51&  10.10&   0.52&CPD -61 282    &C s O   & Y   &H& RN, uvby, UBV    \\  
    7687&   9.47&   0.78&   25.8&  0.060&   6.53&   9.37&   0.79&HD 10166       &C s     & Y   &I& uvby    \\  
    7772&   8.91&   0.38&    9.5&  0.114&   3.80&   8.84&   0.37&HD 10236       &C       & Y   &D&UBV          \\  
    7935&   8.94&   0.48&   12.6&  0.088&   4.43&   8.90&   0.47&HD 10440       &C       & Y   &I&UBV          \\  
    8130&  10.40&   0.50&   11.3&  0.127&   5.67&  10.21&   0.50&G 133-35       &  s O   &     &I& CLLA, uvby         \\  
    8298&  10.14&   0.69&   15.4&  0.123&   6.08&  10.09&   0.69&BD+00 287      &C       &     &D& UBV          \\  
    8389&  10.89&   0.71&   14.0&  0.128&   6.63&  10.69&   0.92&CD -37 689     &C       & Y   &D&UBV          \\  
    8558&   9.05&   0.35&    9.5&  0.081&   3.93&   9.01&   0.37&HD 11569       &C s O     & Y   &H&  RN \\  
    9634&   8.39&   0.44&   13.7&  0.082&   4.08&   8.34&   0.43&HD 12653       &C       & Y   &D&UBV          \\  
   10208&   9.11&   0.44&   12.7&  0.099&   4.63&   9.07&   0.42&HD 13441       &C       & Y   &I&UBV          \\  
   10353&   9.30&   0.46&   12.3&  0.109&   4.76&   9.22&   0.45&HD 13650       &C       & Y   &I&UBV          \\  
   10360&   8.58&   0.32&   12.6$^F$&  0.095&   4.08&   8.56&   0.32&HD 13721       &C s      & Y   &D& blue straggler      \\  
   10375&   9.18&   0.44&    9.7&  0.126&   4.11&   9.11&   0.43&HD 13670       &C       & Y   &I& UBV          \\  
   10385&  11.44&   0.58&   20.3&  0.104&   7.97&  10.87&   1.10&CD -33 758     &C       & Y   &D& K dwarf          \\  
   10529&   9.91&   0.46&   42.5$^F$&  0.059&   8.05&   9.91&       &BD+66 193     &        &     &  & Dbl. C D        \\  
   10637&   9.36&   0.42&    9.9&  0.135&   4.34&   9.28&   0.42&HD 14139       &C       & Y   &I&UBV          \\  
   11435&   8.69&   0.46&   13.9&  0.113&   4.40&   8.64&   0.45&HD 15276       &C       & Y   &D&UBV          \\  
   12579&   9.21&   0.50&   14.5&  0.088&   5.02&   9.16&   0.52&BD+46 610     &  O    &     &I& AFG, CLLA         \\  
   13849&  10.89&   0.60&   13.9&  0.140&   6.60&  10.67&   0.61&               &C       & Y   &D& UBV          \\  
   14192&   9.03&   0.43&    9.5&  0.086&   3.91&   8.94&   0.43&HD 19239       &C       & Y   &D&UBV          \\  
   14594&   8.10&   0.48&   25.8&  0.044&   5.16&   8.04&   0.49&HD 19445       &  s O    & Y   &H& F20, GCC        \\  
   14946&   8.56&   0.39&   10.9&  0.096&   3.73&   8.48&   0.38&HD 19768       &        &     &          \\  
   15756&  11.01&   0.73&   12.1&  0.127&   6.43&  10.90&   0.95&CD -49 942     &C       & Y   &D&UBV          \\  
   15998&  10.41&   0.75&   16.8&  0.107&   6.53&  10.17&   0.85&BD+44 697     &  s     &     &D&uvby          \\  
   16089&   8.81&   0.42&   11.2&  0.115&   4.07&   8.77&   0.42&HD 21515       &C       & Y   &D&UBV          \\  
   16404&   9.97&   0.65&   17.6&  0.087&   6.20&   9.91&   0.67&BD+66 268     &  s O   &     &H&  CLLA, uvby, Dbl. S       \\  
   16479&   8.36&   0.43&   13.6&  0.078&   4.03&   8.30&   0.42&HD 21925       &C       & Y   &D&UBV          \\  
   17085&   9.64&   0.50&   12.6&  0.145&   5.14&   9.58&   0.50&HD 22785       &C       & Y   &D&UBV, F20      \\  
   17241&  10.27&   0.48&    8.2&  0.127&   4.85&  10.25&   0.44&CPD -59 286    &C       & Y   &H&UBV          \\  
   17481&   8.78&   0.36&   10.1&  0.126&   3.80&   8.70&   0.39&HD 23290       &C s     & Y   &D&UBV, uvby, Pleiad          \\  
   17497&   9.04&   0.37&    9.8&  0.132&   3.99&   8.98&   0.40&HD 23289       &  s O     &     &D& Pleiad          \\  
 \hline
 \end{tabular}
 \end{table*}

 \setcounter{table}{0}
 \begin {table*}
  \centering
\caption{Candidate Subdwarf Sample (contd.)}
 \begin{tabular}{rrrrrcrrrcccl} 
  \hline \hline
 HIP & V$_T$ & (B-V)$_T$ & $\pi$ &$ \sigma_\pi \over \pi$ & M$_V$ & V$_l$ & (B-V)$_l$ & ID & Phot & Spec &sd?& Comments \\  
  \hline 
   17844&   9.19&   0.49&   12.9&  0.094&   4.74&   9.15&   0.49&BD+43 817     &        &     &          \\  
   18700&   9.34&   0.55&   17.0&  0.098&   5.49&   9.33&   0.56&HD 286457      &C       & Y   &I&UBV          \\  
   19007&   9.64&   0.75&   24.2&  0.063&   6.56&   9.53&   0.82&HD 25673       &C       & Y   &I&UBV, F20          \\  
   19215&   9.02&   0.38&    9.0&  0.140&   3.80&   8.98&   0.36&HD 25821       &  s O   &     &D& UBV, uvby          \\  
   19637&   8.77&   0.44&   12.4&  0.105&   4.24&   8.71&   0.43&HD 26500       &        & Y   &D& LCO         \\  
   19797&   9.31&   0.37&   12.8&  0.104&   4.85&   9.23&   0.36&HD 284248      &  s O   & Y   &H& GCC, F20         \\  
   20413&  10.38&   0.65&   13.1&  0.148&   5.97&  10.32&   0.67&LP 156-48      &  O     &     &D&UBV          \\  
   20527&  11.43&   0.68&   22.6&  0.123&   8.20&  10.89&   1.29&VR 9           &C O     &     &D& Hyad          \\  
   20895&  11.52&   0.60&   25.0&  0.130&   8.51&  10.92&   1.39&LP 358-300     &  O     &     &D& M0, Hyad, Dbl. C B         \\  
   21000&   9.83&   0.60&   84.8$^{F*}$&  0.056&   9.47&   9.83&       &BD+04 701A   &C       & Y   &D&UBV, F20, Dbl. C A          \\  
   21125&  10.27&   0.68&   14.7&  0.113&   6.11&  10.19&   0.68&CPD -30 618    &C       & Y   &I&UBV          \\  
   21261&  11.14&   0.62&   21.1&  0.105&   7.75&  10.74&   1.20&Leiden 65      &C O     &     &D& Hyades M dwarf      \\  
   21478&   9.60&   0.49&   10.6&  0.146&   4.72&   9.43&   0.57&HD 283749      &  O     & Y   &D& LCO         \\  
   21586&  10.73&   0.61&   15.7&  0.130&   6.70&  10.39&   0.62&G 175-43       &  O     &     &I & F20          \\  
   21609&   9.85&   0.64&   17.0&  0.058&   6.00&   9.85&       &HD 29907       &C s O   & Y   & H & uvby, UBV, F20       \\  
   21767&  10.49&   0.57&   14.7&  0.140&   6.33&  10.40&   0.70&HD 280067      &  O     &     &D& F20          \\  
   22177&  11.42&   0.65&   22.5&  0.103&   8.17&  10.92&   1.28&Leiden 119     &C O     & Y   &D& Hyades M dwarf        \\  
   22246&  10.24&   0.76&   24.1$^F$&  0.129&   7.15&  10.12&   0.80&G 96-1     &  O     &     &D& UBV, F20, CLLA         \\  
   22632&   9.18&   0.47&   15.6&  0.077&   5.14&   9.13&   0.50&HD 31128       &C s O   & Y   &H& UBV, F20     \\  
   23573&  10.08&   0.68&   16.2&  0.069&   6.12&  10.03&   0.74&HD 273011      &C O     & Y   &I& UBV          \\  
   24296&   9.95&   0.54&   11.2&  0.094&   5.21&   9.85&   0.55&HD 273832      &C       & Y   &I& UBV          \\  
   24316&   9.52&   0.46&   14.6&  0.069&   5.33&   9.43&   0.50&HD 34328       &C  s O  & Y   &H& GCC, uvby, UBV     \\  
   24421&   9.41&   0.51&   12.1&  0.119&   4.83&   9.34&   0.52&HD 34165       &C       & Y   &D& UBV          \\  
   24935&   8.84&   0.45&   13.6&  0.060&   4.50&   8.77&   0.45&HD 35138       &C       & Y   &I&UBV          \\  
   25590&   9.01&   0.35&   10.1&  0.134&   4.03&   8.99&   0.37&HD 35707       &        &     &          \\  
   25717&   8.25&   0.43&   13.6&  0.091&   3.91&   8.19&   0.46&HD 36045       &C s     & Y   &D&UBV, uvby          \\  
   26452&   9.65&   0.51&   13.1&  0.118&   5.25&   9.60&   0.51&G 248-45       &  O     &     &I& CLLA, UBV        \\  
   26676&  10.34&   0.66&   14.3&  0.139&   6.12&  10.20&   0.65&G 102-20       &C s O   & Y   &H& RN, CLLA, uvby   \\  
   26688&   7.75&   0.43&   20.6$^F$&  0.043&   4.32&   7.70&   0.42&HD 37792   &C       & Y   &I&UBV, F20          \\  
   27361&   9.22&   0.38&   10.5&  0.150&   4.32&   9.19&   0.37&BD+35 375      &        &     &          \\  
   28122&   9.48&   0.49&   23.6&  0.135&   6.35&   9.48&   0.72&HD 40007       &C       & Y   &D&UBV, Dbl.          \\  
   28188&   9.07&   0.47&   16.6&  0.089&   5.16&   9.00&   0.47&HD 40057       &        & Y   &I& F20          \\  
   29322&  11.59&   0.43&   33.6&  0.126&   9.22&  11.29&   1.42&LHS 1832       &C O     & Y   &D& M dwarf          \\  
   29510&   8.43&   0.38&   11.4&  0.132&   3.72&   8.39&   0.37&HD 42634       &        & Y   &D& LCO          \\  
   30311&   9.24&   0.45&   10.9&  0.126&   4.43&   9.18&   0.43&HD 44191       &        &     &          \\  
   30481&   8.27&   0.40&   12.7&  0.075&   3.79&   8.23&   0.41&HD 44002       &  s O   &     &I& uvby          \\  
   31639&   9.72&   0.65&   17.6&  0.076&   5.94&   9.64&   0.70&CPD -25 1545   &C       & Y   &I&UBV, F20          \\  
   32009&   9.72&   0.69&   18.4&  0.114&   6.05&   9.63&   0.69&G 109-21       &C       & Y   &D&UBV          \\  
   32308&  10.79&   0.79&   18.1&  0.125&   7.08&  10.73&   0.80&BD+15 1305    &C       & Y   &D&UBV          \\  
   33282&  11.48&   0.67&   19.3&  0.129&   7.92&  11.30&   1.34&LP 205-37      &  O     &     &D& M dwarf          \\  
   33283&  11.37&   0.31&   11.4&  0.125&   6.65&  11.10&   0.79&L 59-34        &C       & Y   &I&UBV          \\  
   34145&   9.76&   0.54&   14.0&  0.112&   5.49&   9.62&   0.55&HD 268542      &        & Y   &H?& LCO         \\  
   34146&   8.10&   0.45&   15.7&  0.058&   4.09&   8.05&   0.47&HD 53545       &C s     & Y   &D&UBV, F20          \\  
   34548&   9.11&   0.47&   12.9&  0.112&   4.67&   9.06&   0.46&HD 53871       &        &     &D& F20          \\  
   35163&  10.08&   0.75&   19.0$^F$&  0.142&   6.47&   9.98&   0.73&BD+58 1015 &  O     &     &D?& Dbl. S      \\  
   35232&   9.74&   0.44&    8.1&  0.109&   4.29&   9.69&   0.46&HD 57489       &C       & Y   &I&UBV          \\  
   35560&   8.54&   0.44&   12.7&  0.092&   4.06&   8.50&   0.44&HD 57391       &C       & Y   &I&UBV          \\  
   36491&   8.54&   0.51&   20.0&  0.083&   5.05&   8.48&   0.54&HD 59374       &C s O   & Y   &I& UBV, F20, CLLA, RN \\  
   36818&  10.44&   0.73&   15.3$^F$&  0.090&   6.37&  10.49&   0.61&CPD -45 1588   &C s O   & Y   &I/H& RN, uvby   \\  
   38541&   8.37&   0.60&   35.3&  0.030&   6.11&   8.27&   0.62&HD 64090       &  s O   & Y   &H& F20, GCC, Dbl.         \\  
 \hline
 \end{tabular}
 \end{table*}
 \setcounter{table}{0}
 \begin {table*}
  \centering
\caption{Candidate Subdwarf Sample (contd.(}
 \begin{tabular}{rrrrrcrrrcccl} 
  \hline \hline
 HIP & V$_T$ & (B-V)$_T$ & $\pi$ &$ \sigma_\pi \over \pi$ & M$_V$ & V$_l$ & (B-V)$_l$ & ID & Phot & Spec &sd?& Comments \\  
  \hline
   39911&   9.63&   0.59&   15.0&  0.066&   5.51&   9.60&   0.60&HD 68089       &C O s   & Y   &I&UBV, uvby          \\  
   40408&   9.31&   0.38&    8.6&  0.147&   3.97&   9.30&   0.37&BD+32 1699    &        &     &          \\  
   40778&   9.72&   0.48&   10.4&  0.142&   4.80&   9.73&   0.48&HD 233511      &  O s   &     &H& F20, GCC         \\  
   41563&   8.94&   0.33&    9.5&  0.135&   3.82&   8.83&   0.33&HD 71768       &C       & Y   &D&UBV          \\  
   42278&   7.81&   0.38&   15.1&  0.094&   3.72&   7.79&   0.40&HD 73176       &C       & Y   &D&UBV, Dbl. G          \\  
   43445&   8.70&   0.47&   13.5&  0.085&   4.35&   8.63&   0.48&HD 75596       &C s     & Y   &I& uvby, UBV          \\  
   43490&   9.58&   0.53&   14.1&  0.064&   5.33&   9.55&   0.54&CPD -79 363    &C       & Y   &H& UBV          \\  
   43973&   9.60&   0.64&   18.1&  0.080&   5.89&   9.52&   0.66&BD-8 2534     &C       & Y   &I&UBV, Dbl.           \\  
   44116&   8.54&   0.44&   12.7&  0.098&   4.05&   8.48&   0.45&HD 76910       &C s     & Y   &I& uvby,  F20        \\  
   44124&   9.76&   0.41&   12.4&  0.139&   5.22&   9.66&   0.48&BD-3 2525     &  O s   & Y   &H& F20, RN, CLLA        \\  
   44436&  11.23&   0.39&   14.8&  0.073&   7.09&  11.23&       &HD 78084B?     &      & Y   &D&UBV, uvby, Dbl. C D          \\  
   45162&   9.20&   0.48&   11.7&  0.108&   4.53&   9.18&   0.48&BD+38 2007    &        &     &          \\  
   46051&   9.23&   0.57&   17.5&  0.102&   5.44&   9.16&   0.58&BD+36 1951    &        &     &          \\  
   46120&  10.14&   0.60&   16.5&  0.060&   6.22&  10.11&   0.56&Gl 345         &C O s   & Y   &H& RN, uvby, UBV     \\  
   46250&   8.42&   0.46&   14.4&  0.100&   4.21&   8.38&   0.47&HD 81298       & s       &     &D& uvby          \\  
   46509&   4.92&   0.41&   58.5&  0.065&   3.76&   4.59&   0.41&HD 81997       &  O s   &     &D& uvby, Dbl. X          \\  
   47161&   9.34&   0.42&    8.6&  0.136&   4.02&   9.29&   0.43&HD 83356       &C       & Y   &D&UBV          \\  
   47171&   9.40&   0.49&   11.9&  0.112&   4.78&   9.31&   0.58&BD-3 2726     &C   & Y   &I&UBV          \\  
   47640&   8.88&   0.44&   12.3&  0.094&   4.33&   8.84&   0.42&HD 83888       &        &     &D&F20          \\  
   47948&  10.15&   0.57&   11.9&  0.129&   5.53&  10.08&   0.58&BD+0 2554     &C       & Y   &D&F20, UBV          \\  
   48146&   9.64&   0.50&   14.0$^F$&  0.112&   5.37&   9.57&   0.50&BD+9 2242     &C       & Y   &D&UBV          \\  
   48152&   8.38&   0.40&   12.4&  0.085&   3.85&   8.33&   0.40&HD 84937       &C O s   & Y   &H& F20, GCC, UBV         \\  
   49574&   8.82&   0.48&   12.7&  0.134&   4.35&   8.75&   0.48&HD 87908       &C       & Y   &I&UBV, Dbl. S X          \\  
   49785&   8.56&   0.48&   15.9$^F$&  0.061&   4.57&   8.51&   0.49&HD 88198       &C       & Y   &D&UBV          \\  
   49868&   8.97&   0.76&   47.1&  0.058&   7.34&   9.48&   1.34&BD+75 403     &  O     &     &D& M dwarf, Dbl. C A          \\  
   50153&   9.71&   0.45&    7.5&  0.128&   4.08&   9.69&   0.44&BD+81 327     &        &     &          \\  
   50532&   9.97&   0.49&   10.2&  0.147&   5.01&   9.91&   0.55&BD-9 2044     &C       & Y   &D&UBV          \\  
   50965&   9.93&   0.49&    9.7&  0.146&   4.86&   9.80&   0.58&G 162-51       &C O s   & Y   &I& RN, CLLA, uvby, UBV     \\  
   51156&   9.10&   0.43&   10.4&  0.113&   4.18&   9.06&   0.42&HD 90527       &C       & Y   &D&UBV          \\  
   51298&   9.28&   0.46&   10.4&  0.120&   4.37&   9.24&   0.44&HD 90764       &C       & Y   &D&UBV          \\  
   51300&   9.29&   0.41&    8.0&  0.123&   3.80&   9.21&   0.41&HD 91043       &C W     & Y   &D & UBV, Walraven          \\  
   51769&  10.49&   0.72&   16.2&  0.111&   6.53&  10.50&   0.68&G 162-68       &C O s   &     &I& RN, CLLA, uvby, UBV    \\  
   52285&  10.13&   0.79&   18.7&  0.130&   6.49&   9.89&   0.91&BD+4 2370     &C       & Y   &I&UBV, Dbl. C A          \\  
   53070&   8.26&   0.49&   19.2&  0.059&   4.68&   8.21&   0.50&HD 94028       &C O s   &     &H& F20, GCC, UBV         \\  
   53911&  11.15&   0.70&   17.7&  0.125&   7.39&  10.92&   0.72&TW Hya         &C O     & Y   &D& M dwarf        \\  
   54519&  11.21&   0.75&   16.7&  0.129&   7.32&  11.07&   1.59&BD+19 2423    &C       & Y   &D& M dwarf        \\  
   54641&   8.22&   0.48&   17.8&  0.043&   4.47&   8.16&   0.48&HD 97320       &C O s   & Y   &H& AFG, UBV         \\  
   54768&   9.14&   0.52&   14.0&  0.090&   4.88&   9.11&   0.52&HD 97354       &  O     &     &I& UBV          \\  
   54834&   8.82&   0.46&   15.2&  0.069&   4.73&   8.88&   0.05&CPD -56 4321   &C       & Y   &H&UBV          \\  
   54993&   8.89&   0.39&    9.7&  0.118&   3.81&   8.83&   0.38&HD 97874       &C       & Y   &D&UBV          \\  
   55790&   9.14&   0.47&   11.0&  0.135&   4.34&   9.07&   0.48&HD 99383       &C O s   & Y   &H& AFG, uvby, UBV         \\  
   55978&   9.14&   0.39&    8.3&  0.126&   3.73&   9.07&   0.41&HD 99805       &C s W     & Y   &D&Walraven, uvby, UBV          \\  
   57360&   8.80&   0.43&   12.5&  0.096&   4.27&   8.75&   0.43&HD 102200      &C O s   & Y   &H& RN, AFG, uby, UBV        \\  
   57450&   9.94&   0.56&   13.6&  0.113&   5.61&   9.91&   0.58&BD+51 1696    &  O s   &     &H& CLLA          \\  
   57939&   6.51&   0.75&  109.2&  0.007&   6.70&   6.42&   0.75&HD 103095      &  O s   &     &H& F20, GCC         \\  
   59258&   7.53&   0.36&   17.3&  0.048&   3.73&   7.51&   0.35&HD 105584      &  s     &     &I& uvby          \\  
   59750&   6.11&   0.47&   44.3$^F$&  0.023&   4.34&   6.11&       &HD 106516      &  O C1   &     &I&F20, Dbl. O         \\  
   60251&   9.11&   0.53&   16.5&  0.145&   5.20&   9.00&   0.54&HD 107440      &C W     & Y   &D& UBV, Walraven, Dbl. X S \\  
   60632&   9.71&   0.42&   11.0&  0.118&   4.91&   9.66&   0.44&HD 108177      &C O s   &     &H& GCC, uvby, UBV         \\  
   60783&  12.08&   0.70&   36.8&  0.137&   9.91&  12.08&       &G 148-61       &        &     & & Dbl. X S         \\  
 \hline
 \end{tabular}
 \end{table*}
 \setcounter{table}{0}
 \begin {table*}
  \centering
\caption{Candidate Subdwarf Sample (contd.)}
 \begin{tabular}{rrrrrcrrrcccl} 
  \hline \hline
 HIP & V$_T$ & (B-V)$_T$ & $\pi$ &$ \sigma_\pi \over \pi$ & M$_V$ & V$_l$ & (B-V)$_l$ & ID & Phot & Spec &sd?& Comments \\  
  \hline 
   60852&   8.52&   0.47&   14.4&  0.075&   4.32&   8.48&   0.50&HD 108540      &C s       & Y   &I&UBV, uvby          \\  
   61085&   9.18&   0.45&   10.3&  0.128&   4.24&   9.13&   0.44&HD 108974      &        &     &          \\  
   62261&   9.45&   0.35&    8.9&  0.150&   4.21&   9.42&   0.36&HD 110963      &  s     &     &D& uvby          \\  
   62858&   9.57&   0.50&   11.0&  0.129&   4.77&   9.51&   0.50&BD+16 2432    &C       &     &I&UBV          \\  
   63063&   9.99&   0.73&   19.3$^F$&  0.089&   6.41&   9.93&   0.81&BD+8 2658     &C O s   &     &D & UBV, uvby, CLLA     \\ 
   63553&   8.57&   0.40&   11.1&  0.105&   3.80&   8.51&   0.44&HD 113125      &C s      &     &D&uvby, UBV          \\  
   63781&   9.35&   0.40&    7.9&  0.150&   3.83&   9.30&   0.38&HD 113517      &C       &     &D&UBV          \\  
   63912&   8.96&   0.44&   10.3&  0.092&   4.02&   8.93&   0.41&HD 113862      &  O     &     &D& UBV          \\  
   63970&  10.14&   0.52&   12.7&  0.130&   5.66&  10.07&   0.52&BD+33 2300    &        &     &D& F20, Dbl.           \\  
   64386&   9.90&   0.61&   14.3&  0.134&   5.67&   9.86&   1.06&BD+19 2646    &  O s   &     &I& CLLA, uvby        \\  
   64765&   8.92&   0.35&   10.9&  0.125&   4.11&   8.86&   0.34&HD 115132      &C s     &     &D& uvby, UBV          \\  
   65040&   9.93&   0.64&   15.4&  0.086&   5.87&   9.77&   0.65&BD+7 2634     &  O s   &     &I& CLLA, uvby          \\  
   65201&   8.85&   0.48&   15.5&  0.093&   4.81&   8.80&   0.45&HD 116064      &C O s   &     &H& GCC, uvby, UBV, Dbl. C C    \\  
   65268&   7.71&   0.44&   19.0&  0.050&   4.11&   7.64&   0.49&HD 116316      &  s     &     &I& uvby, Dbl. S          \\  
   65322&   9.21&   0.44&    9.5&  0.106&   4.10&   9.18&   0.43&HD 116530      &        &     &          \\  
   65940&  10.51&   0.78&   16.9&  0.093&   6.65&  10.41&   0.92&BD-3 3488     &C O     &     &D&UBV, K2V - U1          \\  
   66169&  10.07&   0.68&   15.2&  0.121&   5.97&  10.11&   0.70&BD+2 2691     &  C1    &     &D&  UBV       \\  
   66500&   9.62&   0.48&    8.8&  0.148&   4.34&   9.60&   0.48&BD+37 2434    &  s     &     &I& uvby          \\  
   66815&   8.89&   0.54&   17.6&  0.063&   5.11&   8.83&   0.55&HD 119173      &C       &     &I&UBV, F20          \\  
   66828&  11.23&   0.78&   26.5&  0.080&   8.34&  10.89&   1.33&McC 697        &  O     &     &D& M dwarf          \\  
   67189&   9.91&   0.48&    9.6&  0.148&   4.83&   9.85&   0.48&HD 119530      &C       &     &D&UBV          \\  
   67655&   8.03&   0.65&   40.0&  0.025&   6.04&   7.97&   0.66&HD 120559      &  C1 O s   &     &I& uvby UBV         \\  
   68165&  10.04&   0.76&   21.3&  0.090&   6.69&   9.98&   0.88&BD+7 2721     &  C1 O s   & Y   &D& uvby UBV         \\  
   68452&   8.92&   0.47&   11.9&  0.116&   4.30&   8.88&   0.46&HD 122241      &  C1    & Y   &I&  LCO        \\  
   68870&   9.52&   0.44&    9.9&  0.140&   4.49&   9.51&   0.43&HD 123192      &  C1    & Y   &D&  LCO        \\  
   70152&  11.03&   0.73&   15.1&  0.124&   6.93&  10.60&   0.97&BD+9 2879     &  C1 O s   & Y   &D& UBV, uvby          \\  
   70622&   9.86&   0.46&    8.4&  0.129&   4.47&   9.83&   0.46&HD 126913      &        &     &          \\  
   70681&   9.34&   0.58&   19.2&  0.075&   5.75&   9.28&   0.61&HD 126681      &C O s   & Y   &H& F20, GCC, uvby, UBV         \\  
   70689&   8.58&   0.37&   11.7&  0.090&   3.92&   8.53&   0.37&HD 126488      &C       &     &D&UBV, Dbl. G          \\  
   71886&   8.96&   0.40&   11.5&  0.127&   4.27&   8.94&   0.38&HD 129392      &C       & Y   &D&UBV          \\  
   71887&   8.83&   0.47&   13.4&  0.080&   4.47&   8.79&   0.46&HD 129515      &        &     &          \\  
   71939&   8.90&   0.46&   14.3&  0.104&   4.68&   8.81&   0.48&HD 129510      &C       &     &D&F20, UBV          \\  
   72461&   9.82&   0.38&   10.3&  0.138&   4.88&   9.73&   0.44&BD+26 2606    &  O s   &     &H& uvby, GCC         \\  
   72765&   9.14&   0.42&    8.7&  0.118&   3.83&   9.10&   0.42&HD 131448      &        &     &          \\  
   73614&   8.39&   0.42&   12.7&  0.094&   3.92&   8.33&   0.44&HD 133614      &C s     &     &D&UBV          \\  
   73798&  10.06&   0.56&   15.6&  0.111&   6.02&   9.93&   0.63&HD 133338      &C       &     &H&UBV          \\  
   74078&   8.55&   0.45&   13.3&  0.086&   4.18&   8.48&   0.45&HD 133808      &C W     & Y   &D&Walraven, UBV          \\  
   74590&   8.99&   0.37&    8.8&  0.140&   3.72&   8.93&   0.41&HD 135340      &  s      &     &D& uvby          \\  
   74994&   9.12&   0.39&    8.3&  0.139&   3.72&   9.07&   0.38&HD 135860      &  s      & Y   &D&uvby          \\  
   75432&   9.72&   0.53&   13.2&  0.114&   5.33&   9.64&   0.53&HD 137203      &        &     &          \\  
   75473&   9.71&   0.52&   12.5&  0.139&   5.19&   9.66&   0.52&BD+8 3027     &        &     &          \\  
   75475&   9.74&   0.56&   12.9&  0.125&   5.30&   9.63&   0.57&HD 137028      &        &     &          \\  
   75618&   9.80&   0.56&   13.2&  0.122&   5.41&   9.74&   0.56&HD 137593      &C s     &     &I&UBV, uvby          \\  
   76670&   9.34&   0.50&   12.4&  0.122&   4.81&   9.30&   0.51&HD 139781      &        &     &          \\  
   77208&   9.35&   0.49&   11.2&  0.135&   4.59&   9.28&   0.49&HD 140849      &        &     &          \\  
   77326&  10.58&   0.67&   12.9&  0.135&   6.13&  10.48&   0.85&BD-6 4279     &  O     &     &D& K3V - U1         \\  
   77432&   9.04&   0.43&   10.1&  0.141&   4.06&   8.96&   0.43&HD 141011      &C       &     &D&UBV          \\  
   77508&   9.88&   0.49&    9.6&  0.109&   4.79&   9.80&   0.49&BD+37 2678    &        &     &          \\  
   78143&   8.77&   0.40&   10.1&  0.124&   3.78&   8.74&   0.40&HD 142647      &        &     &          \\  
   78195&   8.36&   0.41&   11.9&  0.087&   3.74&   8.30&   0.43&HD 143108      &  s     &     &D&uvby          \\  
 \hline
 \end{tabular}
 \end{table*}
 \setcounter{table}{0}
 \begin {table*}
  \centering
\caption{Candidate Subdwarf Sample (contd.)}
 \begin{tabular}{rrrrrcrrrcccl} 
  \hline \hline
 HIP & V$_T$ & (B-V)$_T$ & $\pi$ &$ \sigma_\pi \over \pi$ & M$_V$ & V$_l$ & (B-V)$_l$ & ID & Phot & Spec &sd?& Comments \\  
  \hline 
   78251&   9.06&   0.61&   20.8&  0.077&   5.66&   8.96&   0.64&HD 143007      &   O     &     &I& UBV        \\  
   78282&   8.87&   0.41&   11.5&  0.122&   4.17&   8.82&   0.39&HD 143293      &        &     &          \\  
   78296&   9.18&   0.48&   14.1&  0.082&   4.92&   9.07&   0.56&HD 143177      &  s     &     &I& uvby          \\  
   78952&   9.94&   0.50&   12.1&  0.130&   5.35&   9.85&   0.51&HD 144454      & C1     &     &I& UBV        \\  
   79139&   7.73&   0.38&   15.8&  0.070&   3.72&   7.67&   0.42&HD 145184      &  s     &     &I& uvby          \\  
   79856&  10.31&   0.46&    8.9&  0.118&   5.06&  10.23&   0.45&BD+77 622     &        &     &          \\  
   80295&  10.52&   0.77&   20.7&  0.105&   7.10&  10.40&   1.01&LTT 6452       &  C1 O     &     &D& K3V - U1       \\  
   80303&   9.84&   0.54&   13.3&  0.084&   5.46&   9.73&   0.54&BD+47 2335    &        &     &          \\  
   80422&   8.57&   0.40&   10.9&  0.126&   3.75&   8.49&   0.44&HD 147685      &  C1 s W     &     &D&uvby, Walraven   \\  
   80448&   7.33&   0.30&   22.0&  0.128&   4.05&   7.33&   0.64&HD 147633      &   C1 O    &     &D& Dbl. C A         \\  
   80781&  10.86&   0.79&   13.9&  0.116&   6.57&  10.74&   0.80&BD+62 1487    &        &     &  & Dbl. G        \\  
   80789&  10.31&   0.57&   11.8&  0.130&   5.67&  10.24&   0.58&G 169-13       &  O     &     &I& CLLA         \\  
   81013&   8.95&   0.42&   10.1&  0.124&   3.97&   8.90&   0.44&BD+12 3035    &        &     &D& VRI          \\  
   81170&   9.72&   0.69&   20.7&  0.072&   6.30&   9.60&   0.74&HD 149414      &  C1 O s   &     &H& uvby, F20, Dbl.     \\  
   81251&  10.58&   0.64&   12.8&  0.149&   6.11&  10.58&   0.66&BD+23 2961    &        &     &  & Dbl. S        \\  
   81350&   8.86&   0.41&    9.6&  0.119&   3.77&   8.81&   0.40&HD 149823      &        &     &          \\  
   81617&   8.62&   0.32&   10.8&  0.108&   3.78&   8.58&   0.32&HD 150147      &  C1 s O W   &     &D&uvby, Walraven    \\  
   82409&   9.54&   0.52&   12.6&  0.147&   5.05&   9.46&   0.53&HD 151854      &  C1      &     &D& UBV          \\  
   82578&  10.86&   0.73&   15.2&  0.087&   6.78&  11.11&   1.55&G 257-43       &        &     &D& M dwarf          \\  
   83070&   8.90&   0.41&    9.5&  0.141&   3.79&   8.81&   0.43&HD 153150      &   C1     &     & & Dbl         \\  
   83226&  10.16&   0.66&   14.3&  0.139&   5.93&  10.05&   0.77&CPD -51 10106  &C       &     &D&UBV          \\  
   83332&   9.55&   0.40&    8.8&  0.106&   4.26&   9.43&   0.39&BD+82 507     &        &     &          \\  
   83443&   9.26&   0.52&   13.5&  0.116&   4.91&   9.25&   0.53&HD 153959      &C s      &     &D&UBV, uvby          \\  
   84754&  10.23&   0.78&   18.4&  0.106&   6.55&  10.04&   0.84&HD 329788      &C       &     &D&UBV          \\  
   85999&   9.31&   0.48&   12.4&  0.113&   4.78&   9.15&   0.52&HD 159175      &C W     &     & D&Walraven, UBV           \\  
   86183&   9.37&   0.40&    7.4&  0.109&   3.73&   9.38&   0.33&HD 234462      &  O     &     & D & UBV         \\  
   86393&   9.60&   0.50&   11.3&  0.137&   4.87&   9.51&   0.51&HD 159750      &C       &     &D&UBV          \\  
   86536&   9.23&   0.42&   11.0&  0.117&   4.44&   9.15&   0.40&HD 159753      &C       &     &D&UBV          \\  
   88066&   8.87&   0.40&    9.8&  0.141&   3.83&   8.83&   0.39&HD 163398      &C       &     &D&UBV          \\  
   88084&   9.34&   0.56&   18.1&  0.077&   5.62&   9.19&   0.62&HD 164139      &  C1      &     &I&UBV          \\  
   88231&  10.15&   0.52&   11.1&  0.149&   5.38&   9.98&   0.52&BD-18 4758    &   C1   &     &D&UBV          \\  
   88648&  10.28&   0.71&   15.4&  0.133&   6.22&  10.21&   0.61&HD 321320      &  C1 O s   &     &H & uvby          \\  
   88955&   9.63&   0.58&   15.1&  0.096&   5.52&   9.42&   0.58&HD 165898      &    C1    &     &D & Dbl. S, UBV         \\  
   89053&  10.49&   0.71&   16.1&  0.116&   6.53&  10.29&   0.84&CPD -50 10562  &C       &     &D&UBV          \\  
   89215&  10.37&   0.75&   17.0&  0.112&   6.52&  10.37&       &BD+5 3640     &  C1 O s   &     &H& CLLA, uvby         \\  
   89396&   8.05&   0.43&   21.0&  0.054&   4.66&   7.99&   0.43&HD 167038      &  C1      &     & D & Dbl. S, UBV       \\  
   89554&   8.26&   0.44&   16.1&  0.065&   4.29&   8.22&   0.44&HD 166913      &C s O     &     &H&  GCC, UBV, uvby       \\  
   89734&  11.10&   0.76&   17.0&  0.084&   7.24&  10.98&   0.78&LTT 18473      &        &     &          \\  
   89877&  10.72&   0.65&   15.3&  0.077&   6.64&  10.56&   0.65&G 259-28       &        &     &          \\  
   89932&   9.19&   0.47&   11.1&  0.119&   4.42&   9.13&   0.47&HD 168375      &C W     &     & D&Walraven, UBV           \\  
   90616&  10.34&   0.60&   11.8&  0.129&   5.70&  10.28&   0.62&BD+4 3763     &   C1     &     & I &UBV         \\  
   90724&   9.61&   0.54&   34.3$^*$&  0.145&   7.28&   9.09&   0.26&HD 170368      &C       &     &D& UBV, Sp=A7, Dbl. C C         \\  
   91489&  10.85&   0.68&   37.5&  0.046&   8.72&  10.85&       &LP 355-13      &        &     & & Dbl. S         \\  
   92277&  10.59&   0.62&   14.1&  0.137&   6.33&  10.34&   0.70&G 21-25        &C O s   &     &D&UBV, CLLA          \\  
   93031&   9.31&   0.48&   12.7&  0.136&   4.84&   9.28&   0.49&HD 175479      &C W     &     &D&Walraven, UBV            \\  
   93341&  10.22&   0.61&   15.6&  0.117&   6.19&  10.10&   0.70&HD 230409      &  O s   &     &I& uvby          \\  
   93556&   8.96&   0.42&   11.1&  0.064&   4.19&   8.89&   0.40&HD 177672      &        &     &          \\  
   93725&   9.30&   0.42&    8.5&  0.118&   3.93&   9.23&   0.41&BD+31 3455    &        &     &          \\  
   94347&   7.31&   0.44&   22.3&  0.087&   4.05&   7.26&   0.44&HD 174930      &C s      &     &I&UBV, uvby, Dbl. X          \\  
   94704&  11.10&   0.74&   11.4&  0.148&   6.38&  11.09&   0.66&G 207-23       &  O     &     &H& UBV          \\  
 \hline
 \end{tabular}
 \end{table*}
 \setcounter{table}{0}
 \begin {table*}
  \centering
\caption{Candidate Subdwarf Sample (contd.)}
 \begin{tabular}{rrrrrcrrrcccl} 
  \hline \hline
 HIP & V$_T$ & (B-V)$_T$ & $\pi$ &$ \sigma_\pi \over \pi$ & M$_V$ & V$_l$ & (B-V)$_l$ & ID & Phot & Spec &sd?& Comments \\  
  \hline 
   95031&  10.02&   0.76&   21.7&  0.069&   6.70&   9.90&   0.88&BD-3 4564     &C O     &     &D&UBV. K3V - U1          \\  
   95190&   9.57&   0.49&   11.6&  0.147&   4.91&   9.45&   0.56&HD 181376      &C       &     &H&UBV          \\  
   95341&  10.19&   0.35&    6.7&  0.142&   4.33&  10.08&   0.34&BD+67 1148    &        &     &          \\  
   95429&  11.31&   0.80&   33.1&  0.091&   8.90&  11.19&   0.80&               &        &     & & Dbl. S         \\  
   95800&   8.81&   0.42&   11.0&  0.105&   4.02&   8.78&   0.41&HD 182163      &C O     & Y   &I&UBV          \\  
   95924&   9.46&   0.47&   10.1&  0.136&   4.49&   9.36&   0.46&HD 183705      &        &     &  & Dbl. G        \\  
   95996&  10.27&   0.47&    9.6&  0.136&   5.17&  10.22&   0.51&BD+35 3659    &  O s   &     &H& uvby         \\  
   96043&   9.80&   0.57&   13.8&  0.121&   5.49&   9.63&   0.58&HD 183639      &C       & Y   &D&UBV, Dbl. G          \\  
   97110&   8.75&   0.45&   12.3&  0.105&   4.20&   8.70&   0.44&HD 186533      &        &     &          \\  
   97127&   9.26&   0.40&    8.5&  0.140&   3.91&   9.22&   0.40&HD 186214      &C       & Y   &D&UBV          \\  
   97174&  10.40&   0.76&   18.7&  0.087&   6.76&  10.32&   0.77&G 125-48       &        &     &          \\  
   97463&   9.67&   0.54&   13.4&  0.120&   5.30&   9.57&   0.62&HD 186755      &C       & Y   &I&UBV          \\  
   98020&   8.92&   0.56&   25.3&  0.046&   5.94&   8.83&   0.60&HD 188510      &  O s   &     &H& F20, GCC, CLLA, uvby      \\  
   98322&  10.27&   0.69&   17.2&  0.062&   6.45&  10.25&   0.70&BD+80 642     &        &     &          \\  
   99267&  10.15&   0.51&   12.0&  0.094&   5.55&  10.11&   0.51&BD+42 3607    &  O    &     &H& GCC         \\  
  100207&   8.85&   0.42&   10.7&  0.125&   3.99&   8.77&   0.42&HD 192863      &C       & Y   &D&UBV          \\  
  100568&   8.72&   0.54&   22.9&  0.054&   5.51&   8.65&   0.55&HD 193901      &C O s   &     &H& F20, GCC, UBV        \\  
  101103&   9.47&   0.44&    9.7&  0.149&   4.41&   9.46&   0.38&HD 194702      &C O    & Y   &I&UBV          \\  
  101650&   9.36&   0.44&    9.0&  0.146&   4.14&   9.33&   0.44&HD 196048      &C       & Y   &D&UBV          \\  
  101814&  10.34&   0.74&   16.2&  0.062&   6.39&  10.25&   0.74&BD+76 810     &        &     &          \\  
  101892&   9.46&   0.47&    9.3&  0.144&   4.30&   9.43&   0.47&HD 196263      &C       & Y   &D&UBV          \\  
  101987&   9.22&   0.43&    9.9&  0.138&   4.20&   9.16&   0.43&HD 196682      &C       & Y   &D&UBV          \\  
  101989&  10.66&   0.73&   13.4&  0.126&   6.30&  10.60&   0.74&BD+40 4272    &        &     &          \\  
  103269&  10.37&   0.57&   14.2&  0.102&   6.14&  10.28&   0.59&G 212-7        &  O    &     &H& AFG, F20, CLLA         \\  
  103287&   9.21&   0.73&   26.8&  0.103&   6.35&   9.07&   0.75&BD+13 4571    &        & Y   &D  & Dbl. X, LCO        \\  
  103714&  10.19&   0.54&   12.8&  0.137&   5.73&  10.12&   0.55&CPD -34 8839   &C       & Y   &D&UBV          \\  
  104289&  10.29&   0.65&   14.6&  0.126&   6.11&  10.25&   0.68&HD 200855      &C O s    & Y   & I &uvby, UBV         \\  
  105773&   7.99&   0.46&   17.5&  0.048&   4.20&   7.95&   0.45&HD 204093      &   s     &     &D& uvby          \\  
  105849&  10.86&   0.80&   14.9&  0.080&   6.73&  10.60&   0.80&LP 75-16       &        &     &          \\  
  106204&  10.81&   0.78&   22.9&  0.082&   7.61&  10.67&   1.22&CD -24 16689   &  C1 O     & Y   &D& K7V - U1         \\  
  106904&  10.29&   0.64&   13.0&  0.146&   5.86&  10.19&   0.64&HD 205682      &C       & Y   &I&UBV          \\  
  106924&  10.53&   0.55&   15.2&  0.080&   6.44&  10.36&   0.55&BD+59 2407    &  O s   &     &H& GCC, CLLA         \\  
  107873&   9.36&   0.48&   10.1&  0.138&   4.38&   9.31&   0.48&HD 207691      &C       & Y   &D&UBV          \\  
  108006&   9.18&   0.41&    9.3$^F$&  0.139&   4.01&   9.12&   0.41&HD 207792      &C       & Y   &D&UBV          \\  
  108095&   8.61&   0.52&   19.7&  0.058&   5.09&   8.52&   0.53&HD 208068      &C s     & Y   &I& UBV, uvby          \\  
  108598&   9.73&   0.71&   23.0$^F$&  0.060&   6.53&   9.57&   0.71&HD 208740      &C s     & Y   &I& UBV, uvby          \\  
  108655&   8.92&   0.40&   10.1&  0.138&   3.93&   8.86&   0.39&BD+2 4457     &C       & Y   &D&UBV          \\  
  108836&  11.25&   0.62&   24.0&  0.119&   8.15&  10.99&   1.14&LTT 8824       &C       &     &D& K dwarf          \\  
  109067&   9.61&   0.66&   21.5$^F$&  0.074&   6.27&   9.55&   0.67&BD+11 4725    &C O s   & Y   &I/H& F20, CLLA, uvby    \\  
  109869&   9.72&   0.52&   13.8&  0.111&   5.43&   9.57&   0.52&HD 211127      &C O     & Y   &D&UBV          \\  
  109945&   9.23&   0.42&    9.4&  0.138&   4.09&   9.18&   0.41&BD+17 4720    &C       & Y   &D&UBV          \\  
  110621&   9.06&   0.42&    9.7&  0.130&   3.99&   9.01&   0.41&HD 212457      &        & Y   &I&LCO          \\  
  110776&   9.80&   0.79&   21.7$^F$&  0.067&   6.48&   9.70&   0.82&HD 212753      &C O s   & Y   &I& CLLA, uvby         \\  
  111374&   7.90&   0.41&   15.1&  0.067&   3.80&   7.86&   0.42&HD 213763      &C s     & Y   &D&uvby, UBV          \\  
  111426&   9.39&   0.46&   11.0&  0.109&   4.59&   9.35&   0.46&HD 213670      &C       & Y   &D&UBV          \\  
  111871&  10.53&   0.74&   15.5&  0.117&   6.49&  10.47&   1.31&BD+10 4791    &C O s   &     &D& UBV, uvby, CLLA         \\  
  112384&   9.06&   0.43&    9.6&  0.142&   3.97&   9.02&   0.43&HD 215437      &C       & Y   &I&UBV          \\  
  112389&  11.05&   0.80&   26.0&  0.082&   8.12&  10.76&   1.59&AC+18 1061     &C O     &     &D& M dwarf          \\  
  113430&   8.10&   0.46&   16.8&  0.052&   4.22&   8.05&   0.48&HD 216924      &C s     & Y   &D&UBV, uvby          \\  
  113542&   8.79&   0.40&   10.8&  0.101&   3.97&   8.75&   0.39&HD 217337      &s O     & Y   &D& uvby          \\  
 \hline
 \end{tabular}
 \end{table*}
 \setcounter{table}{0}
 \begin {table*}
  \begin{center}
\caption{Candidate Subdwarf Sample (contd.)}
 \begin{tabular}{rrrrrcrrrcccl} 
  \hline \hline
 HIP & V$_T$ & (B-V)$_T$ & $\pi$ &$ \sigma_\pi \over \pi$ & M$_V$ & V$_l$ & (B-V)$_l$ & ID & Phot & Spec &sd?& Comments \\  
  \hline 
  113868&  10.18&   0.68&   14.6&  0.123&   6.01&  10.10&   0.68&HD 217740      &C       & Y   &D&UBV, Dbl. S          \\  
  114125&  10.00&   0.57&   13.4&  0.119&   5.63&   9.89&   0.57&BD+33 4648    &        &     &          \\  
  114271&   8.30&   0.44&   14.3&  0.084&   4.08&   8.25&   0.41&HD 219181      &C       & Y   &H& F20, UBV          \\  
  114299&   9.20&   0.42&    9.6&  0.131&   4.12&   9.16&   0.41&BD+43 4402    &        &     &          \\  
  114487&   8.47&   0.40&   12.0&  0.094&   3.85&   8.40&   0.39&HD 218810      &C       & Y   &I&UBV          \\  
  114627&   8.91&   0.47&   14.0&  0.105&   4.64&   8.86&   0.47&HD 219101      &C       &     &D&UBV          \\  
  114735&   8.57&   0.40&   11.5&  0.111&   3.88&   8.51&   0.40&HD 219242      &C       & Y   &I&UBV          \\  
  114837&   9.96&   0.47&   10.1&  0.147&   4.98&   9.89&   0.58&HD 219369      &C s     & Y   &I& UBV, uvby          \\  
  115031&   8.35&   0.47&   15.7$^F$&  0.085&   4.33&   8.28&   0.47&HD 219678      &C       & Y   &D&UBV, Dbl. O         \\  
  115194&   8.97&   0.77&   31.5&  0.038&   6.47&   8.87&   0.81&HD 219953      &  s     & Y   &I& uvby          \\  
  115361&   8.17&   0.44&   15.3&  0.072&   4.10&   8.13&   0.44&HD 220164      &C s     & Y   &D&uvby, UBV          \\  
  116437&   9.94&   0.40&    9.3&  0.145&   4.79&   9.80&   0.57&CPD -43 9761   &C       & Y   &D&UBV          \\  
  117121&  11.60&   0.58&   19.9&  0.143&   8.10&  11.60&       &BD-21 6469    &C       & Y   &D& M dwarf          \\  
  117242&   8.84&   0.43&   10.8&  0.110&   4.01&   8.80&   0.40&HD 236211      &  O     &     & D & UBV         \\  
  117823&   9.42&   0.47&   11.0&  0.119&   4.62&   9.33&   0.50&HD 223923      &C       & Y   &I&UBV          \\  
  117882&   9.27&   0.48&   11.9$^*$&  0.122&   4.64&   9.24&   0.45&HD 224040      &C s      & Y   &D&uvby, UBV          \\  
  118165&   9.19&   0.42&    9.2&  0.141&   4.02&   9.14&   0.42&HD 224463      &C       & Y   &D&UBV          \\  
 \hline
 \end{tabular}
\end{center}
$^F$ in column 5 indicates that the goodness of fit statistic, $|F|$, listed in the Hipparcos
catalog exceeds 2.5, and that the astrometric solution is not reliable. \\
$^*$ in column 5 indicates that the parallax given in the Hipparcos catalogue has been revised (see \S 2.2). \\
Column 10 lists the available photometry: s, Str\"omgren; W, Walraven; O, UBV(RI) (literature); C \& C1, 
UBVRI (SAAO), C1 indicating single epoch observations. \\ 
Column 12 (sd?) gives our final abundance classification: \\
D $\equiv$ [m/H] $\ge -0.3$; I $\equiv$ -0.3 $>$ [m/H] $\ge -1$; H $\equiv$ [m/H] $\le -1$ \\
Column 13 identifies known double stars and gives the  basis for the abundance classification: \\
UBV - $\delta_{0.6}$/[Fe/H]  calibration (\S4); uvby - Str\"omgren data (\S3.1); Walraven photometry
(\S3.2); \\
or spectroscopic measurements, referenced as follows: \\
AFG - Axer {\sl et al.}, 1994; CLLA - Carney {\sl et al.}, 1994; F20 - Fulbright, 2000; \\
GCC - Gratton {\sl et al.}, 1997a;  LCO - this paper; RN - Ryan \& Norris, 1991; U1 - Upgren, 1972. \\
Stars classed as double in the Hipparcos catalogue are flagged as Dbl. \\
Dbl. G indicates the presence of an acceleration term in the Hipparcos solution, interpreted as motion in
an unresolved binary system; \\
Dbl. C identifies separately resolved components, solutions with quality A-D; \\
Dbl. O indicates full orbital solutions; \\
Dbl. X indicates problems with both single-star and binary Hipparcos solutions; \\
Dbl. S flags suspected double stars. See Hipparcos catalogue for full details.
 \end{table*}

For present purposes, since we are concerned with relatively blue stars, we
ignore the systematic colour difference between the Tycho and Johnson systems. Interpolating
between the M5 and 47 Tuc CMDs, we define an (M$_V$, (B-V)) relation corresponding
approximately to [Fe/H] = -1. We have identified all Hipparcos stars
with ${\sigma_pi\ \over \pi} < 0.15$ lying blueward
of that relation, truncating the sample at (B-V)$_T = 0.8$. We also impose a cutoff
at M$_V < 4$, where the evolved halo colour-magnitude
relations approach the disk main sequence. Figure 2 plots the Tycho photometry for the 317 stars which
match these criteria\footnote {The Hipparcos catalogue includes a number of stars lacking Tycho
photometry. Those stars are obviously not included in our present exercise. Nor
have we eliminated stars with significant formal errors in the Tycho photometry.}
Our aim is to sift through this sample, using additional data from the 
literature, together with our own observations, 
to eliminate interlopers and identify {\sl bona fide} subdwarfs
for future detailed study. 

Table 1 lists the Hipparcos astrometry and photometry for the  candidate subdwarfs plotted in 
Figure 2. We list both the Hipparcos catalogue number and a more conventional designation. In
addition to the Tycho photometry, we include the BV literature data cited in the catalogue. A
cursory inspection shows that several of the latter measurements are in significant disagreement with the
Tycho data. Indeed, as discussed further below, 
the majority of those stars prove to have entered the present sample due
to errors in V$_T$, B$_T$.

The sample outlined in Table 1 forms the starting point for the current investigation.
A number of stars have spectroscopic abundance measurements, as described in 
\S3.
Subsequent sections outline supplementary photometric and spectroscopic observations,
drawn both from the literature (based on the extensive cross-referencing in
the SIMBAD database) and from our own observations. Column 10 in Table 1 indicates which stars have
accurate ground-based photometry (in addition to the literature-derived V, (B-V) values cited in the Hipparcos
catalogue, which we denote as V$_l$ and (B-V)$_l$ in Table 1): 
s indicates Str\"omgren photometry (\S 4.1); W, Walraven photometry (\S 4.2); 
O, Johnson/Cousins UBVRI literature data (\S 5.2); and C \& C1, Johnson/Cousins UBVRI photometry from
SAAO observations, where C1 indicates observations at only one epoch (\S 5.3). 
Column 11 flags stars with spectroscopic observations from Las Campanas (\S 6). 
Column 12 summarises our main conclusions by identifying those stars confirmed as likely to
have [m/H]$< -1$ (`H', halo subdwarfs); stars of probable intermediate abundance, $-1 > {\rm [m/H]} > -0.3$ (`I');
and stars likely to have near-solar abundances (`D', disk dwarfs). Finally, column 13 cites the relevant
references for the adopted classification.

\begin{figure}
\psfig{figure=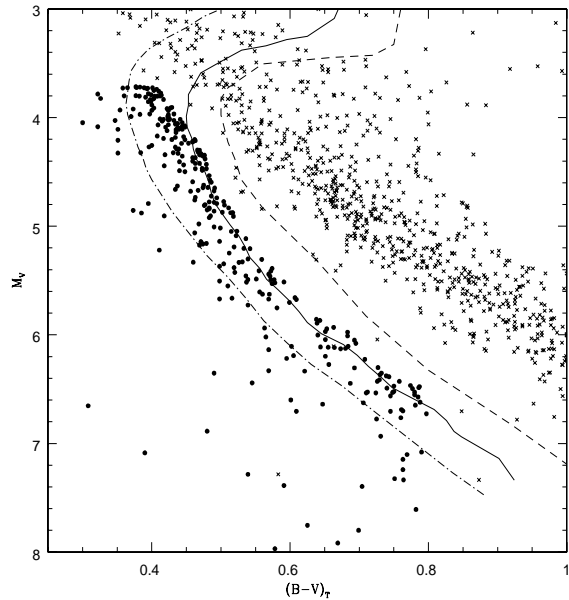,height=8cm,width=8cm
,bbllx=8mm,bblly=57mm,bburx=205mm,bbury=245mm,rheight=9cm}
\caption{The (M$_V$, (B-V)$_T$) colour-magnitude diagram: crosses mark
stars from the Hipparcos catalogue with ${\sigma_\pi \over \pi} < 15\%$ and $\pi > 32$ mas. 
The dashed line outlines the schematic main-sequence and giant branch
in the metal-rich globular, 47 Tucanae (at (m-M)=13.55); the solid line marks the relation for the intermediate
abundance cluster, M5 (at (m-M)=14.5); and the dashed line plots the fiducial relation for the metal-poor 
cluster, NGC 6397 (at (m-M)=12.14). Solid points mark candidate metal-poor subdwarfs.}
\end{figure}

\subsection { {\sl A priori} exclusions}

The stars listed in Table 1 have been selected based solely on parameters given in the
published Hipparcos catalogue. However, auxiliary criteria can be used to eliminate
a number of candidates at the outset. First, five stars are known to be members of
nearby stars clusters: HIP 17481 (HD 23290) and 17497 (HD 23289) are both main-sequence
members of      the Pleiades cluster; HIP 20527 (VR 9),
21261 (Leiden 65) and 22177 (Leiden 119) are known to be M dwarfs in
the Hyades cluster. In the case of the Pleiades stars, the Hipparcos 
parallax is overestimated by $\sim 2$ mas, while   
the (B-V)$_T$ colours measured for all three Hyads are all $\sim 0.6$ magnitudes too blue.

Second, the formal Hipparcos astrometric solutions are known to be unreliable
if the target star is a close double. Thirty-nine stars from Table 1 are identified as double stars or
suspected binaries in the Hipparcos catalogue. Of the systems with resolved components (flagged
as Dbl. C in Table 1), two are listed as having solutions of quality C (65201 and 90216), and two
as quality D (10529 and 44436).  We have not excluded those systems from consideration, but will
interpret the results accordingly. 

In some cases, more accurate astrometry can be derived by re-analysing the individual
Hipparcos measurements. Both Falin \& Mignard (1999) and Fabricius \& Makarov (2000a)
have undertaken such an exercise, and their results show that trigonometric parallaxes
of at least two stars listed in Table 1 require revision: re-analysing data for HIP 21000
(BD +4 701A) indicates that $\pi_{Hip}$ is overestimated by 76 mas, while the
catalogued parallax of HIP 90724 (HD 170368) should be reduced by 27 mas. 
In both cases, the revised parallaxes are $\sim 7$ mas, with uncertainties of $\sim15\%$.
As discussed later, accurate UBVRI photometry is consistent with the larger distances
and higher luminosities, and both stars are clearly ruled out as possible metal-poor subdwarfs.

Finally, the Hipparcos catalogue cites a  goodness of fit statistic (which we denote $|F|$) 
for the astrometric solution 
obtained for most stars (stars flagged as Dbl. X are exceptions). Solutions where
$|F|$ is greater than 2.5 can generally be regarded as suspect. Seventeen stars
in Table 1 exceed that limit, and these stars can be eliminated as potential subdwarf
calibrators. 

\section {Previous spectroscopic observations}

A sizeable minority of the sample listed in Table 1 are known metal-poor subdwarfs or
subgiants, with previous spectroscopically-based abundance determinations. 
Those stars are useful benchmarks in assessing the accuracy of both photometric abundance
estimates and of our own spectrophotometry (\S6). Table 2 summarises those
data, drawn from seven main sources: low-resolution spectroscopy by Ryan \& Norris (1991 - RN); 
Carney {\sl et al.'s} (1994 - CLLA) analysis of high-resolution data centred on the Mgb feature;
and conventional high-resolution spectroscopic analysis by Axer {\sl et al.} (1994 - AFG), 
Gratton {\sl et al.} (1997a - GCC), Ryan \& Deliyannis (1998 - RD),
Clementini {\sl et al.} (1998 - Cl1) and Fulbright (2000 - F20).
Several well known halo stars, such as HD 19445, 64090 and 84937, have numerous
abundance determinations; our tabulation is representative, rather than exhaustive. 
As discussed by Reid (1998) and Clementini {\sl et al.} (1999), systematic differences exist
between the abundance scales used in some of these analyses; in particular, CLLA and RN tend to
derive lower abundances for metal-poor ([m/H]$<-1$) stars. Given our current aims, 
however, we have not attempted to adjust all measurements onto a single, self-consistent scale. 

 \begin {table}
 \begin{center}
\caption{Stars with spectroscopic abundance estimates}
 \begin{tabular}{rrlrrl}
  \hline \hline 
 HIP & [m/H] & ref. & HIP & [m/H] & ref. \\
\hline

    5004&  -1.02& RN  &   50965&  -0.39& RN   \\
    7459&  -1.17& RN  &   51769&  -0.65& CLLA \\
    8130&  -0.64& CLLA&   51769&  -1.25& RN   \\
    8558&  -1.10& RN  &   53070&  -1.38& GCC  \\
   12579&  -0.86& CLLA&   53070&  -1.55& F20  \\
   12579&  -0.78& AFG &   53070&  -1.34& Cl1  \\
   14594&  -1.87& AFG &   54641&  -1.04& AFG  \\
   14594&  -1.89& GCC &   54641&  -1.01& RN   \\
   14594&  -2.13&  F20&   55790&  -1.56& AFG  \\
   14594&  -1.97&  Cl1&   57360&  -1.20& AFG  \\
   16404&  -1.92& AFG &   57360&  -1.22& RN   \\
   16404&  -1.92& GCC &   57450&  -1.26& GCC  \\
   17085&  -0.22& F20 &   57939&  -1.22& GCC  \\
   19007&  -0.62& F20 &   57939&  -1.46& F20  \\
   19797&  -1.33& AFG &   57939&  -1.30& Cl1  \\
   19797&  -1.57& GCC &   59750&  -0.78& F20  \\
   19797&  -1.68& F20 &   60632&  -1.38& AFG  \\
   21000&  -0.16& F20 &   60632&  -1.55& GCC  \\
   21586&  -0.91& F20 &   60632&  -1.65& F20  \\
   21609&  -1.76& F20 &   63063&  -0.48& CLLA \\
   21767&  -0.44& F20 &   63970&  -0.09& F20  \\
   22246&  -0.22& CLLA&   64386&  -0.84& CLLA \\
   22246&  -0.38& F20 &   65040&  -0.82& CLLA \\
   22632&  -1.59& F20 &   65201&  -1.86& GCC  \\
   24316&  -1.44& GCC &   66815&  -0.64& F20  \\
   24316&  -1.71& F20 &   70681&  -1.25& F20  \\
   26452&  -0.89& CLLA&   71886&  -0.40& F20  \\
   26676&  -1.17& CLLA&   71887&  -0.49& F20  \\
   26676&  -1.02& RN  &   71939&  -0.37& F20  \\
   26688&  -0.60& F20 &   72461&  -2.07& AFG  \\
   28188&  -0.62& F20 &   72461&  -2.29& GCC  \\
   31639&  -0.62& F20 &   80789&  -0.96& CLLA \\
   34146&  -0.40& F20 &   81170&  -1.26& F20  \\
   34548&  -0.46& F20 &   81170&  -1.14& Cl1  \\
   36491&  -0.81& CLLA&   89215&  -1.36& CLLA \\
   36491&  -1.02& AFG &   89554&  -1.44& AFG  \\
   36491&  -0.85& RN  &   89554&  -1.44& GCC  \\
   36491&  -0.93& F20 &   92277&   0.01& CLLA \\
   36491&  -0.88& Cl1 &   98020&  -1.62& AFG  \\
   36818&  -0.75& RN  &   98020&  -1.38& GCC  \\
   38541&  -1.69& AFG &   98020&  -1.67& F20  \\
   38541&  -1.60& GCC &   99267&  -2.01& AFG  \\
   38541&  -1.79& F20 &  100568&  -1.00& GCC  \\
   38541&  -1.54& Cl1 &  100568&  -1.17& F20  \\
   40778&  -1.70& F20 &  103269&  -1.78& CLLA \\
   44116&  -0.58& F20 &  103269&  -1.60& AFG  \\
   44124&  -1.90& CLLA&  103269&  -1.81& F20  \\
   44124&  -2.20& RN  &  106924&  -1.91& CLLA \\
   44124&  -1.96& F20 &  106924&  -1.62& Cl1 \\
   46120&  -2.10& RD  &  109067&  -0.97& F20  \\
   47640&  -0.08& F20 &  109067&  -0.95& CLLA \\
   48146&  -0.05& F20 &  110776&  -0.46& CLLA \\
   48152&  -2.07& GCC &  111871&  -0.50& CLLA \\
   48152&  -2.08& F20 &  114271&  -1.80& F20  \\

\hline
 \end{tabular}
 \end{center}
See text for references
 \end{table}

\section {Intermediate-band photometric observations}

\subsection {Str\"omgren photometry}

\begin{figure}
\psfig{figure=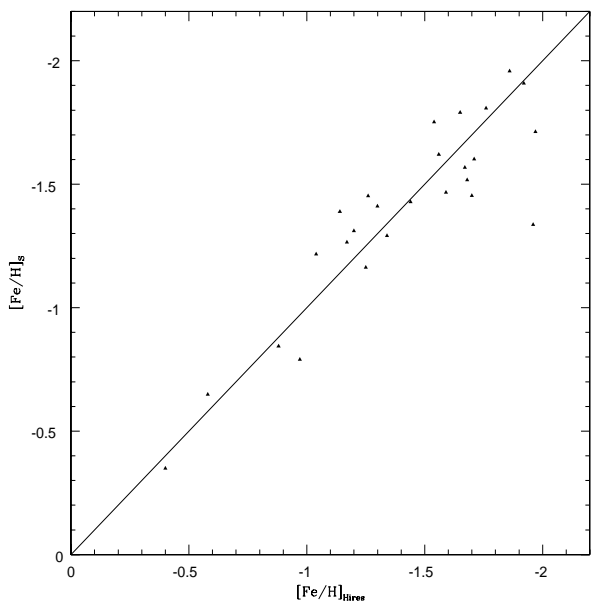,height=8cm,width=8cm
,bbllx=8mm,bblly=57mm,bburx=205mm,bbury=245mm,rheight=9cm}
\caption{A comparison of abundances derived using Schuster \& Nissen's calibration of Str\"omgren
photometry against results from conventional, high-resolution spectroscopic analysis. }
\end{figure}
\begin{figure}
\psfig{figure=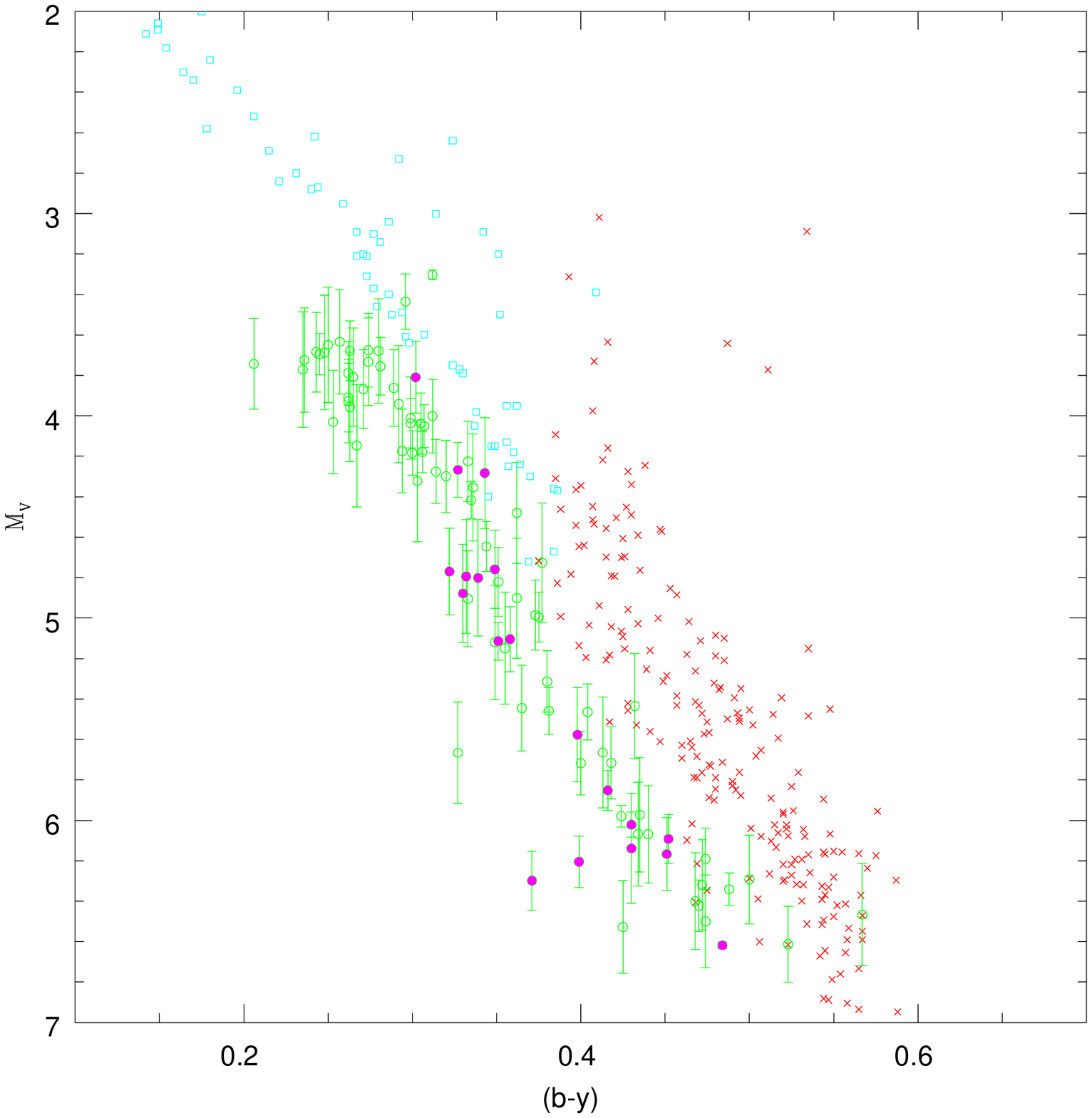,height=8cm,width=8cm
,bbllx=8mm,bblly=57mm,bburx=205mm,bbury=245mm,rheight=9cm}
\caption{The (M$_V$, (b-y) diagram: crosses mark stars with [m/H]$< -0.25$ from Schuster \& Nissen (1988)
and Schuster {\sl et al.} (1993); points with errorbars plot data for stars from Table 3, where
solid points identify stars with [m/H]$< -1.3$. 
Open squares  mark stars in the Hyades cluster (Crawford \& Perry, 1969). }
\end{figure}
The {\sl uvby} system devised by Str\"omgren (1966) provides an effective means of 
estimating the physical properties of F- and G-type stars. The ($b-y$) colour is correlated
with effective temperature, while the $m_1$ index, defined as
\begin{displaymath}
m_1 \ = \ (v-b) \ - \ (b-y) 
\end{displaymath}
measures metallicity by determining the relative line-blanketing in blue and 
ultraviolet passbands. Finally, the $c_1$ index, defined as
\begin{displaymath}
c_1 \ = \ (u-v) \ - \ (v-b)
\end{displaymath}
is gravity sensitive, allowing separation of main-sequence dwarfs and subgiants.

We have located  Str\"omgren photometry of
97 stars from Table 1, notably from  Hauck \& Mermilliod's (1998) {\sl uvby} catalogue.
Table 3 lists the relevant data. together with 
the source of the photometry. We have used the
relations derived by Schuster \& Nissen (1989) to derive abundance estimates for
those stars with measured $m_1$ and $c_1$ indices. As discussed by Reid (1998), there are
systematic offsets between high-resolution spectral analyses and this calibration, partially tied to the
revision in the value for the solar iron abundance (see Bi\'emont {\sl et al.}, 1991). However, the 
discrepancies are generally less than 0.2 dex., as illustrated in Figure 3.

\begin{figure}
\psfig{figure=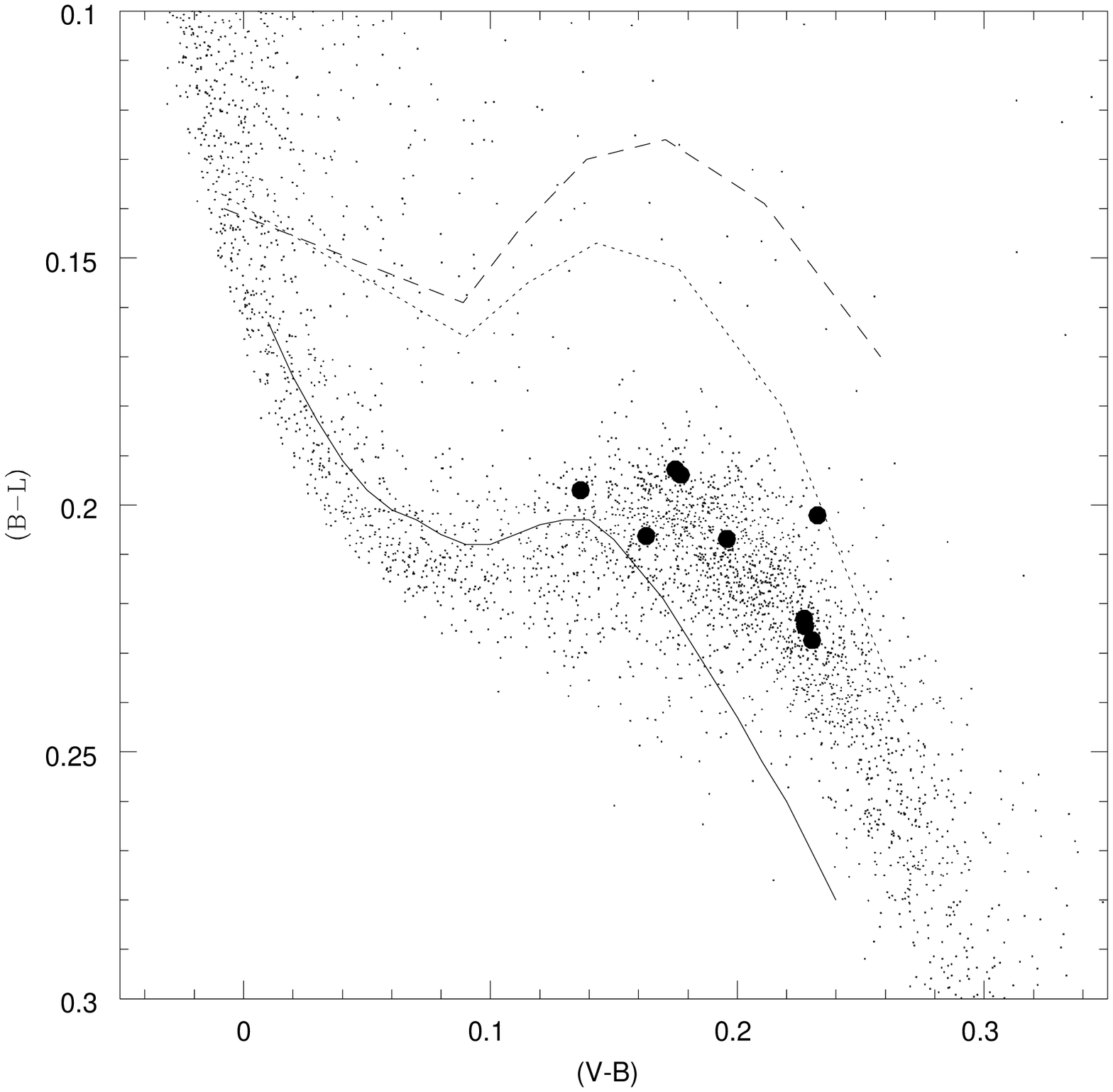,height=8cm,width=8cm
,bbllx=8mm,bblly=57mm,bburx=205mm,bbury=245mm,rheight=9cm}
\caption{The Walraven ((B-L), (V-B)) two-colour diagram from de Geus {\sl et al.} (1991). The
solid line plots the Hyades sequence, the dotted line marks the predicted location
of [m/H]$=-1.0$ subdwarfs, and the dash-dot line plots the expected [m/H]$=-2.0$ sequence
(from Lub \& Pel, 1977). 
The nine stars listed in Table 4 are plotted as solid points, and none lie above the
[m/H]=-1.0 sequence outlined by the stellar models. }
\end{figure}

We have combined Hipparcos parallax measurements with the V-band
data listed in Table 3 to derive absolute visual magnitudes. Figure 4 plots
the resulting (M$_V, (b-y)$) colour-magnitude diagram. Since we are considering each 
programme star individually, we have not applied Lutz-Kelker corrections. 
As a reference disk sequence, we plot data for stars with Str\"omgren abundances [m/H]$ > -0.25$ 
(from Schuster \& Nissen, 1988), together with observations of members of the 
Hyades cluster (Crawford \& Perry, 1969). 
Distances for the latter stars are
derived from their proper motions and the Hipparcos determination of
the average convergent point (Perryman {\sl et al.}, 1998).

Several stars deserve special mention: 
the bluest star listed in Table 3 is HIP 10360, which is 
classified as a field blue straggler by Bond \& MacConnell (1980); the  goodness of fit statistic,
$|F|$, is 3.51, indicating unreliable Hipparcos astrometry. 
Two other stars listed in Table 4, HIP 36818 and 109067, are moderately metal-poor, but lack accurate parallax
data.  The bluest star plotted in Figure 4 is HIP 81617, which  lies
close to the Galactic Plane and has near-solar abundance. As 
discussed further in section 4.2, 
this is probably a reddened, early-type main-sequence star. 
Finally, the Hipparcos catalogue lists identical parallax measurements for HIP 44436 and 44435.
We include Str\"omgren photometry of HIP 44435 (HD 78084) in Table 3; however, 
as discussed further in \S6, spectroscopy shows that the two stars are unrelated.

Considering the other stars in Table 3, nineteen have photometric abundances [Fe/H]$_S < -1.4$. 
The latter are plotted as solid points in Figure 4, and many lie closer to the disk main sequence
than might be expected. Several 
are known binaries, notably HIP 16404 (BD +66 268), 21609 (HD 29907) and HIP 38541 (HD 64090).
The four stars which fall significantly below the main group of subdwarfs
are HIP 24316, 46120, 109067 and 117882. Only HIP 24316 (HD 34328) and
HIP 46120 (Gl 345) have photometric abundances [Fe/H]$_S< -1.4$. 
Both Str\"omgren photometry and our spectroscopic observations (\S6.2) 
of the last star, HIP 117882 (HD 224040; (b-y)=0.327), indicate 
near-solar abundance. The star is  double (CCDM 23546-2302), and the Hipparcos double-star annex
lists a parallax of 8.6 mas for each component (as opposed to 11.9 mas in the main catalogue), 
but it seems likely that even this value is an overestimate. 

 \setcounter{table}{2}
 \begin {table*}
 \begin{center}
\caption{Str\"omgren photometry of candidate subdwarfs}
 \begin{tabular}{rrrrrcrclll} 
  \hline \hline
 HIP & V & b-y & m$_1$ & c$_1$ & M$_V$ & [m/H] &ref &Comments \\  
  \hline 
     435&  9.290 &  0.336 &  0.100 &  0.387 &  4.354 & -0.97& O1  \\ 
    3139&  9.269 &  0.292 &  0.130 &  0.412 &  3.941 & -0.49& HM  \\ 
    4981&  9.110 &  0.380 &  0.147 &  0.356 &  5.313 & -0.58& HM  \\ 
    5004& 10.258 &  0.472 &  0.248 &  0.169 &  6.319 & -1.13& S1  \\ 
    5896&  4.856 &  0.312 &  0.168 &  0.444 &  3.303 &  0.02& HM  \\  
    7459& 10.123 &  0.365 &  0.092 &  0.231 &  5.445 & -1.09& S1  \\ 
    7687&  9.364 &  0.470 &  0.309 &  0.250 &  6.422 & -0.34& Ol1 \\ 
    8130& 10.170 &  0.432 &  0.203 &  0.270 &  5.435 & -0.47& S2  \\ 
   10360&  8.530 &  0.198 &  0.175 &  0.648 &   &  & HM & blue straggler \\ 
   14594&  8.056 &  0.351 &  0.058 &  0.208 &  5.114 & -1.71& HM  \\ 
   15998& 10.165 &  0.500 &  0.395 &  0.307 &  6.292 & -0.08& S2  \\ 
   16404&  9.939 &  0.451 &  0.089 &  0.122 &  6.167 & -1.91& HM & binary \\ 
   17481&  8.702 &  0.236 &  0.159 &  0.627 &  3.724 & -0.05& Ol2 \\ 
   17497&  9.000 &  0.263 &  0.158 &  0.525 &  3.956 & -0.06& HM  \\ 
   19215&  9.000 &  0.235 &  0.169 &  0.639 &  3.771 &  0.09& HM  \\ 
   19797&  9.234 &  0.322 &  0.071 &  0.292 &  4.770 & -1.52& HM  \\ 
   21609&  9.940 &  0.452 &  0.106 &  0.132 &  6.092 & -1.81& S1 & binary \\ 
   22632&  9.138 &  0.358 &  0.068 &  0.244 &  5.104 & -1.47& Ol2 \\ 
   24316& 10.476 &  0.371 &  0.060 &  0.205 &  6.298 & -1.60& S1  \\ 
   25717&  8.194 &  0.289 &  0.150 &  0.470 &  3.862 & -0.19& HM  \\ 
   26676& 10.195 &  0.435 &  0.171 &  0.155 &  5.972 & -1.29& S1  \\ 
   30481&  8.270 &  0.262 &  0.127 &  0.451 &  3.789 & -0.55& HM  \\ 
   34146&  8.058 &  0.299 &  0.140 &  0.402 &  4.037 & -0.35& HM  \\ 
   36491&  8.480 &  0.373 &  0.117 &  0.282 &  4.985 & -0.84& Ol2 \\ 
   36818& 10.566 &  0.384 &  0.143 &  0.203 &   & -0.92& S1  \\ 
   38541&  8.282 &  0.430 &  0.109 &  0.116 &  6.021 & -1.75& HM  & binary\\ 
   39911&  9.584 &  0.404 &  0.144 &  0.236 &  5.464 & -0.92& Ol2 \\ 
   40778&  9.716 &  0.339 &  0.071 &  0.258 &  4.801 & -1.45& HM  \\ 
   42278&  7.790 &  0.243 &  0.157 &  0.519 &  3.685 & -0.08& HM  \\ 
   43445&  8.647 &  0.320 &  0.121 &  0.355 &  4.299 & -0.65& F1  \\ 
   44116&  8.492 &  0.299 &  0.120 &  0.394 &  4.011 & -0.65& F1  \\ 
   44124&  9.653 &  0.349 &  0.076 &  0.250 &  5.120 & -1.34& HM  \\ 
   44435&  7.390 &  0.237 &  0.158 &  0.528 &  3.241 & -0.06& T1  \\ 
   46120& 10.118 &  0.399 &  0.086 &  0.116 &  6.205 & -1.72& S1  \\ 
   46250&  8.382 &  0.294 &  0.157 &  0.448 &  4.174 & -0.10& HM  \\ 
   46509&  4.599 &  0.296 &  0.164 &  0.451 &  3.435 &  0.00& HM  \\ 
   48152&  8.342 &  0.302 &  0.054 &  0.369 &  3.809 & -2.24& HM  \\ 
   50965&  9.793 &  0.377 &  0.145 &  0.305 &  4.727 & -0.59& S2  \\ 
   51769& 10.479 &  0.425 &  0.202 &  0.206 &  6.527 & -0.76& S1  \\ 
   53070&  8.229 &  0.344 &  0.079 &  0.258 &  4.646 & -1.29& HM  &binary\\ 
   54641&  8.165 &  0.335 &  0.084 &  0.300 &  4.417 & -1.22& Ol2 \\ 
   55790&  9.076 &  0.343 &  0.063 &  0.275 &  4.283 & -1.62& S1  & binary?\\ 
   55978&  9.039 &  0.257 &  0.167 &  0.555 &  3.634 &  0.07& HM  \\ 
   57360&  8.740 &  0.333 &  0.079 &  0.297 &  4.225 & -1.31& S1  & binary?\\ 
   57450&  9.909 &  0.398 &  0.099 &  0.179 &  5.577 & -1.45& HM  \\ 
   57939&  6.427 &  0.484 &  0.222 &  0.155 &  6.618 & -1.41& HM  \\ 
   59258&  7.506 &  0.245 &  0.132 &  0.490 &  3.696 & -0.48& Ol3 \\ 
   60632&  9.671 &  0.330 &  0.059 &  0.287 &  4.878 & -1.79& S1  & binary?\\ 
   60852&  8.483 &  0.314 &  0.128 &  0.368 &  4.275 & -0.54& HM  \\ 
   62261&  9.400 &  0.267 &  0.149 &  0.469 &  4.147 & -0.20& HM  \\ 
   63063&  9.920 &  0.475 &  0.325 &  0.264 &  6.348 & -0.25& S2  \\ 
   63553&  8.506 &  0.274 &  0.161 &  0.401 &  3.733 & -0.02& HM  \\ 
   64386&  9.888 &  0.413 &  0.170 &  0.236 &  5.665 & -0.74& S1  \\ 
   64765&  8.843 &  0.253 &  0.152 &  0.566 &  4.030 & -0.15& F2  \\ 

 \hline
 \end{tabular}
 \end{center}
 \end{table*}

 \setcounter{table}{2}
 \begin {table*}
 \begin{center}
\caption{Str\"omgren photometry of candidate subdwarfs}
 \begin{tabular}{rrrrrcrrllc} 
  \hline \hline
 HIP & V & b-y & m$_1$ & c$_1$ & M$_V$ & [m/H] &ref &Comments \\  
  \hline

   65040&  9.778 &  0.418 &  0.174 &  0.245 &  5.716 & -0.71& S1  \\ 
   65201&  8.807 &  0.349 &  0.050 &  0.278 &  4.759 & -1.96& S1  & binary?\\ 
   65268&  7.656 &  0.307 &  0.111 &  0.384 &  4.050 & -0.79& K1  \\ 
   66500&  9.600 &  0.303 &  0.140 &  0.405 &  4.322 & -0.35& HM  \\ 
   67655&  7.969 &  0.424 &  0.173 &  0.207 &  5.979 & -0.92& S1  \\ 
   68165&  9.970 &  0.523 &  0.417 &  0.275 &  6.612 & -0.23& S1  \\ 
   70152& 10.571 &  0.567 &  0.527 &  0.280 &  6.466 & -0.03& S1  \\ 
   70681&  9.300 &  0.400 &  0.130 &  0.191 &  5.717 & -1.16& S1  \\ 
   72461&  9.730 &  0.332 &  0.052 &  0.289 &  4.794 & -2.00& HM  \\ 
   72765&  9.110 &  0.265 &  0.151 &  0.538 &  3.808 & -0.17& HM  \\ 
   73614&  8.349 &  0.271 &         &         &  3.868 &       & Ol2 \\ 
   74590&  8.927 &  0.250 &  0.163 &  0.579 &  3.649 &  0.01& HM  \\ 
   74994&  9.092 &  0.248 &  0.170 &  0.656 &  3.687 &  0.11& HM  \\ 
   75618&  8.877 &  0.362 &  0.140 &  0.501 &  4.480 & -0.56& HM  \\ 
   78195&  8.298 &  0.274 &  0.144 &  0.483 &  3.676 & -0.27& Ol2 \\ 
   78296&  9.075 &  0.351 &  0.090 &  0.209 &  4.821 & -1.11& M1  \\ 
   79139&  7.685 &  0.263 &  0.124 &  0.477 &  3.678 & -0.61& HM  \\ 
   80422&  8.492 &  0.280 &  0.141 &  0.479 &  3.679 & -0.32& HM  \\ 
   81170&  9.611 &  0.474 &  0.202 &  0.159 &  6.191 & -1.39& HM  & binary\\ 
   81617&  8.576 &  0.206 &  0.164 &  0.721 &  3.743 &  0.17& HM  \\ 
   83443&  9.252 &  0.333 &  0.160 &  0.339 &  4.904 & -0.17& M1  \\ 
   88648& 10.201 &  0.430 &  0.078 &  0.159 &  6.139 & -1.80& S2 & binary? \\ 
   89215& 10.348 &  0.474 &  0.261 &  0.141 &  6.500 & -1.34& S1  \\ 
   89554&  8.233 &  0.327 &  0.074 &  0.309 &  4.267 & -1.43& HM  \\ 
   93341& 10.103 &  0.440 &  0.202 &  0.210 &  6.069 & -0.83& HM  \\ 
   94347&  7.259 &  0.312 &  0.107 &  0.366 &  4.001 & -0.85& HM  \\ 
   95996& 10.238 &  0.355 &  0.073 &  0.231 &  5.149 & -1.38& S1  & binary? \\ 
   98020&  8.836 &  0.416 &  0.100 &  0.163 &  5.852 & -1.57& S1  & binary?\\ 
  100568&  8.660 &  0.381 &  0.103 &  0.217 &  5.459 & -1.27& HM  \\ 
  104289& 10.246 &  0.434 &  0.198 &  0.231 &  6.068 & -0.70& HM  \\ 
  105773&  7.964 &  0.306 &  0.144 &  0.379 &  4.179 & -0.30& HM  \\ 
  108095&  8.523 &  0.375 &  0.139 &  0.297 &  4.995 & -0.64& HM  \\ 
  108598&  9.564 &  0.486 &  0.274 &  0.254 &   & -0.52& Ol1 \\ 
  109067&  9.556 &  0.423 &  0.180 &  0.223 &   & -0.79& S1  \\ 
  110776&  9.670 &  0.492 &  0.332 &  0.256 &   & -0.36& S1  \\ 
  111374&  7.860 &  0.281 &  0.151 &  0.465 &  3.755 & -0.17& Ol2 \\ 
  111871& 10.448 &  0.468 &  0.302 &  0.286 &  6.400 & -0.14& S1  \\ 
  113430&  8.056 &  0.300 &  0.149 &  0.415 &  4.183 & -0.22& T1  \\ 
  113542&  8.760 &  0.262 &  0.156 &  0.506 &  3.927 & -0.09& HM  \\ 
  114837&  9.879 &  0.362 &  0.151 &  0.350 &  4.901 & -0.44& O1  \\ 
  115194&  8.849 &  0.488 &  0.319 &  0.249 &  6.341 & -0.41& S1  \\ 
  115361&  8.115 &  0.305 &  0.138 &  0.405 &  4.038 & -0.39& Ol1 \\ 
  117882& 10.288 &  0.327 &  0.195 &  0.454 &  5.666 &  0.37& HM  \\ 

 \hline
 \end{tabular}
\end{center}
Stars lacking M$_V$ estimates have $|F| > 2.5$ (see \S2.2 and Table 1). \\
References: F1 - Ferro {\sl et al.}, 1991; F2 - Franco, 1994; HM - Hauck \& Mermilliod, 1998; 
K1 - Knude, 1981;
 M1 - Manfroid {\sl et al.}, 1987;  O1 - Oblak, 1991;  Ol1 - Olsen, 1994a; Ol2 - Olsen, 1994b;  \\
S1 - Schuster \& Nissen, 1988; S2 - Schuster {\sl et al.}, 1993; 
T1 - Twarog, 1980.
 \end{table*}
 
 \begin {table*}
 \begin{center}
\caption{Walraven photometry of candidate subdwarfs} 
 \begin{tabular}{rrrrrcrrrcccl}
  \hline \hline
 HIP & Name & V & V-B & B-U & U-W & B-L & \\
\hline
51300 & HD  91043 &   -0.9313&  0.1771 & 0.3142&  0.1669 & 0.1939    \\ 
55978 &  HD  99805&    -0.8785&  0.1631 &  0.3482 & 0.1845&  0.2063   \\
60251 & HD 107440 &   -0.8755&   0.2303 & 0.3064 & 0.1899 & 0.2274    \\ 
74078 & HD 133808 &   -0.6397 &  0.1958 &  0.3083 &  0.1814&  0.2069    \\
80422 & HD 147685 &   -0.6591 &  0.1749 &  0.3131 & 0.1728 & 0.1928    \\
81617&  HD 150147 &   -0.6795 &  0.1365 &  0.3838 & 0.1678&  0.1970    \\
85999 & HD 159175 &   -0.9128 &  0.2325 &  0.2714 & 0.1935&  0.2021   \\ 
89932 & HD 168375 &   -0.9010 &  0.2275 &  0.2980 & 0.1890&  0.2246    \\
93031 & HD 175479 &   -0.9679 &  0.2271 & 0.2977  &0.1939 & 0.2230    \\
\hline
 \end{tabular}
\end{center}
 \end{table*}

\subsection {Walraven photometry}

The Walraven photometric system covers the wavelength range from 5700 \AA\
to the ultraviolet atmospheric cutoff with a series of five intermediate-
and narrow-band filters (Lub \& Pel, 1977). The ((B-L), (V-B)) two-colour 
diagram is particularly sensitive to abundance variations (and insensitive to changes
in gravity). Nine stars from Table 1, lying near the Galactic Plane, have Walraven
photometry by de Geus {\sl et al.} (1991). All are expected to be early-type disk dwarfs
at distances exceeding 100 parsecs. 

Figure 5 plots data for the five stars, superimposed
on the full catalogue of AFG dwarfs observed in the de Geus {\sl  et al.} survey. 
As the figure shows, 
subdwarfs  with [m/H]$ < -1$  lie over 0.05 magnitudes blueward of the
disk main-sequence in (V-B). None of the stars listed in Table 4 falls above
the [m/H]=-1.0 sequence, and only HIP 60251 (HD 107440) appears likely to have an
abundance significantly below the solar value. Our UBV observations of the last-mentioned star,
discussed in the following section,  reveal only a modest ultraviolet excess, while
spectroscopy (\S6.2) shows linestrengths consistent
with an abundance within a factor of two of the solar value. We conclude that none of the stars
listed in Table 4 is a halo subdwarf. 

 \begin {table}
 \begin{center}
\caption{Abundance standards for $\delta_{0.6}$ calibration}
 \begin{tabular}{lcccc} 
  \hline \hline
 Name & $\delta_{0.6}$ & (U-B) & (B-V) & [Fe/H] \\  
  \hline 
  HD 3567    &   0.195&  -0.160&   0.460&  -1.17 \\
  HD 16031   &   0.257&  -0.210&   0.440&  -1.66 \\
  HD 19445   &   0.284&  -0.240&   0.460&  -1.88 \\
  HD 22879   &   0.163&  -0.080&   0.540&  -0.76 \\
  HD 25329   &   0.211&   0.380&   0.860&  -1.69 \\
  HD 30649   &   0.103&   0.020&   0.590&  -0.46 \\
  HD 59374   &   0.175&  -0.110&   0.520&  -1.02 \\
  HD 63077   &   0.173&  -0.070&   0.570&  -0.78 \\
  HD 64090   &   0.262&  -0.120&   0.610&  -1.60 \\
  HD 64606   &   0.140&   0.170&   0.730&  -0.93 \\
  HD 69611   &   0.135&  -0.020&   0.580&  -0.55 \\
  HD 74000   &   0.281&  -0.230&   0.430&  -1.52 \\
  HD 84937   &   0.275&  -0.200&   0.370&  -2.04 \\
  HD 91324   &   0.055&  -0.020&   0.500&  -0.23 \\
  HD 94028   &   0.213&  -0.170&   0.470&  -1.38 \\
  HD 103095  &   0.194&   0.160&   0.750&  -1.22 \\
  HD 108177  &   0.270&  -0.220&   0.430&  -1.55 \\
  HD 114762  &   0.141&  -0.080&   0.520&  -0.67 \\
  HD 118659  &   0.136&   0.090&   0.680&  -0.59 \\
  HD 134169  &   0.154&  -0.070&   0.550&  -0.68 \\
  HD 134439  &   0.203&   0.170&   0.760&  -1.57 \\
  HD 134440  &   0.310&   0.360&   0.870&  -1.57 \\
  HD 136352  &   0.118&   0.060&   0.640&  -0.21 \\
  HD 140283  &   0.282&  -0.220&   0.490&  -2.38 \\
  HD 148816  &   0.141&  -0.070&   0.530&  -0.68 \\
  HD 157089  &   0.103&  -0.010&   0.560&  -0.51 \\
  HD 158226A &   0.182&  -0.040&   0.610&  -0.63 \\
  HD 158226B &   0.144&   0.130&   0.710&  -0.63 \\
  HD 184499  &   0.125&  -0.010&   0.580&  -0.53 \\
  HD 188510  &   0.292&  -0.150&   0.610&  -1.37 \\
  HD 193901  &   0.243&  -0.150&   0.560&  -1.00 \\
  HD 194598  &   0.249&  -0.190&   0.490&  -1.03 \\
  HD 201891  &   0.220&  -0.160&   0.510&  -0.94 \\
  BD+17 4708 &   0.222&  -0.190&   0.450&  -1.65 \\
  G74-5      &   0.191&  -0.090&   0.570&  -0.99 \\
  G78-1      &   0.152&  -0.090&   0.520&  -0.78 \\
  G246-38    &   0.275&  -0.080&   0.650&  -1.92 \\
  G194-22    &   0.242&  -0.190&   0.480&  -1.49 \\
  G165-53    &   0.232&  -0.150&   0.550&  -1.26 \\
  G166-45    &   0.281&  -0.230&   0.430&  -2.29 \\
  G207-5     &   0.130&  -0.050&   0.540&  -0.46 \\
  G125-64    &   0.288&  -0.220&   0.510&  -2.01 \\
\hline
 \end{tabular}
 \end{center}
 \end{table}

\begin{figure}
\psfig{figure=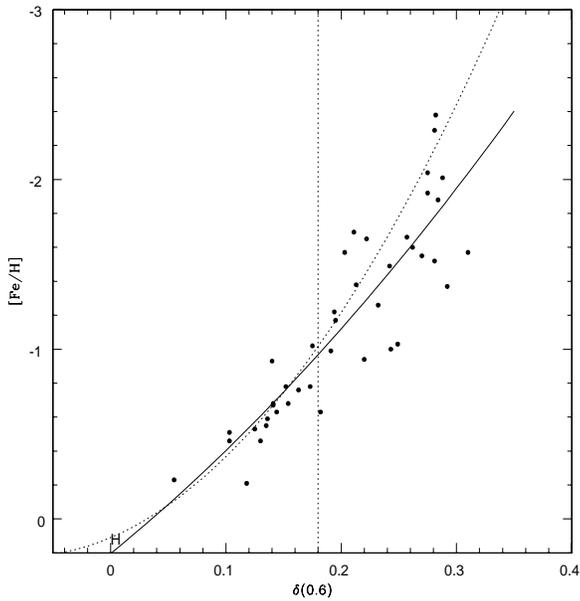,height=8cm,width=8cm
,bbllx=8mm,bblly=57mm,bburx=205mm,bbury=245mm,rheight=9cm}
\caption{The calibration of ultraviolet excess, $\delta_{0.6}$, as a 
function of metal abundance. The solid line shows the present calibration; the
dotted relation plots Carney's (1979) relation; H marks the location of Hyades stars, 
the reference point of the calibration; and the dotted vertical line shows $\delta_{0.6}$=0.18, the
criterion we adopt to segregate candidate halo subdwarfs.}
\end{figure}

\begin{figure}
\psfig{figure=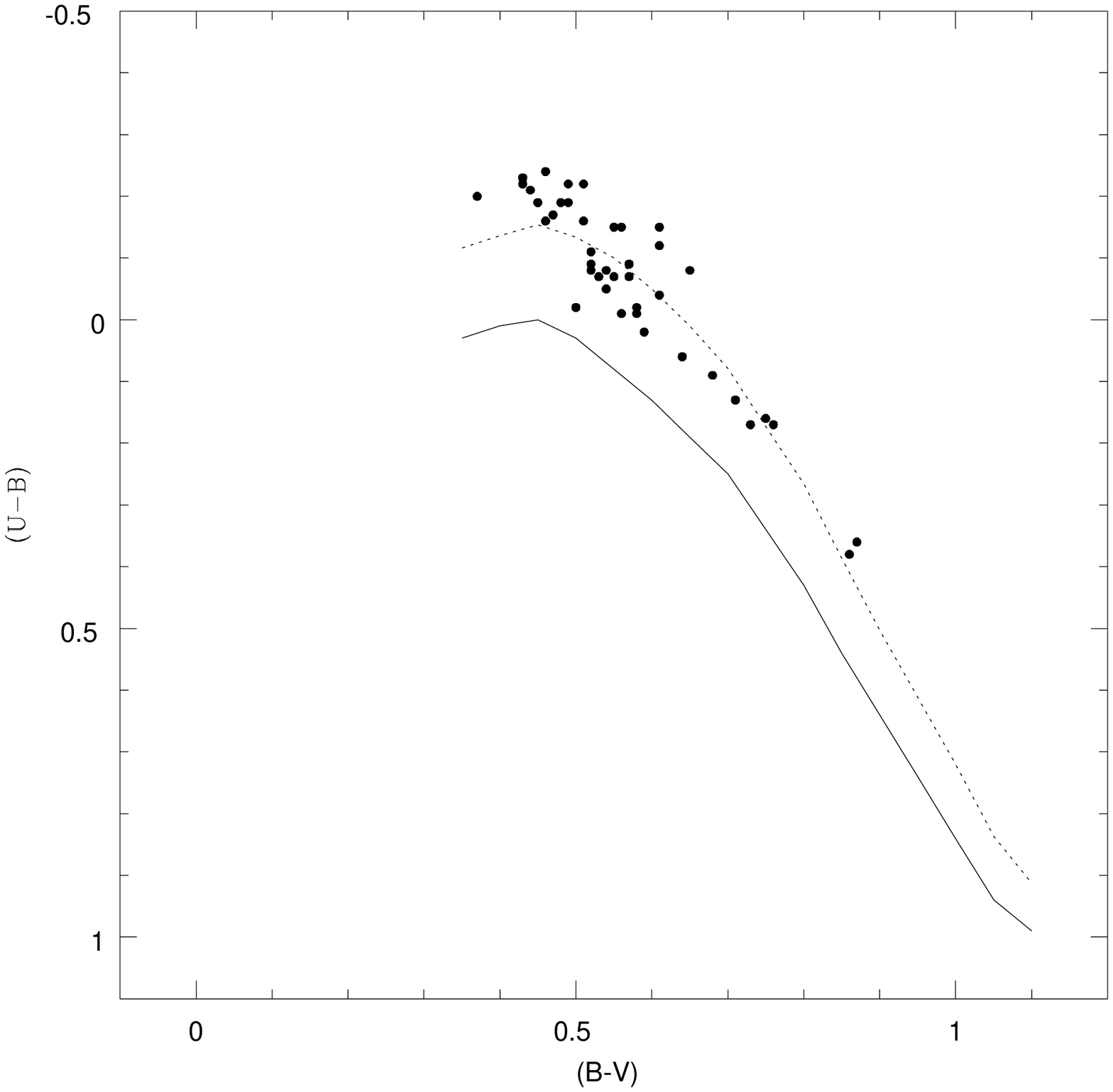,height=8cm,width=8cm
,bbllx=8mm,bblly=57mm,bburx=205mm,bbury=245mm,rheight=9cm}
\caption{The (U-B)/(B-V) distribution of the stars which calibrate the $\delta_{0.6}$/[Fe/H]
relation. The solid line plots the Hyades (U-B)/(B-V) relation (from Sandage, 1969), while
the dotted line plots the two-colour relation for $\delta_{0.6} = 0.18$ magnitudes. }
\end{figure}
\section {Broadband photometric observations}

\subsection {Ultraviolet excess and stellar abundances}

The Johnson-Cousins UBVRI system is, by far, the most widely used broadband
photometric system. The main characteristics  are described by
Bessell (1979, 1983). The U-band covers the wavelength range from the
atmospheric cutoff ($\sim3200\AA$) to $\sim3950\AA$. As result, the total
flux in U depends strongly on the extent of line blanketing, and hence the
abundance of heavy elements. Shortly after the inception of the UBV system, 
Wallerstein \& Carlson (1960) showed that ultraviolet excess, defined as
\begin{displaymath}
\delta(U-B) \ = \ (U-B)_{Hyades} \ - \ (U-B)_{obs}
\end{displaymath}
could be calibrated against stellar metallicity, [Fe/H] (where
Fe is taken as representative of all metals). 

The degree of excess ultraviolet flux at a given chemical abundance depends 
on the effective temperature, with the maximum variation occurring at
(B-V)=0.6 magnitudes. Sandage (1969) took this variation into
account by using differential blanketing calculations by Wildey {\sl et al.} (1962)
to compute the correction factors required to scale an observed $\delta$(U-B) to
the appropriate value for a star at (B-V)=0.6, $\delta_{0.6}$.  Carney (1979) compiled
data for stars with high-resolution spectroscopic abundance analyses and derived
a relation between $\delta_{0.6}$ and [Fe/H]. As noted above, there have been changes in both
the zeropoint and scale of abundance determinations since Carney's analysis, so we
have re-computed the calibration. Using UBV photometry from Carney (1979) and Carney {\sl et al.} (1994), we
have calculated $\delta_{0.6}$ for 42 stars with [Fe/H] measurements by either Gratton {\sl et al.}
(1997a) or Axer {\sl et al.} (1994). The relevant data are listed in Table 5. The best-fit 
second order polynomial is
\begin{displaymath}
[Fe/H] \ = \ 0.203 - 5.517 \delta_{0.6} - 5.512 \delta_{0.6}^2 
\end{displaymath}
Figure 6 compares our revised calibration against Carney's original relation. The 
main discrepancies lie at abundances below [Fe/H]=-1.0.  

The dispersion about the mean relation is significant: $\sigma_{[Fe/H]} = 0.26$ dex, with
the uncertainties increasing with decreasing abundance. Moreover, Figure 7 shows that the majority
of the calibrating stars fall within a relatively restricted range in (B-V) colour. 
However, our main purpose is identifying candidate subdwarfs, 
rather than deriving accurate abundances. Figure 6 shows that stars with halo-like abundances
([Fe/H]$ < -1$) can be expected to 
have ultraviolet excess values of $\delta_{0.6} > 0.18$
([Fe/H]=-0.97 dex from our calibration). We adopt this as our primary selection criterion in
identifying new subdwarf candidates from UBV data.

\begin{figure}
\psfig{figure=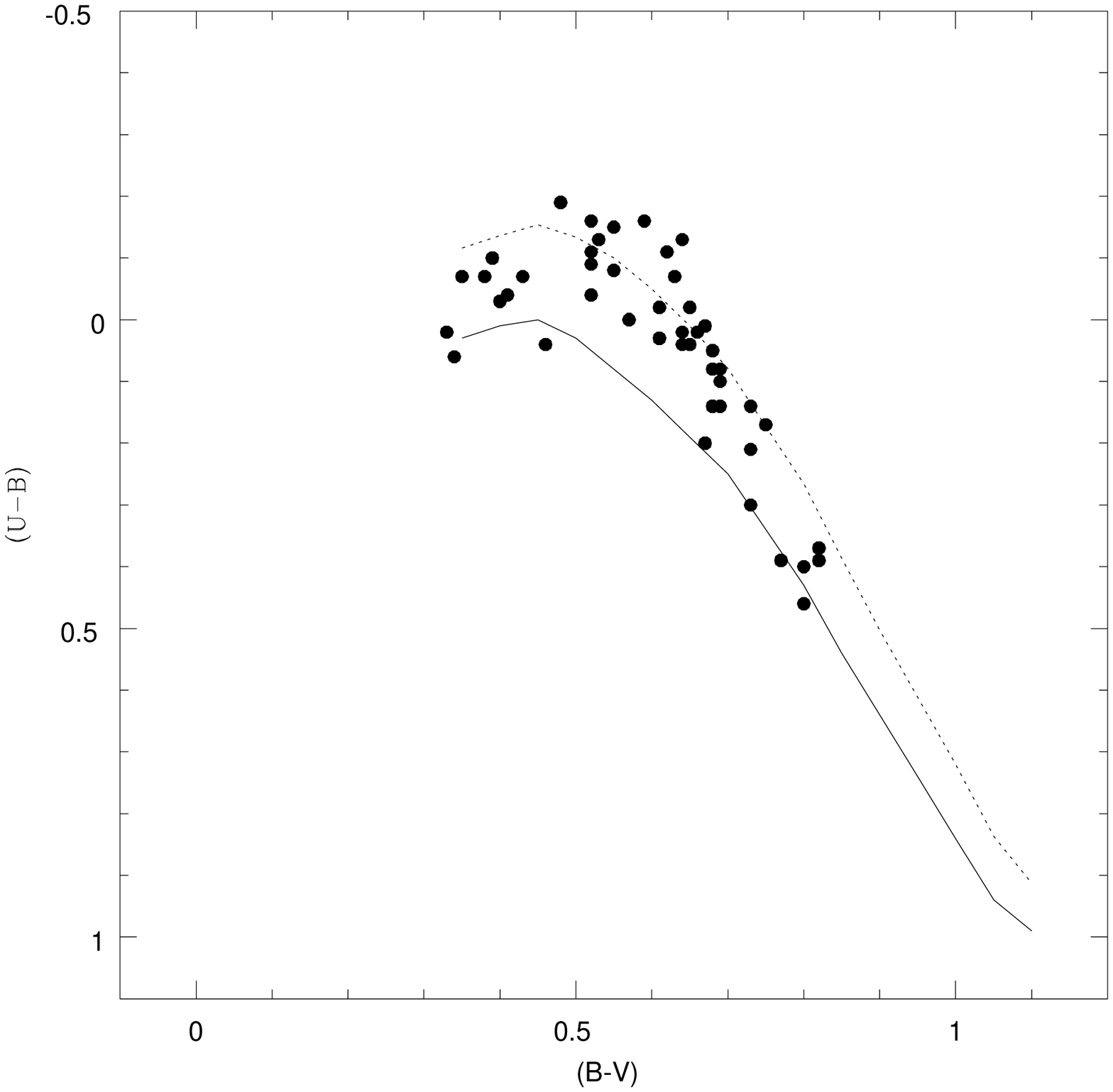,height=8cm,width=8cm
,bbllx=8mm,bblly=57mm,bburx=205mm,bbury=245mm,rheight=9cm}
\caption{The  (U-B)/(B-V) distribution for the stars listed in Table 6. As in Figure 7, 
the solid line and dotted line are the Hyades and $\delta_{0.6}$=0.18 sequences.}
\end{figure}

 \begin {table*}
 \begin{center}
\caption{Published UBVRI photometry of candidate subdwarfs}
 \begin{tabular}{rrrrrrrrrrcc}
  \hline \hline 
 HIP & U-B & B-V & V & V-R & V-I &$\delta_{U-B}$ & $\delta_{0.6}$ & [Fe/H]$_{0.6}$ & M$_V$ & ref \\
\hline
    3855&       &       &  10.56&   0.61&   1.14&       &       &       &    7.01&   W1     \\
    5004&   0.17&   0.75&  10.24&   0.44&   0.86&   0.17&   0.18&   -1.0&    6.30&   RN1    \\
    7459&  -0.16&   0.52&  10.12&   0.33&   0.68&   0.21&   0.23&   -1.4&    5.44&   RN1    \\
    8130&   0.14&   0.68&  10.52&       &       &   0.09&   0.09&   -0.3&    5.79&   CLLA   \\
    8558&       &   0.40&   9.02&   0.18&   0.40&       &       &       &    3.91&   E1     \\
   12579&  -0.09&   0.52&   9.16&       &       &   0.14&   0.15&   -0.8&    4.97&   CLLA   \\
   14594&  -0.24&   0.46&   8.06&       &       &   0.25&   0.28&   -1.8&    5.12&   C1     \\
   16404&  -0.08&   0.65&   9.91&       &       &   0.27&   0.28&   -1.8&    6.14&   CLLA \\ 
   19215&   0.06&   0.34&   8.97&       &       &   0.14&   0.00&    0.2&    3.74&   O1     \\
   20413&   0.20&   0.67&  10.32&   0.40&   0.76&   0.01&   0.01&    0.1&    5.91&   F1     \\
   20527&       &   1.29&  10.90&   0.73&   1.46&       &       &       &    7.67&   R1     \\
   20895&   1.26&   1.37&  10.99&   0.83&   1.63&       &       &       &    7.98&   W2     \\
   21261&       &   1.20&  10.75&   0.71&   1.33&       &       &       &    7.37&   R1     \\
   21478&       &   0.61&   9.43&       &       &       &       &       &    4.56&   K1     \\
   21586&   0.39&   0.77&  10.36&   0.49&   0.94&  -0.01&  -0.01&    0.3&    6.34&   R2     \\
   21609&  -0.13&   0.64&   9.84&       &       &   0.31&   0.31&   -2.0&    5.99&   RN1    \\
   22177&       &   1.28&  10.91&   0.76&   1.44&       &       &       &    7.67&   R1     \\
   22246&   0.46&   0.80&  10.24&       &       &  -0.03&  -0.03&    0.4&        &   CLLA   \\
   22632&       &   0.51&   9.18&   0.26&   0.56&       &       &       &    5.15&   E1     \\
   24316&       &   0.55&   9.45&   0.30&   0.61&       &       &       &    5.27&   E1     \\
   26452&  -0.08&   0.55&   9.57&       &       &   0.16&   0.16&   -0.8&    5.16&   CLLA   \\
   26676&  -0.02&   0.65&  10.22&       &       &   0.21&   0.21&   -1.2&    6.00&   CLLA   \\
   26676&   0.01&   0.67&  10.20&   0.41&   0.82&   0.20&   0.20&   -1.2&    5.98&   RN1    \\
   30481&  -0.07&   0.35&   8.30&       &       &   0.10&   0.12&   -0.5&    3.82&   O3     \\
   33282&   1.32&   1.34&  11.27&   0.80&   1.51&       &       &       &    7.70&   F1     \\
   35163&   0.30&   0.73&  10.02&   0.38&   0.74&   0.00&   0.00&    0.2&        &   F1     \\
   36491&  -0.11&   0.52&   8.49&       &       &   0.16&   0.17&   -0.9&    5.00&   CLLA   \\
   36491&  -0.13&   0.53&   8.44&   0.31&   0.64&   0.19&   0.21&   -1.2&    4.95&   RN1    \\
   36818&  -0.02&   0.61&  10.50&       &       &   0.16&   0.16&   -0.8&        &   RN1    \\
   38541&  -0.12&   0.61&   8.26&       &       &   0.26&   0.26&   -1.6&    6.00&   C1     \\
   40778&  -0.19&   0.48&   9.76&       &       &   0.21&   0.24&   -1.5&    4.85&   CLLA   \\
   44124&  -0.19&   0.48&   9.66&       &       &   0.21&   0.24&   -1.5&    5.13&   CLLA   \\
   44124&  -0.19&   0.48&   9.67&       &       &   0.21&   0.24&   -1.5&    5.14&   RN1    \\
   46120&  -0.16&   0.59&  10.10&       &       &   0.28&   0.29&   -1.8&    6.19&   RN1    \\
   46509&   0.04&   0.46&   4.59&       &       &  -0.03&  -0.04&    0.4&    3.43&   C2     \\
   50965&   0.00&   0.57&   9.80&       &       &   0.10&   0.10&   -0.4&    4.73&   CLLA   \\
   50965&   0.00&   0.57&   9.80&       &       &   0.10&   0.10&   -0.4&    4.73&   RN1    \\
   51769&   0.05&   0.68&  10.52&       &       &   0.18&   0.18&   -0.9&    6.57&   CLLA   \\
   51769&   0.08&   0.69&  10.47&   0.40&   0.78&   0.16&   0.16&   -0.8&    6.52&   RN1    \\
   54768&  -0.04&   0.52&   9.14&       &       &   0.09&   0.10&   -0.4&    4.87&   O1     \\
   57450&  -0.15&   0.55&   9.92&       &       &   0.23&   0.23&   -1.4&    5.59&   CLLA   \\
   57939&   0.17&   0.75&   6.43&       &       &   0.17&   0.18&   -0.9&    6.62&   CLLA   \\
   63063&   0.40&   0.80&   9.93&       &       &   0.03&   0.03&    0.0&        &   CLLA   \\
   63912&  -0.04&   0.41&   8.93&       &       &   0.05&   0.05&   -0.1&    3.99&   O2     \\
   64386&   0.04&   0.64&   9.89&       &       &   0.14&   0.14&   -0.7&    5.67&   CLLA   \\
   65040&   0.02&   0.64&   9.78&       &       &   0.16&   0.16&   -0.8&    5.72&   CLLA   \\
   66828&       &   1.33&  10.89&   0.78&   1.52&       &       &       &    8.01&   W2     \\
   70152&   1.29&   1.35&  10.58&   0.82&   1.57&       &       &       &    6.47&   RN1    \\
   72461&  -0.24&   0.42&   9.74&       &       &   0.25&   0.29&   -1.8&    4.80&   C4 \\
   78251&       &   0.64&   8.96&       &       &       &       &       &    5.46&   CS \\
   80448&   0.16&   0.64&   7.33&   0.38&   0.73&   0.02&   0.02&    0.0&    4.10$^*$&  C3 \\
   80789&   0.03&   0.61&  10.24&       &       &   0.11&   0.11&   -0.5&    5.60&   CLLA   \\
   81013&       &       &   8.77&   0.25&   0.52&       &       &       &    3.79&   M1     \\
   81170&   0.10&   0.75&   9.63&       &       &   0.24&   0.28&   -1.8&    6.21&   SK \\

\hline
 \end{tabular}
 \end{center}
 \end{table*}

\setcounter{table}{5}
 \begin {table*}
 \begin{center}
\caption{Published UBVRI photometry of candidate subdwarfs contd.)}
 \begin{tabular}{rrrrrrrrrrcc}
  \hline \hline 
 HIP & U-B & B-V & V & V-R & V-I &$\delta_{U-B}$ & $\delta_{0.6}$ & [Fe/H]$_{0.6}$ & M$_V$ & ref \\
\hline
   86183&   0.02&   0.33&   9.38&       &       &   0.18&   0.00&    0.2&    3.73&   O2     \\
   88648&  -0.11&   0.62&  10.24&   0.39&   0.79&   0.26&   0.26&   -1.6&    6.18&   RN1    \\
   89215&   0.21&   0.73&  10.43&       &       &   0.09&   0.10&   -0.4&    6.58&   CLLA   \\
   92277&   0.14&   0.73&  10.35&       &       &   0.16&   0.17&   -0.9&    6.10&   CLLA   \\
   92277&   0.14&   0.69&  10.33&   0.38&   0.68&   0.10&   0.10&   -0.4&    6.08&   RN1    \\
   93341&   0.10&   0.69&  10.07&   0.40&   0.79&   0.14&   0.14&   -0.7&    6.04&   RN1    \\
   94704&   0.02&   0.66&  11.29&   0.40&   0.84&   0.18&   0.18&   -1.0&    6.57&   R2     \\
   95800&  -0.07&   0.43&   8.79&   0.26&   0.51&   0.07&   0.08&   -0.3&    4.00&   D1     \\
   95996& -0.12 &   0.48&  10.25&   0.33&   0.65&  0.14  & 0.16  & -0.7 &    5.16& R3,  W3  \\
   98020&  -0.15&   0.61&   8.82&       &       &   0.29&   0.29&   -1.8&    5.84&   C1   \\
   99267&  -0.22&   0.51&  10.11&       &       &   0.26&   0.29&   -1.8&    5.51&   CLLA \\
  101103&  -0.10&   0.39&   9.46&   0.24&   0.49&   0.11&   0.13&   -0.6&    4.39&   D1     \\
  103269&  -0.11&   0.62&  10.27&       &       &   0.26&   0.26&   -1.6&    6.03&   CLLA   \\
  104289&   0.08&   0.68&  10.23&   0.39&   0.78&   0.15&   0.15&   -0.7&    6.05&   RN1    \\
  106924&  -0.07&   0.63&  10.34&       &       &   0.24&   0.24&   -1.4&    6.25&   CLLA   \\
  109067&   0.04&   0.65&   9.55&       &       &   0.15&   0.15&   -0.7&        &   CLLA   \\
  110776&   0.39&   0.82&   9.71&       &       &   0.08&   0.09&   -0.3&        &   CLLA   \\
  111871&   0.37&   0.82&  10.44&       &       &   0.10&   0.11&   -0.5&    6.39&   CLLA   \\
  112389&       &   1.24&  10.67&   0.74&   1.40&       &       &       &    7.74&   W2     \\
  113542&  -0.07&   0.38&   8.75&       &       &   0.09&   0.10&   -0.4&    3.92&   G1     \\
  117242&  -0.03&   0.40&   8.80&       &       &   0.04&   0.04&   -0.1&    3.97&   O2     \\

\hline
 \end{tabular}
 \end{center}
*: HIP 80448 is an unresolved X-ray binary. Fabricius \& Makarov (2000b) derive
individual magnitudes of V$_T$=8.13 and 8.32. \\
Stars lacking M$_V$ values have unreliable Hipparcos parallax measurements. \\
References: C1 - Carney, 1979; C2 - Celis, 1975; C3 - Cutispoto {\sl et al.}, 1991;
C4 - Carney, 1983;  CS - Cousins \& Stoy, 1962; 
CLLA - Carney {\sl et al.}, 1994; D1 - Dean, 1981; E1 - Eggen, 1990; 
F1 - Figueras et al, 1990;  G1 - Guetter, 1980; K1 - Kenyon {\sl et al.}, 1994;  O1 - Oja, 1986;
O2 - Oja, 1985; O3 - Oja, 1991; R1 - Reid, 1993; R2 - Rossello {\sl et al.}, 1988; 
R3 - Roman, 1955 (UBV only); 
W1 - Weis, 1986; W2 - Weis, 1993; W3 - Weis, 1996.
 \end{table*}

\subsection {Published UBVRI observations of candidate subdwarfs}

Using the SIMBAD database, we have located seventy-six UBVRI photometric observations
of sixty-nine stars from Table 1. Those data are listed in Table 6, where we list R- and
I-magnitudes on the Cousins system, using the relations given by Bessell (1979, 1983) and
Bessell \& Weis (1987) to transform data where necessary. We note that several subdwarfs
(eg HD 19445) have Johnson RI photometry from the 1960s: we have not included those
data in the current compilation.
 
Figure 8 plots the (U-B)/(B-V)
two-colour diagram; and Figure 9 shows the location of these stars on the (M$_V$, (V-I)) and
(M$_V$, (B-V)) colour-magnitude planes. 
As with the Str\"omgren and Walraven datasets, the majority
have photometric properties consistent with abundances [Fe/H]$> -1$. Indeed, several
stars are identified as late-type K or early-type M dwarfs with highly-discrepant Tycho
photometric colours. Of the relatively small number of stars in Table 6 identified as having
halo-like abundances, most are well known subdwarfs. Only HIP 94704 (G207-23)
 stands out as a possible addition to
current samples, and that star has a $\delta_{0.6}$ abundance of [Fe/H] = -1.0, close to the upper boundary
of the halo distribution.

\begin{figure*}
\psfig{figure=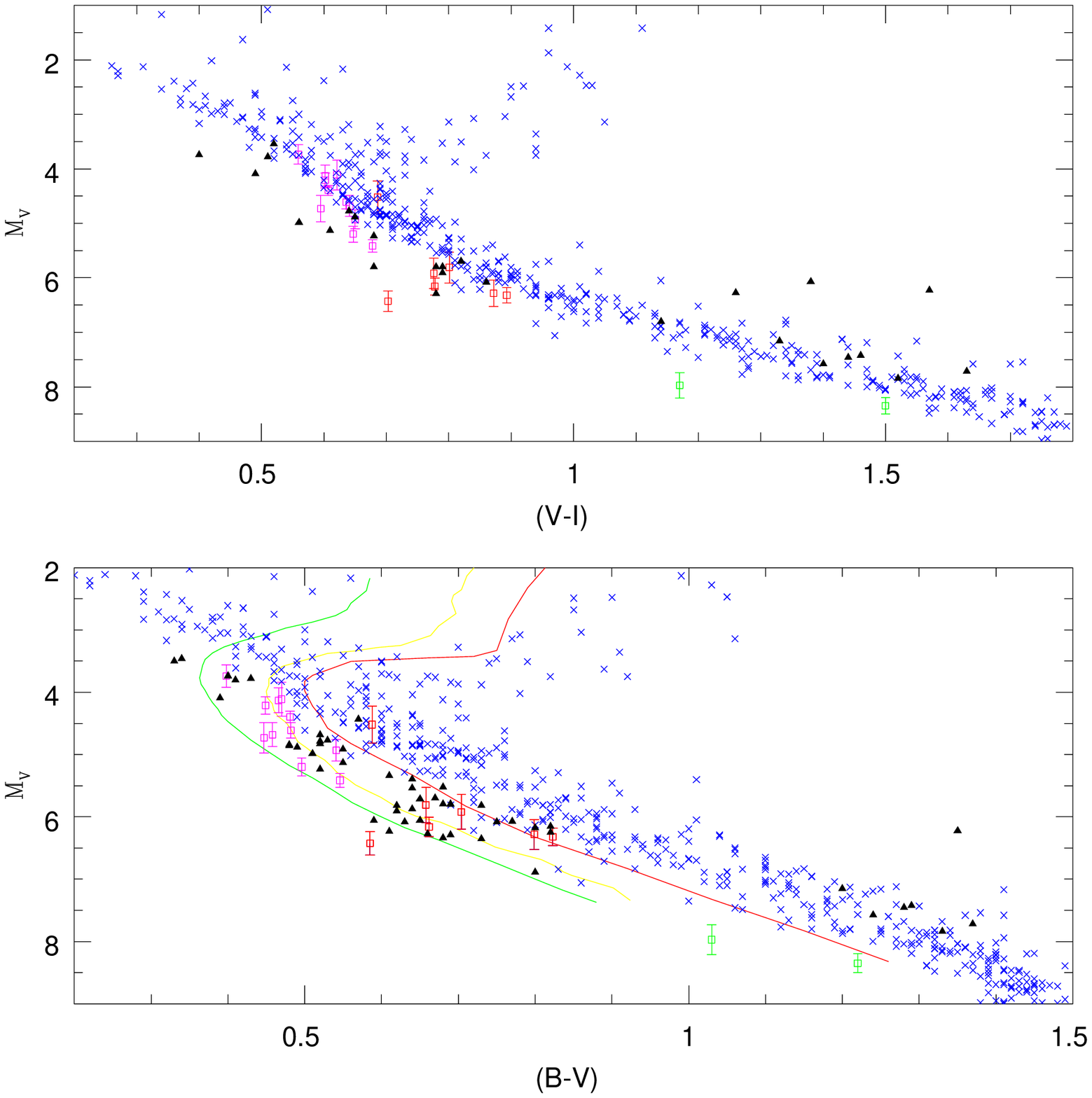,height=13.5cm,width=15cm
,bbllx=8mm,bblly=57mm,bburx=205mm,bbury=245mm,rheight=14.5cm}
\caption{The  (M$_V$, (V-I)) and (M$_V$, (B-V)) colour-magnitude diagrams outlined by the
candidate subdwarfs listed in Table 6. The disk main sequence (crosses) is delineated  by stars
from the Gliese/Jahreiss Nearby Star Catalogue with both accurate photometry (Bessell, 1991; Leggett, 1992)
and accurate trigonometric parallaxes (ESA, 1997); known subdwarfs are plotted as open squares with errorbars; 
the mean colour-magnitude relations for 47 Tuc, M5 and NGC 6397 are superimposed on the
(M$_V$, (B-V)) diagram; and the candidate subdwarfs are plotted as filled triangles.}
\end{figure*}

\subsection {SAAO UBVRI observations}

In addition to compiling literature photometry, we have undertaken a programme
of new observations using the facilities at the Sutherland station of the South African Astronomical
Observatory. Between two and four photometric measurements have been obtained of 175 stars from Table 1.
In addition, UBVRI data have been obtained for thirteen known metal-poor subdwarfs.
Combined with the literature data discussed 
above, these measurements provide the first extensive, reliable Cousins R- and I-band 
data for metal-poor dwarfs with accurate trigonometric parallax measurements. 

The observations were made between July 1998 and December 1999 using the 
modular photometer on the 0.5-metre telescope. The photometer employs a Hamamatsu R943-02 (GaAs)
photomultiplier, and a Johnson-Cousins UBVRI filter set (Kilkenny {\sl et al.}, 1998). 
The data were reduced using standard techniques, and calibrated through observations of E-region standard
stars (Cousins, 1973; Menzies {\sl et al.}, 1989). Full details on the techniques employed are
given by Kilkenny {\sl et al.} (1998). Table 7 lists the derived colours and magnitudes for the
Hipparcos candidate subdwarfs; twenty-three stars were observed at only one epoch, and those data
are listed separately at the end of the table. 
Table 8 presents SAAO data for additional subdwarfs, where
we also list Hipparcos astrometry (G113-26 was not observed by Hipparcos)
and, if appropriate, inferred absolute magnitudes.

All of the programme stars are bright, and, as a result, the photometric
uncertainties are typically less than 0.01 magnitude in V and in each colour. 
Six stars have V-band photometric uncertainties exceeding 
0.015 magnitudes. These 
include the known variable TW Hydrae, HIP 53911, a 10 Myr-old K7 T Tauri star, and the
early-type M dwarf, AC +18 1061 (HIP 112389). 

Comparing the SAAO photometry against the literature data included in the Hipparcos
catalogue (i.e. V$_l$, not V$_T$), we find
\begin{displaymath}
\Delta V \ = \ (V_{SAAO} - V_l) \ = \ -0.007 \pm 0.042
\end{displaymath}
The largest discrepancy is for HIP 117121, where V$_{SAAO} = 11.12$, 0.48 magnitudes
brighter than the value listed in the Hipparcos catalogue. Since no (B-V)$_l$ measurement is
given, the SAAO data are clearly more reliable. Eliminating that point gives
\begin{displaymath}
\Delta V \ = \ -0.005 \pm 0.021
\end{displaymath}

Figure 10 plots the UBV two-colour diagram outlined by the stars with SAAO data; 
ninety percent of the sample fall between the Hyades sequence and the $\delta_{0.6}$
calibration. The star lying well above the main-sequence, at (B-V)=0.88, (U-B)=-0.45, is
TW Hydrae, while HIP 90724 (HD 170368) has an unusually red (U-B) colour and falls below
the Hyades sequence. Neither star has unusual colours in (V-R) or (V-I). SIMBAD lists a spectral
type of A7V for HIP 90724, and, as noted above, the  true parallax is less than 10 mas. 
The location on the UBV plane is consistent with its being a distant A dwarf, 
reddened by E$_{B-V} \sim 0.3$ magnitudes. The star lies towards the Galactic Bulge, albeit
at a modest distance from the Plane ( l = 358$^o$, b= -12$^o$), and patchy foreground
reddening at the observed level is not unreasonable.

We have used the $\delta_{0.6}$ calibration outlined above to estimate abundances for
stars with (B-V) colours between 0.35 and 1.10 magnitudes, and those estimates are
listed in Tables 7 and 8. Twenty-nine stars have [Fe/H]$_{0.6} \le -1.0$.
Of these, twenty-two were previously known to be halo stars; seven (HIP 4750, 17241, 43490, 
54834, 73798, 95190, 114271\footnote{ HIP 114271's inclusion  in Fulbright's (2000) sample 
was prompted by its location in Figure 2.}) are additions to the list of metal-poor
calibrators. 

\begin{figure}
\psfig{figure=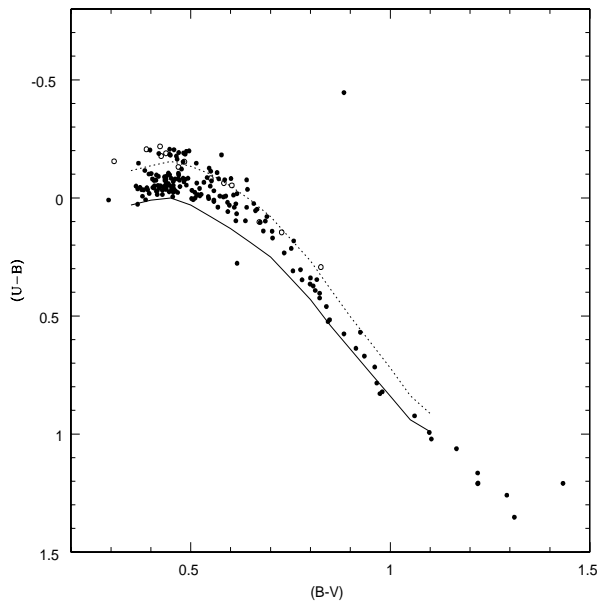,height=8cm,width=8cm
,bbllx=8mm,bblly=57mm,bburx=205mm,bbury=245mm,rheight=9cm}
\caption{The  (U-B)/(B-V) distribution for Hipparcos stars with SAAO photometry (Table 7). Additional
observations of known subdwarfs (from Table 8) are plotted as open circles. As in Figure 6, 
the solid line and dotted line are the Hyades and $\delta_{0.6}$ sequences.}
\end{figure}

\begin{figure}
\psfig{figure=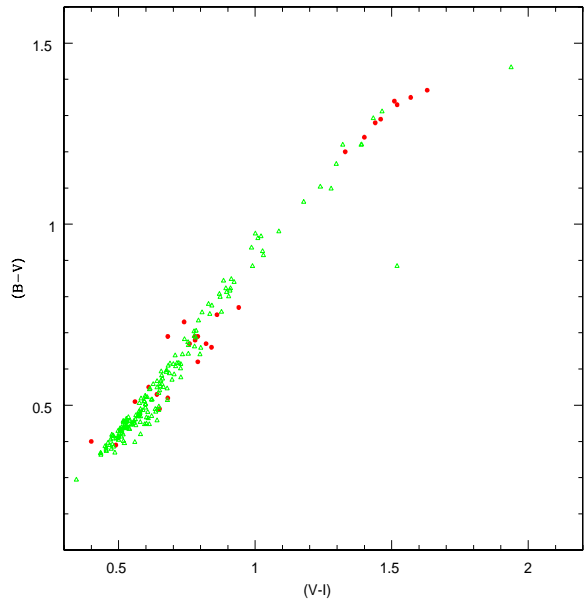,height=8cm,width=8cm
,bbllx=8mm,bblly=57mm,bburx=205mm,bbury=245mm,rheight=9cm}
\caption{The  (B-V)/(V-I) distribution for Hipparcos stars from Tables 6 and 7. Literature
data are plotted as solid points; SAAO photometry, as open triangles. The
two datasets produce sequences in good agreement.}
\end{figure}

\begin{figure}
\psfig{figure=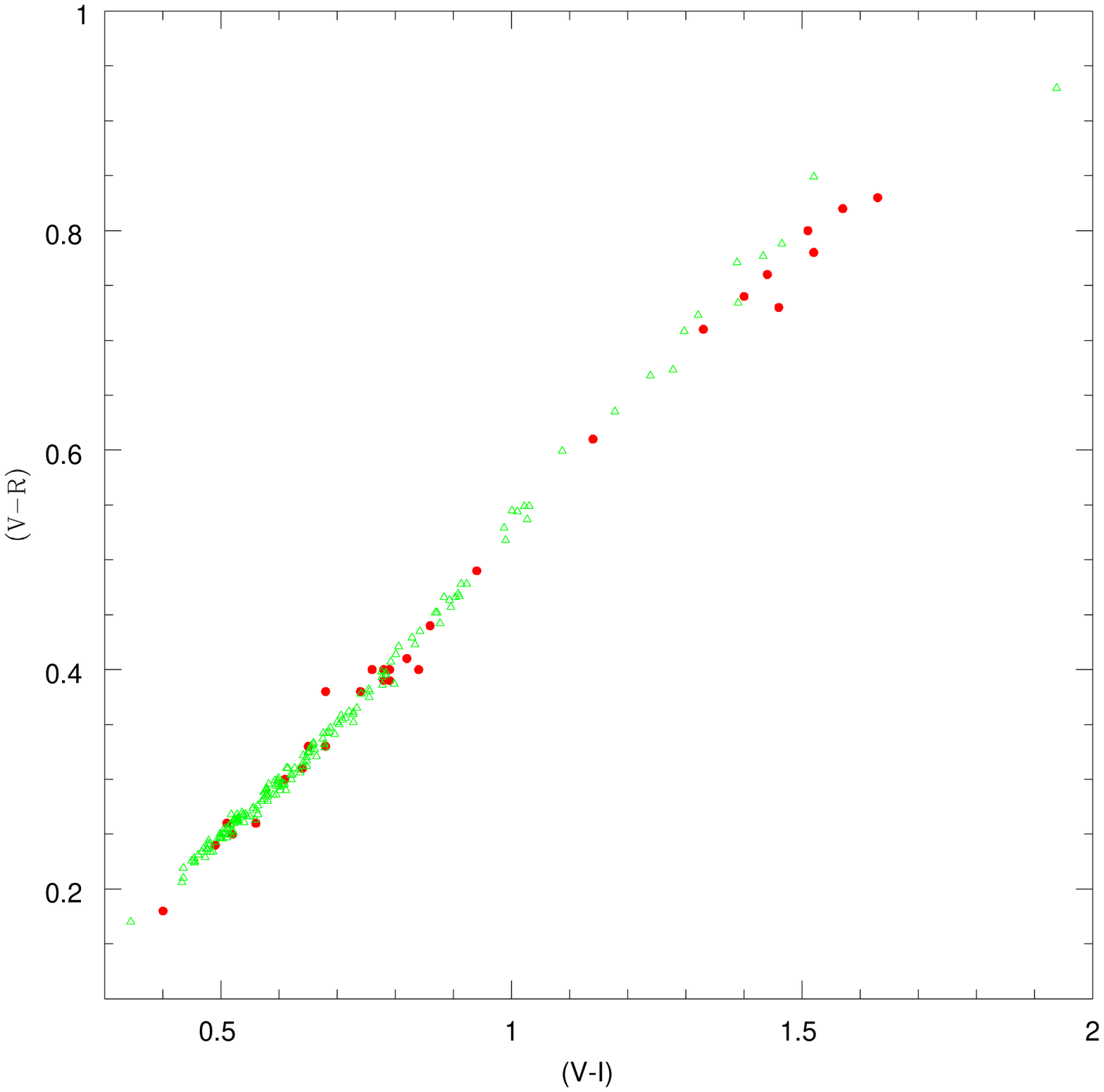,height=7.5cm,width=8cm
,bbllx=8mm,bblly=57mm,bburx=205mm,bbury=245mm,rheight=8.5cm}
\caption{The  (V-R)/(V-I) two-colour diagram for stars from Tables 6 and 7; the
symbols have the same meaning as in Figure 11.  There is a
systematic offset between the two datasets for K and M dwarfs. Given the good agreement
evident in Figure 11, this offset likely arises from differences in the R bandpass employed
in different observations. }
\end{figure}

The VR$_C$I$_C$ observations listed in Table 6 are drawn from a variety of literature
sources. The (B-V)/(V-I) and (V-R)/(V-I) two-colour diagrams plotted in Figures 11 and 12
allow us to assess how well those observations match the well-defined SAAO Cousins system.
Figure 11 shows that the BVI data are in excellent agreement. Discrepancies are more
evident in the VRI plane, both individual (HIP 20527, HIP 92277) and systemic: the
literature data for M dwarfs ( (V-I)$>$1) lies blueward of the SAAO sequence in
(V-R). This is not unexpected, since the extended red tail of the Cousins R-band
is difficult to reproduce exactly, and different filter/detector combinations lead to
significant colour terms.

Finally, Figure 13 plots the distribution of the candidate subdwarfs listed in Table 7 in
the (M$_V$, (B-V)) and (M$_V$, (V-I)) planes. In most cases, the locations of individual
stars are consistent with abundances inferred from the $\delta_{0.6}$ ultraviolet excess, 
and the overwhelming majority are mildly metal-poor disk dwarfs. 

\begin{figure*}
\psfig{figure=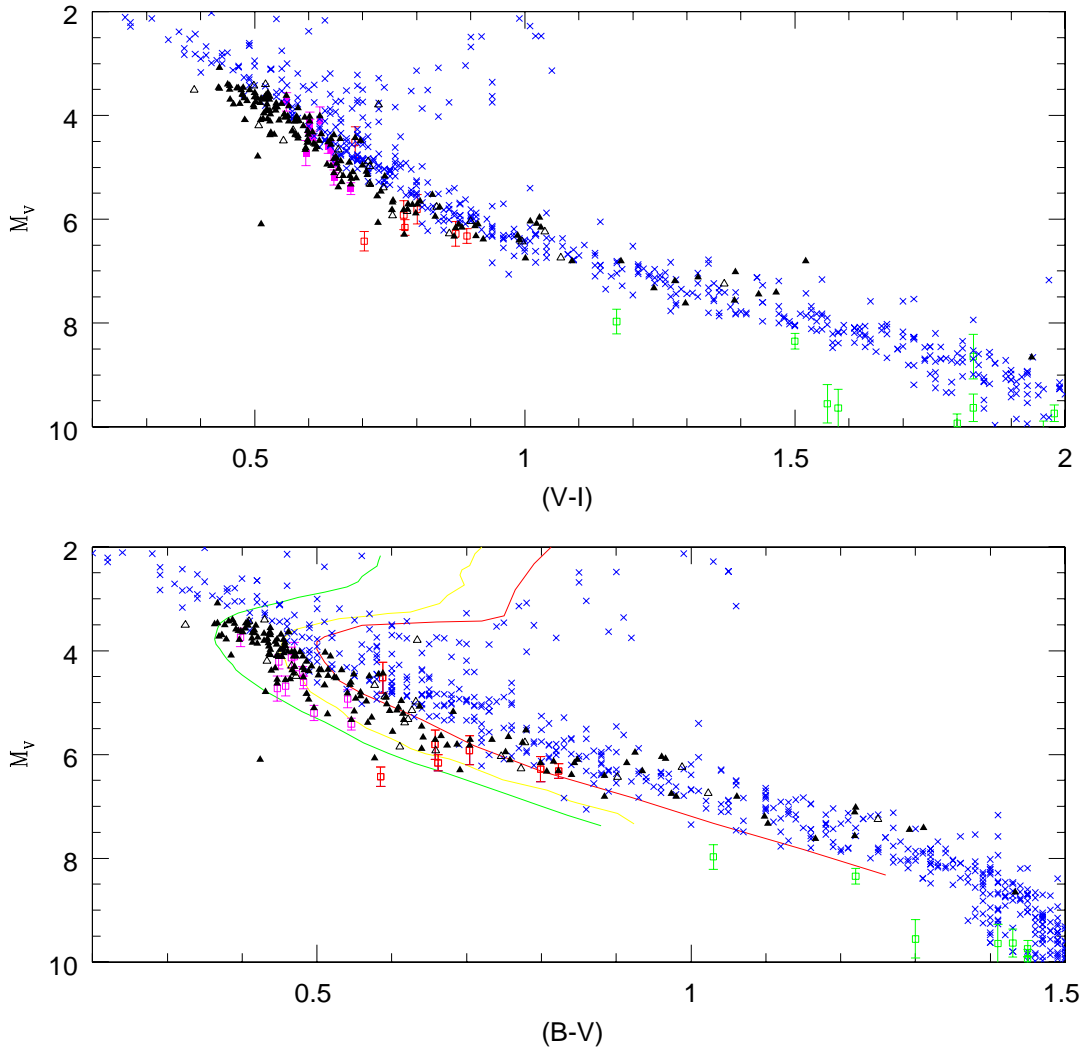,height=14cm,width=15cm
,bbllx=8mm,bblly=57mm,bburx=205mm,bbury=245mm,rheight=15cm}
\caption{The  (M$_V$, (V-I)) and (M$_V$, (B-V)) colour-magnitude diagrams for 
stars listed in Table 7. As in Figure 9, known subdwarfs are plotted as open squares with
error-bars; solid triangles are stars with multiple-epoch observations; open triangles
mark stars with single-epoch photometry,}
\end{figure*}
One star stands out from the main body of data: HIP 28122, at (M$_V$=6.37, (B-V)=0.42). 
While the Hipparcos catalogue notes no duplicity problems, 
inspection of the Palomar plates in the Digital Sky Survey show
that this star (HD 40007, or BD +10 936) lies $\sim 20$ arcseconds from another star
of similar brightness.  That star is BD +10 936B (V=10.02$\pm0.03$, (B-V)=0.45$\pm0.02$:
Kilkenny, priv. comm.), star 28121 in the Hipparcos input catalogue, but unobserved in the
survey itself.  The goodness of fit statistic for the astrometry of HIP 28122 
is $|F|$=2.48, barely within our adopted limits. As discussed further below, spectroscopy
indicates that both stars are of near-solar abundance, with no evidence for peculiarities.
It seems likely that the proximity of the bright companion has affected the Hipparcos
analysis, and the parallax has been overestimated.

 \begin {table*}
 \begin{center}
\caption{SAAO UBVRI photometry}
 \begin{tabular}{rrrrrrrrrrrrrrrrcc}
  \hline \hline 
 HIP & U-B & B-V & V & V-R & V-I &$\sigma_{U-B}$ &$\sigma_{B-V}$ &$\sigma_V$ &$\sigma_{V-R}$ &$\sigma_{V-I}$ &
N$_{obs}$&$\delta_{U-B}$ & $\delta_{0.6}$ & [Fe/H] & M$_V$ \\
\hline
    435&  -0.081&   0.469&   9.309&   0.282&   0.574&  5&  2&  1&  1&  2&  4&  0.092&  0.107&   -0.4&    4.37 \\
    1051&  -0.089&   0.434&   8.750&   0.261&   0.540&  0&  0& 12&  2&  0&  2&  0.092&  0.106&   -0.4&    3.98 \\
    1719&   0.460&   0.840&   9.653&   0.478&   0.923&  2&  6&  6&  4&  3&  4&  0.058&  0.058&   -0.1&    6.48 \\
    3139&  -0.078&   0.438&   9.309&   0.263&   0.527&  6&  7& 20&  6& 11&  4&  0.080&  0.091&   -0.3&    3.98 \\
    3531&   1.208&   1.220&  10.895&   0.734&   1.390&  5& 22& 12&  4&  5&  4& & &   &    7.27 \\
    3855&   0.923&   1.061&  10.566&   0.635&   1.178& 14& 30& 11&  8& 13&  4&  0.028&  0.031&    0.0&    7.02 \\
    4450&   0.008&   0.506&   9.642&   0.290&   0.576&  2&  4&  5&  3&  6&  4&  0.028&  0.031&    0.0&    4.62 \\
    4576&  -0.038&   0.436&   9.511&   0.259&   0.516&  1&  5&  5&  1&  0&  4&  0.041&  0.045&   -0.1&    4.33 \\
    4750&  -0.107&   0.566&  10.087&   0.315&   0.641&  3&  2&  5&  3&  3&  5&  0.203&  0.203&   -1.1&    5.27 \\
    4981&   0.031&   0.597&   9.110&   0.342&   0.676&  3&  2&  6&  3&  4&  3&  0.096&  0.098&   -0.4&    5.31 \\
    5004&   0.182&   0.758&  10.266&   0.442&   0.877&  2&  7&  5&  2&  3&  4&  0.172&  0.186&   -1.0&    6.33 \\
    5097&  -0.053&   0.456&   9.300&   0.274&   0.556&  2&  5&  4&  0&  3&  4&  0.057&  0.062&   -0.2&    4.30 \\
    6251&  -0.040&   0.522&   8.417&   0.297&   0.605&  2& 10& 13&  3&  0&  3&  0.092&  0.100&   -0.4&     \\
    6758&   0.067&   0.615&   9.701&   0.347&   0.688&  9&  6& 17&  4&  8&  5&  0.081&  0.081&   -0.3&    5.45 \\
    7303&  -0.012&   0.607&   9.845&   0.354&   0.709&  0&  2&  4&  4&  6&  3&  0.150&  0.150&   -0.8&    5.41 \\
    7459&  -0.147&   0.514&  10.107&   0.329&   0.679&  2&  7&  3&  0&  1&  4&  0.191&  0.210&   -1.2&    5.43 \\
    7687$^1$&   0.373&   0.807&   9.397&   0.452&   0.869&  9&  8& 28&  8& 12&  4&  0.072&  0.074&   -0.2&    6.46 \\
    7772&  -0.044&   0.405&   8.817&   0.237&   0.470&  1&  2&  4&  3&  2&  3&  0.053&  0.058&   -0.1&    3.71 \\
    7935&  -0.079&   0.486&   8.886&   0.286&   0.590&  6&  8&  8&  2&  4&  3&  0.101&  0.117&   -0.5&    4.39 \\
    8298&   0.304&   0.775&  10.079&   0.435&   0.842&  3& 11&  3&  2&  3&  3&  0.081&  0.084&   -0.3&    6.02 \\
    8389&   0.637&   0.914&  10.688&   0.549&   1.030&  9& 13&  9&  2&  2&  3&  0.031&  0.031&    0.0&    6.42 \\
    8558&  -0.147&   0.369&   9.002&   0.234&   0.486&  4&  9&  5&  4&  5&  3&  0.169&  0.208&   -1.2&    3.89 \\
    9634&  -0.075&   0.435&   8.335&   0.267&   0.537&  1&  2&  6&  6&  6&  4&  0.078&  0.088&   -0.3&    4.02 \\
   10208&  -0.092&   0.439&   9.065&   0.261&   0.528&  1&  4&  5&  1&  0&  3&  0.094&  0.109&   -0.5&    4.58 \\
   10353&  -0.104&   0.469&   9.211&   0.284&   0.580&  3&  3&  5&  0&  4&  3&  0.115&  0.136&   -0.6&    4.66 \\
   10360&   0.009&   0.294&   8.546&   0.170&   0.345&  2&  3&  6&  2&  2&  4&       &       &       &     \\
   10375&  -0.098&   0.447&   9.098&   0.266&   0.550&  4&  4&  7&  3&  5&  4&  0.099&  0.115&   -0.5&    4.03 \\
   10385&   0.993&   1.098&  10.872&   0.673&   1.278&  6& 24&  3&  6&  6&  3& -0.005& -0.005&    0.2&    7.41 \\
   10637&  -0.083&   0.451&   9.274&   0.274&   0.555&  1& 12&  5&  3&  1&  3&  0.084&  0.095&   -0.4&    4.25 \\
   11435&  -0.033&   0.453&   8.635&   0.262&   0.528&  3&  4&  3&  1&  4&  4&  0.035&  0.038&    0.0&    4.35 \\
   13849&   0.784&   0.966&  10.650&   0.549&   1.022&  6&  4&  7&  3&  6&  3& -0.012& -0.012&    0.3&    6.37 \\
   14192&  -0.029&   0.462&   8.931&   0.268&   0.528&  4&  1&  3&  5&  2&  3&  0.036&  0.040&    0.0&    3.82 \\
   15756&   0.716&   0.961&  10.887&   0.544&   1.010& 12&  9&  3&  6&  2&  4&  0.046&  0.046&   -0.1&    6.30 \\
   16089&  -0.019&   0.396&   8.774&   0.231&   0.461&  3&  6&  2&  1&  3&  5&  0.031&  0.034&    0.0&    4.02 \\
   16479&  -0.032&   0.429&   8.296&   0.250&   0.498&  0&  1&  2&  1&  2&  3&  0.036&  0.040&    0.0&    3.96 \\
   17085&  -0.047&   0.431&   9.585&   0.255&   0.506&  2&  3&  4&  5&  6&  3&  0.051&  0.056&   -0.1&    5.09 \\
   17241&  -0.184&   0.447&  10.247&   0.290&   0.612&  2&  4&  1&  4&  5&  4&  0.185&  0.226&   -1.3&    4.82 \\
   17481&   0.027&   0.367&   8.700&   0.206&   0.433&  3&  5&  7& 13&  1&  3& -0.004& -0.004&    0.2&    3.72 \\
   18700&  -0.008&   0.573&   9.335&   0.327&   0.661&  2& 10&  7&  6&  4&  3&  0.111&  0.115&   -0.5&    5.49 \\
   19007&   0.404&   0.823&   9.532&   0.467&   0.910&  8&  9& 10&  8& 13&  3&  0.077&  0.079&   -0.3&    6.45 \\
   20527&   1.352&   1.311&  10.892&   0.788&   1.465&  4& 35&  7&  1&  7&  2&        &        &       &    7.66 \\
   21000&   0.097&   0.613&   9.841&   0.361&   0.728&  2&  4&  8&  3&  4&  3&  0.049&  0.049&   -0.1&     \\
   21125&   0.104&   0.674&  10.217&   0.382&   0.754&  1&  8&  9&  4&  8&  3&  0.115&  0.115&   -0.5&    6.05 \\
   21261&   1.165&   1.219&  10.709&   0.723&   1.321& 12& 22&  1&  4&  4&  2&        &        &       &    7.33 \\
   21609&  -0.076&   0.640&   9.855&   0.387&   0.798&  5&  7&  4&  3&  1&  4&  0.254&  0.254&   -1.6&    6.01 \\
   22177&   1.259&   1.292&  10.901&   0.777&   1.433&  3& 39&  2&  0& 13&  2&        &        &       &    7.66 \\
   22632&  -0.197&   0.489&   9.135&   0.309&   0.634&  1&  3&  6&  1&  3&  3&  0.220&  0.256&   -1.6&    5.10 \\
   23573&   0.214&   0.752&  10.046&   0.423&   0.834&  7&  7&  5&  3&  6&  3&  0.130&  0.140&   -0.7&    6.09 \\
   24296&   0.019&   0.592&   9.854&   0.337&   0.675&  8&  6& 14&  1&  5&  3&  0.103&  0.106&   -0.4&    5.10 \\
   24316&  -0.199&   0.496&   9.427&   0.312&   0.647&  3&  4&  8&  1&  2&  3&  0.227&  0.263&   -1.6&    5.25 \\
   24421&  -0.009&   0.523&   9.340&   0.301&   0.599&  2&  4&  8&  3&  8&  3&  0.062&  0.068&   -0.2&    4.75 \\
   24935&  -0.097&   0.470&   8.767&   0.280&   0.570&  2&  3&  8&  1&  1&  3&  0.109&  0.128&   -0.6&    4.43 \\
\hline
 \end{tabular}
 \end{center}
 \end{table*}

\setcounter{table}{6}
 \begin {table*}
 \begin{center}
\caption{SAAO UBVRI photometry (contd.)}
 \begin{tabular}{rrrrrrrrrrrrrrrrcc}
  \hline \hline 
 HIP & U-B & B-V & V & V-R & V-I &$\sigma_{U-B}$ &$\sigma_{B-V}$ &$\sigma_V$ &$\sigma_{V-R}$ &$\sigma_{V-I}$ &
N$_{obs}$&$\delta_{U-B}$ & $\delta_{0.6}$ & [Fe/H] & M$_V$ \\
\hline

   25717&  -0.014&   0.429&   8.192&   0.254&   0.512& 19& 13& 16&  6&  6&  3&  0.018&  0.020&    0.1&    3.86 \\
   26676&   0.024&   0.658&  10.217&   0.414&   0.801& 48& 11& 54& 38& 32&  5&  0.176&  0.176&   -0.9&    5.99 \\
   26688&  -0.091&   0.419&   7.697&   0.249&   0.517&  4&  6&  7&  4&  4&  3&  0.097&  0.113&   -0.5&     \\
   28122&  -0.051&   0.424&   9.507&   0.250&   0.512&  6&  5&  8&  5&  5&  4&  0.056&  0.062&   -0.2&     \\
   29322&   1.209&   1.433&  11.282&   0.930&   1.938&  5& 22&  2&  5&  2&  2&        &        &       &  8.91   \\
   31639&   0.170&   0.705&   9.645&   0.397&   0.784&  3& 10&  4&  1&  5&  3&  0.089&  0.091&   -0.3&    5.87 \\
   32009&   0.233&   0.734&   9.628&   0.407&   0.792&  7& 16&  5&  1&  5&  3&  0.078&  0.079&   -0.3&    5.95 \\
   32308&   0.829&   0.974&  10.718&   0.545&   1.001&  9& 54&  5&  4&  2&  3& -0.041& -0.041&    0.4&    7.01 \\
   33283&   0.346&   0.816&  11.104&   0.469&   0.908& 10& 15&  1&  3&  3&  3&  0.119&  0.129&   -0.6&    6.39 \\
   34146&  -0.069&   0.456&   8.047&   0.268&   0.539&  5&  6&  9&  3&  4&  3&  0.073&  0.082&   -0.3&    4.03 \\
   35232&  -0.080&   0.461&   9.696&   0.272&   0.561&  4&  6&  1&  5&  5&  3&  0.087&  0.099&   -0.4&    4.24 \\
   35560&  -0.097&   0.414&   8.511&   0.247&   0.499&  9&  2&  3&  1&  6&  3&  0.104&  0.122&   -0.6&    4.03 \\
   36491&  -0.086&   0.541&   8.485&   0.324&   0.650&  7&  3&  6&  8&  6&  3&  0.157&  0.171&   -0.9&    4.99 \\
   36818&  -0.077&   0.585&  10.573&   0.350&   0.703&  2&  2&  5&  3&  2&  3&  0.192&  0.193&   -1.1&     \\
   39911&  -0.018&   0.616&   9.591&   0.356&   0.714&  4&  2&  4&  1&  4&  3&  0.167&  0.167&   -0.9&    5.47 \\
   41563&  -0.034&   0.382&   8.813&   0.228&   0.454&  2&  6&  2&  2& 12&  3&  0.051&  0.056&   -0.1&    3.70 \\
   42278&  -0.050&   0.363&   7.783&   0.219&   0.435&  2&  2&  6&  6&  1&  3&  0.075&  0.084&   -0.3&    3.68 \\
   43445&  -0.083&   0.481&   8.634&   0.286&   0.578&  5&  2&  6&  3&  9&  3&  0.102&  0.119&   -0.5&    4.29 \\ 
   43490&  -0.113&   0.550&   9.554&   0.321&   0.664& 15&  7& 19& 17& 23&  3&  0.193&  0.194&   -1.1&    5.30 \\
   43973&   0.048&   0.666&   9.520&   0.380&   0.756&  6&  4&  5&  2&  6&  3&  0.161&  0.161&   -0.8&    5.81 \\
   44124&  -0.185&   0.486&   9.648&   0.317&   0.647&  3&  2&  0&  2&  4&  3&  0.207&  0.241&   -1.4&    5.12 \\
   44436&  -0.038&   0.367&   7.382&   0.210&   0.435&  4&  4&  4&  3&  4&  3&  0.061&  0.068&   -0.2&    3.23 \\
   46120&  -0.182&   0.577&  10.108&   0.352&   0.728&  5& 12&  9&  6&  7&  4&  0.289&  0.297&   -1.9&    6.20 \\
   47161&  -0.039&   0.436&   9.271&   0.251&   0.505&  2&  5&  1&  1&  3&  3&  0.042&  0.046&   -0.1&    3.94 \\
   47171&  -0.004&   0.582&   9.303&   0.332&   0.658&  5&  4&  5&  4&  3&  3&  0.116&  0.120&   -0.5&    4.68 \\
   47948&   0.140&   0.682&  10.061&   0.378&   0.741&  4&  5&  6&  5&  1&  3&  0.088&  0.088&   -0.3&    5.44 \\
   48146$^2$&   0.009&   0.558&   9.595&   0.310&   0.627& 26& 10& 34& 14&  2&  3&  0.079&  0.080&   -0.3&     \\
   48152&  -0.203&   0.398&   8.329&   0.263&   0.559&  1&  2&  2&  5&  5&  3&  0.214&  0.264&   -1.6&    3.80 \\
   49574$^3$&  -0.122&   0.477&   8.777&   0.291&   0.578&  2& 29& 25& 30& 23&  4&  0.138&  0.163&   -0.8&    4.30 \\
   49785&  -0.054&   0.489&   8.512&   0.287&   0.580&  4&  5&  4&  3&  6&  4&  0.077&  0.088&   -0.3&     \\
   50532&  -0.005&   0.544&   9.906&   0.310&   0.613&  2&  7&  2&  4&  7&  3&  0.079&  0.086&   -0.3&    4.95 \\
   50965&  -0.008&   0.588&   9.794&   0.343&   0.686&  3&  3&  2&  3&  5&  3&  0.126&  0.131&   -0.6&    4.73 \\
   51156&  -0.052&   0.436&   9.059&   0.260&   0.523&  4&  3&  1&  2&  4&  3&  0.055&  0.060&   -0.1&    4.14 \\
   51298&  -0.039&   0.438&   9.243&   0.256&   0.511&  5&  3&  1&  3&  2&  3&  0.041&  0.046&   -0.1&    4.33 \\
   51300&  -0.042&   0.408&   9.196&   0.244&   0.479&  4&  1&  3&  0&  1&  3&  0.050&  0.055&   -0.1&    3.71 \\
   51769&   0.079&   0.691&  10.479&   0.393&   0.777&  6&  8&  3&  2&  4&  3&  0.160&  0.160&   -0.8&    6.53 \\
   52285&   0.569&   0.925&   9.869&   0.537&   1.027&  8&  2&  5&  2&  3&  3&  0.121&  0.136&   -0.6&    6.23 \\
   53070&  -0.190&   0.482&   8.218&   0.306&   0.636&  5&  3&  2&  1&  4&  3&  0.209&  0.243&   -1.5&    4.63 \\
   53911$^4$&  -0.446&   0.884&  10.825&   0.849&   1.520&147&184&227& 96&127&  2&  1.054&  1.486&  &    7.06 \\
   54519&   0.821&   0.980&  10.957&   0.599&   1.087& 14& 10&  6& 21& 29&  2& -0.021& -0.021&    0.3&    7.07 \\
   54641&  -0.153&   0.481&   8.161&   0.295&   0.607&  4&  4&  7&  1&  1&  3&  0.172&  0.201&   -1.1&    4.41 \\
   54834&  -0.177&   0.464&   8.782&   0.296&   0.609&  2&  4&  1&  2&  1&  3&  0.185&  0.217&   -1.3&    4.69 \\
   54993&  -0.053&   0.418&   8.818&   0.236&   0.475& 10&  2&  0&  7&  3&  2&  0.059&  0.066&   -0.2&    3.75 \\
   55790&  -0.192&   0.470&   9.080&   0.300&   0.621&  3&  6&  5&  4&  2&  3&  0.204&  0.238&   -1.4&    4.29 \\
   55978&   0.008&   0.387&   9.065&   0.226&   0.450&  3&  2&  1&  4&  1&  3&  0.007&  0.008&    0.2&    3.66 \\
   57360&  -0.164&   0.466&   8.736&   0.293&   0.602&  4&  4&  1&  4&  2&  3&  0.174&  0.203&   -1.1&    4.22 \\
   60251$^5$&   0.015&   0.558&   9.007&   0.325&   0.650& 15& 13& 34& 10& 24&  3&  0.073&  0.074&   -0.2&    5.09 \\
   60632&  -0.206&   0.447&   9.661&   0.286&   0.595&  4& 10&  3&  5&  3&  3&  0.207&  0.251&   -1.5&    4.87 \\
   60852&  -0.075&   0.472&   8.477&   0.276&   0.563&  1&  2&  5&  5&  4&  3&  0.088&  0.101&   -0.4&    4.27 \\
   62858&  -0.050&   0.549&   9.510&   0.322&   0.641&  7&  0&  0&  7&  7&  3&  0.129&  0.140&   -0.7&    4.72 \\
   63063&   0.392&   0.812&   9.934&   0.457&   0.896&  1&  2&  2&  2&  2&  3&  0.064&  0.065&   -0.2&     \\
   63553&  -0.074&   0.408&   8.504&   0.243&   0.491&  2&  1&  0&  1&  3&  3&  0.082&  0.094&   -0.4&    3.73 \\
   63781&  -0.043&   0.390&   9.290&   0.234&   0.467&  2& 13&  0&  4&  4&  2&  0.057&  0.063&   -0.2&    3.78 \\
   64765&  -0.007&   0.379&   8.846&   0.229&   0.473&  2&  1&  5&  5&  6&  3&  0.025&  0.028&    0.0&    4.03 \\
\hline
 \end{tabular}
 \end{center}
 \end{table*}

\setcounter{table}{6}
 \begin {table*}
 \begin{center}
\caption{SAAO UBVRI photometry (contd.)}
 \begin{tabular}{rrrrrrrrrrrrrrrrcc}
  \hline \hline 
 HIP & U-B & B-V & V & V-R & V-I &$\sigma_{U-B}$ &$\sigma_{B-V}$ &$\sigma_V$ &$\sigma_{V-R}$ &$\sigma_{V-I}$ &
N$_{obs}$&$\delta_{U-B}$ & $\delta_{0.6}$ & [Fe/H] & M$_V$ \\
\hline

   65201&  -0.204&   0.458&   8.803&   0.312&   0.641&  5&  7&  4&  0&  4&  2&  0.209&  0.243&   -1.5&    4.75 \\
   65940&   0.670&   0.935&  10.367&   0.529&   0.987&  5& 13&  1&  1&  9&  3&  0.040&  0.040&    0.0&    6.51 \\
   66815&  -0.072&   0.552&   8.828&   0.325&   0.652&  4&  4&  0&  1&  5&  3&  0.154&  0.158&   -0.8&    5.06 \\
   67189&   0.003&   0.502&   9.862&   0.296&   0.598&  2& 12&  1&  0&  1&  2&  0.029&  0.032&    0.0&    4.77 \\
   70681&  -0.082&   0.601&   9.296&   0.359&   0.727&  6&  4&  0&  1&  2&  2&  0.213&  0.213&   -1.2&    5.71 \\
   70689&  -0.044&   0.373&   8.540&   0.225&   0.455&  7&  5&  2&  4&  1&  2&  0.065&  0.072&   -0.2&    3.88 \\
   71886&  -0.064&   0.438&   8.934&   0.265&   0.529&  9&  7&  2&  0&  4&  2&  0.066&  0.074&   -0.2&    4.24 \\
   71939&  -0.068&   0.483&   8.809&   0.288&   0.575&  3&  4&  2&  3&  2&  2&  0.088&  0.101&   -0.4&    4.59 \\
   73614&  -0.028&   0.414&   8.323&   0.250&   0.500&  1& 10&  9&  0&  0&  2&  0.035&  0.039&    0.0&    3.84 \\
   73798&  -0.036&   0.642&   9.933&   0.375&   0.755&  1&  9&  0&  3&  2&  2&  0.216&  0.216&   -1.2&    5.90 \\
   74078&  -0.034&   0.458&   8.466&   0.264&   0.528&  3&  1&  1&  2&  4&  3&  0.039&  0.043&    0.0&    4.09 \\
   75618&   0.040&   0.609&   9.739&   0.342&   0.682&  8&  3& 16&  1&  5&  3&  0.101&  0.101&   -0.4&    5.34 \\
   77432&  -0.038&   0.452&   8.954&   0.268&   0.518&  8&  2& 10&  2&  3&  2&  0.039&  0.043&    0.0&    3.98 \\
   83226&   0.347&   0.779&  10.035&   0.429&   0.829&  9& 18& 11&  3&  6&  4&  0.045&  0.045&   -0.1&    5.81 \\
   83443&  -0.024&   0.510&   9.248&   0.294&   0.594&  3&  1&  4&  0&  0&  2&  0.064&  0.070&   -0.2&    4.90 \\
   84754&   0.524&   0.844&  10.057&   0.466&   0.884&  6&  4&  5&  7& 15&  3&  0.003&  0.003&    0.2&    6.38 \\
   85999&  -0.066&   0.533&   9.141&   0.320&   0.647&  2&  8&  6&  2&  5&  3&  0.129&  0.140&   -0.7&    4.61 \\
   86393&  -0.011&   0.515&   9.497&   0.304&   0.620&  3& 10&  8&  4&  2&  3&  0.056&  0.062&   -0.2&    4.76 \\
   86536&  -0.026&   0.446&   9.132&   0.266&   0.545&  2&  2&  2&  1&  7&  2&  0.027&  0.029&    0.0&    4.34 \\
   88066&  -0.035&   0.377&   8.832&   0.224&   0.454&  6&  5&  3&  3&  3&  3&  0.054&  0.060&   -0.1&    3.79 \\
   89053&   0.516&   0.848&  10.300&   0.478&   0.913&  2&  2&  3&  5&  7&  3&  0.020&  0.020&    0.1&    6.33 \\
   89554&  -0.181&   0.449&   8.216&   0.290&   0.602&  6&  4&  6&  3&  6&  3&  0.181&  0.222&   -1.3&    4.25 \\
   89932&  -0.014&   0.518&   9.124&   0.296&   0.582&  1&  5&  3&  4&  6&  3&  0.062&  0.068&   -0.2&    4.35 \\
   90724$^6$&   0.277&   0.616&   9.032&   0.362&   0.720&  2&  5&  8&  2&  3&  4& -0.128& -0.128&    0.8&    \\
   92277&   0.141&   0.704&  10.355&   0.398&   0.776&  3&  0&  1&  4& 26&  2&  0.116&  0.121&   -0.5&    6.10 \\
   93031&  -0.015&   0.526&   9.285&   0.299&   0.597&  5&  5& 12& 13& 18&  5&  0.071&  0.078&   -0.3&    4.80 \\
   94347&  -0.105&   0.444&   7.259&   0.268&   0.563&  1&  9&  0&  9&  3&  2&  0.106&  0.125&   -0.6&    4.00 \\
   95031&   0.576&   0.884&   9.868&   0.518&   0.990&  4&  5&  2&  4&  8&  2&  0.032&  0.032&    0.0&    6.55 \\
   95190&  -0.081&   0.570&   9.466&   0.341&   0.696&  4& 10& 13&  9&  8&  5&  0.181&  0.183&   -1.0&    4.79 \\
   95800&  -0.075&   0.413&   8.771&   0.247&   0.510&  3&  0&  3&  7&  4&  2&  0.082&  0.094&   -0.4&    3.98 \\
   96043&   0.097&   0.637&   9.625&   0.358&   0.707&  1&  6&  4&  2&  9&  3&  0.077&  0.077&   -0.3&    5.32 \\
   97127&  -0.036&   0.438&   9.200&   0.262&   0.525&  1&  4&  5&  2&  1&  3&  0.038&  0.042&    0.0&    3.85 \\
   97463&   0.025&   0.613&   9.573&   0.352&   0.700&  3&  4&  3&  6&  2&  3&  0.121&  0.121&   -0.5&    5.21 \\
  100207&  -0.049&   0.438&   8.758&   0.262&   0.525&  1&  1&  4&  4&  1&  3&  0.051&  0.057&   -0.1&    3.90 \\
  100568&  -0.126&   0.546&   8.642&   0.332&   0.678&  0&  2&  6&  1&  3&  3&  0.202&  0.222&   -1.3&    5.44 \\
  101103&  -0.116&   0.385&   9.456&   0.234&   0.482&  3&  2&  7&  3&  2&  3&  0.132&  0.158&   -0.8&    4.39 \\
  101650&  -0.005&   0.455&   9.311&   0.264&   0.523&  2&  0&  9&  1&  3&  4&  0.008&  0.009&    0.2&    4.08 \\
  101892&  -0.024&   0.456&   9.410&   0.258&   0.517&  2&  1&  5&  2&  2&  4&  0.028&  0.030&    0.0&    4.25 \\
  101987&  -0.040&   0.457&   9.140&   0.268&   0.542&  3&  9&  4&  0&  2&  3&  0.044&  0.049&   -0.1&    4.12 \\
  103714&   0.039&   0.568&  10.128&   0.328&   0.655& 13& 11& 18&  5& 14&  3&  0.059&  0.059&   -0.1&    5.66 \\
  104289&   0.098&   0.687&  10.243&   0.395&   0.783&  3&  7&  2&  1&  5&  3&  0.136&  0.136&   -0.7&    6.06 \\
  106904&   0.040&   0.640&  10.186&   0.365&   0.734&  2&  4&  4&  2&  2&  3&  0.138&  0.138&   -0.7&    5.76 \\
  107873&   0.000&   0.511&   9.293&   0.299&   0.593&  4&  3&  2&  2&  2&  3&  0.041&  0.045&   -0.1&    4.31 \\
  108006&  -0.051&   0.409&   9.108&   0.246&   0.498&  5&  2&  1&  1&  3&  3&  0.059&  0.065&   -0.2&     \\
  108095&  -0.030&   0.558&   8.527&   0.333&   0.659&  3&  2&  8&  5&  6&  3&  0.118&  0.123&   -0.6&    5.00  \\
  108598&   0.339&   0.800&   9.574&   0.466&   0.903&  4&  9&  0&  3&  6&  3&  0.091&  0.095&   -0.4&     \\
  108655&  -0.034&   0.392&   8.848&   0.236&   0.477&  9&  1&  2&  3&  8&  3&  0.047&  0.052&   -0.1&    3.87 \\
  108836&   1.062&   1.166&  10.966&   0.708&   1.297& 11& 12&  4&  4&  4&  2&        &        &       &    7.87 \\
  109067&   0.054&   0.662&   9.543&   0.386&   0.778&  6&  9&  4&  4&  3&  3&  0.150&  0.150&   -0.8&     \\
  109869&   0.058&   0.593&   9.554&   0.331&   0.656&  2& 15&  0&  1&  6&  3&  0.065&  0.065&   -0.2&    5.25 \\
  109945&  -0.042&   0.421&   9.154&   0.246&   0.502& 11&  7&  7&  3&  8&  3&  0.048&  0.053&   -0.1&    4.02 \\
  110776&   0.424&   0.823&   9.680&   0.463&   0.893&  9& 13&  4&  6& 10&  3&  0.057&  0.057&   -0.1&     \\
  111374&  -0.030&   0.416&   7.847&   0.241&   0.478&  6&  4&  8&  5&  6&  3&  0.037&  0.040&    0.0&    3.74 \\
  111426&  -0.060&   0.444&   9.348&   0.262&   0.530&  3&  2&  4&  4&  7&  4&  0.061&  0.068&   -0.2&    4.55 \\
  111871&   0.365&   0.799&  10.453&   0.452&   0.872&  7&  7& 19&  3& 13&  3&  0.063&  0.064&   -0.2&    6.40 \\
\hline
 \end{tabular}
 \end{center}
 \end{table*}

\setcounter{table}{6}
 \begin {table*}
 \begin{center}
\caption{SAAO UBVRI photometry (contd.)}
 \begin{tabular}{rrrrrrrrrrrrrrrrcc}
  \hline \hline 
 HIP & U-B & B-V & V & V-R & V-I &$\sigma_{U-B}$ &$\sigma_{B-V}$ &$\sigma_V$ &$\sigma_{V-R}$ &$\sigma_{V-I}$ &
N$_{obs}$&$\delta_{U-B}$ & $\delta_{0.6}$ & [Fe/H] & M$_V$ \\
\hline

  112384&  -0.103&   0.401&   9.020&   0.253&   0.518&  1&  2&  4&  2&  1&  4&  0.113&  0.133&   -0.6&    3.93 \\
  112389$^7$&   1.210&   1.219&  10.663&   0.771&   1.388& 34&  9& 60& 17& 21&  2&        &        &       &    7.74 \\
  113430&  -0.046&   0.437&   8.054&   0.263&   0.524&  4&  2&  6&  6&  4&  3&  0.049&  0.053&   -0.1&    4.18 \\
  113868&   0.309&   0.756&  10.088&   0.421&   0.806&  4&  3&  2&  3&  2&  4&  0.042&  0.042&    0.0&    5.91 \\
  114271&  -0.188&   0.420&   8.256&   0.280&   0.580&  4&  3&  4&  1&  3&  4&  0.194&  0.237&   -1.4&    4.03 \\
  114487&  -0.101&   0.395&   8.408&   0.252&   0.521&  6&  3&  4&  1&  1&  3&  0.113&  0.134&   -0.6&    3.80 \\
  114627&  -0.032&   0.503&   8.851&   0.294&   0.599&  4&  6&  7&  6&  5&  4&  0.065&  0.071&   -0.2&    4.58 \\
  114735&  -0.080&   0.404&   8.493&   0.248&   0.500&  3&  9&  6&  3&  8&  3&  0.089&  0.103&   -0.4&    3.80 \\
  114837&   0.003&   0.545&   9.882&   0.310&   0.615&  1&  3&  1&  3&  0&  2&  0.072&  0.079&   -0.3&    4.90 \\
  115031&  -0.048&   0.473&   8.275&   0.288&   0.574&  1&  7&  3&  2&  6&  3&  0.062&  0.069&   -0.2&     \\
  115361&  -0.053&   0.446&   8.106&   0.263&   0.523&  2&  3&  8&  3&  7&  2&  0.054&  0.059&   -0.1&    4.03 \\
  116437&  -0.063&   0.515&   9.813&   0.304&   0.623&  1&  2&  2&  3&  0&  2&  0.108&  0.117&   -0.5&    4.66 \\
  117121&   1.021&   1.103&  11.123&   0.668&   1.239&  6& 19& 11&  4&  7&  4&        &        &       &    7.62 \\
  117823&  -0.083&   0.493&   9.334&   0.299&   0.598&  5&  2&  8&  7& 16&  5&  0.109&  0.128&   -0.6&    4.54 \\
  117882&  -0.023&   0.467&   9.242&   0.270&   0.536&  2&  3& 10&  4&  4&  4&  0.033&  0.037&    0.0&    4.62 \\
  118165&  -0.014&   0.416&   9.129&   0.241&   0.481&  3&  6&  2&  2&  3&  4&  0.021&  0.023&    0.1&    3.95 \\
 \\
   59750&  -0.121&   0.466&   6.104&   0.286&   0.571&&&&&&  1&  0.131&  0.154&   -0.8&    4.34 \\
   66169&   0.341&   0.778&  10.108&   0.437&   0.837&&&&&&  1&  0.049&  0.049&   -0.1&    6.02 \\
   67655&   0.040&   0.659&   7.971&   0.376&   0.755&&&&&&  1&  0.161&  0.161&   -0.8&    5.98 \\
   68165&   0.601&   0.902&   9.981&   0.523&   0.995&&&&&&  1&  0.043&  0.043&    0.0&    6.62 \\
   68452&  -0.073&   0.452&   8.877&   0.266&   0.554&&&&&&  1&  0.074&  0.084&   -0.3&    4.25 \\
   68870&  -0.018&   0.433&   9.506&   0.255&   0.508&&&&&&  1&  0.021&  0.024&    0.1&    4.48 \\
   70152&   0.858&   0.988&  10.600&   0.566&   1.037&&&&&&  1& -0.042& -0.042&    0.4&    6.49 \\
   78952&   0.027&   0.632&   9.830&   0.362&   0.713&&&&&&  1&  0.141&  0.141&   -0.7&    5.24 \\
   80295&   0.883&   1.023&  10.383&   0.585&   1.067&&&&&&  1&  0.003&  0.003&    0.2&    6.96 \\
   80422&  -0.034&   0.409&   8.500&   0.248&   0.499&&&&&&  1&  0.042&  0.046&   -0.1&    3.69 \\
   80448&   0.100&   0.634&   7.344&   0.370&   0.729&&&&&&  1&  0.071&  0.071&   -0.2&    4.06 \\
   81170&   0.120&   0.746&   9.607&   0.447&   0.900&&&&&&  1&  0.213&  0.225&   -1.3&    6.19 \\
   81617&   0.052&   0.324&   8.565&   0.189&   0.388&&&&&&  1&       &       &       &    3.73 \\
   82409&   0.047&   0.577&   9.456&   0.328&   0.655&&&&&&  1&  0.060&  0.060&   -0.1&    4.96 \\
   83070&  -0.070&   0.430&   8.799&   0.260&   0.520&&&&&&  1&  0.074&  0.083&   -0.3&    3.69 \\
   88084&   0.040&   0.622&   9.192&   0.349&   0.713&&&&&&  1&  0.116&  0.116&   -0.5&    5.48 \\
   88231&   0.031&   0.591&   9.960&   0.357&   0.710&&&&&&  1&  0.090&  0.092&   -0.3&    5.19 \\
   88648&  -0.144&   0.611&  10.184&   0.387&   0.782&&&&&&  1&  0.287&  0.287&   -1.8&    6.12 \\
   88955&   0.098&   0.627&   9.456&   0.344&   0.658&&&&&&  1&  0.064&  0.064&   -0.2&    5.35 \\
   89215&   0.188&   0.773&  10.351&   0.435&   0.860&&&&&&  1&  0.193&  0.209&   -1.2&    6.50 \\
   89396&  -0.049&   0.472&   7.992&   0.279&   0.553&&&&&&  1&  0.062&  0.069&   -0.2&    4.60 \\
   90616&   0.014&   0.617&  10.294&   0.370&   0.738&&&&&&  1&  0.136&  0.136&   -0.7&    5.65 \\
  106204&   1.246&   1.250&  10.620&   0.751&   1.369&&&&&&  1&       &       &       &    7.42 \\
\hline
 \end{tabular}
 \end{center}
Notes: \\
1. HIP 7687 = HD 10166, $\sigma_V = 0.028$ mag. \\
2. HIP 48146 = BD+9 2242, $\sigma_V = 0.034$ mag. \\
3. HIP 49574 = HD 87908, $\sigma_V = 0.025$ mag. \\
4. HIP 53911 = TW Hydrae, $\sigma+V = 0.227$ mag., known variable \\
5. HIP 60251 = HD 107490, $\sigma_V = 0.034$ mag. \\
6. HIP 90724 = HD 170368, reddened A star \\
7. HIP 112389 = AC +18 1061, $\sigma_V = 0.060$ mag., M dwarf \\
Stars with no listed M$_V$ have unreliable parallax measurements (see \S2.2)
 \end{table*}

 \begin {table*}
 \begin{center}
\caption{SAAO UBVRI photometry of known subdwarfs}
 \begin{tabular}{rrrrrrrrrccc}
  \hline \hline 
 Name & U-B & B-V & V & V-R & V-I &N$_{obs}$&$\delta_{U-B}$ & $\delta_{0.6}$ & [Fe/H]$_{0.6}$ & [Fe/H]$_{ref}$ \\
 HIP & $\pi$ (mas) &$\sigma_\pi$ & M$_V$ \\
\hline 
  -12 2669&  -0.155$\pm$0.003&   0.308$\pm$0.003&  10.232$\pm$0.008&   0.205$\pm$0.001&   0.430$\pm$0.001&  3&       &       &      &-1.49$^1$   \\
43099 & 5.76 & 1.50 & $4.03\pm0.50$ & \\
  G48-29&  -0.206$\pm$0.005&   0.389$\pm$0.010&  10.467$\pm$0.001&   0.259$\pm$0.001&   0.550$\pm$0.007&  3&  0.220&  0.272&   -1.7& -2.66$^1$  \\
 47480 & 1.20 & 4.79 & \\
  G80-15&  -0.083$\pm$0.003&   0.550$\pm$0.010&   6.679$\pm$0.005&   0.325$\pm$0.001&   0.662$\pm$0.006&  3&  0.163&  0.166&   -0.9& -0.85$^1$  \\
 17147 & 41.07 & 0.86 & 4.75$\pm0.05$ & \\
  G112-36&   0.293$\pm$0.004&   0.826$\pm$0.007&   9.239$\pm$0.007&   0.462$\pm$0.001&   0.932$\pm$0.005&  3&  0.194&  0.217&   -1.3& -0.82$^2$      \\
 37335 & 3.53 & 1.46 & \\
  G112-43&  -0.153$\pm$0.004&   0.484$\pm$0.007&   9.853$\pm$0.002&   0.306$\pm$0.003&   0.634$\pm$0.003&  3&  0.173&  0.203&   -1.1& -1.51$^1$ \\
 37671 & 3.08 & 5.4 & \\
  G112-54&   0.146$\pm$0.006&   0.728$\pm$0.001&   7.437$\pm$0.005&   0.430$\pm$0.001&   0.864$\pm$0.004&  3&  0.154&  0.161&   -0.8& -0.95$^1$\\
 38625 & 52.01 & 1.85 & 6.02$\pm0.08$ & \\
  G113-22&  -0.053$\pm$0.018&   0.603$\pm$0.010&   9.704$\pm$0.013&   0.360$\pm$0.003&   0.736$\pm$0.010&  3&  0.187&  0.187&   -1.0& -1.30$^1$ \\
 & \\
   G157-93&   0.102$\pm$0.005&   0.672$\pm$0.002&  10.116$\pm$0.001&   0.380$\pm$0.001&   0.763$\pm$0.002&  2&  0.114&  0.114&   -0.5&   -0.99$^1$ \\
 117041 & 8.46 & 1.76 & $4.75\pm0.40$ & \\
  G159-50&  -0.063$\pm$0.001&   0.583$\pm$0.002&   9.084$\pm$0.003&   0.341$\pm$0.001&   0.696$\pm$0.002&  2&  0.176&  0.178&   -1.0& -0.77$^3$      \\
 10449 & 16.17 & 1.34 & $5.12\pm0.17$ & \\
  G270-23&  -0.131$\pm$0.000&   0.469$\pm$0.002&   9.247$\pm$0.000&   0.291$\pm$0.004&   0.609$\pm$0.004&  2&  0.142&  0.167&   -0.9&  -1.30$^3$     \\
 3026 & 9.57 & 1.38 & $4.15\pm0.30$ & \\
  G271-62&  -0.177$\pm$0.002&   0.426$\pm$0.004&  10.361$\pm$0.005&   0.285$\pm$0.001&   0.607$\pm$0.003&  2&  0.182&  0.223&   -1.3& -2.62$^1$      \\
 8572 & 3.22 & 1.75 & \\
  HD 16031&  -0.189$\pm$0.007&   0.438$\pm$0.000&   9.777$\pm$0.003&   0.285$\pm$0.001&   0.600$\pm$0.004&  2&  0.191&  0.234&   -1.4& -1.66$^4$ \\
 11952 & 8.67 & 1.81 & $4.47\pm0.41$ & \\
  HD 74000&  -0.218$\pm$0.001&   0.423$\pm$0.002&   9.665$\pm$0.003&   0.279$\pm$0.007&   0.580$\pm$0.004&  3&  0.223&  0.269&   -1.7&  -1.52$^4$\\
 42592 & 7.26 & 1.32 & $3.97\pm0.36$ & \\
\hline
 \end{tabular}
 \end{center}
References: 1.  Carney {\sl et al.}, 1994; 2. - Clementini {\sl et al.}, 2000; 
3. -  Axer {\sl et al.}, 1994; 4. - Gratton {\sl et al.}, 1997.
 \end{table*}

\subsection {Summary}

We have catalogued the available photometric observations of the 317 stars listed in
Table 1: ninety-seven stars have Str\"omgren photometry; two hundred and twenty-two stars have Johnson/Cousins
UBV(RI) measurements, including 201 from our SAAO observing program; and nine stars
have Walraven photometry. Combined, these observations provide accurate 
colours and magnitudes for 251 stars, or
75\% of the sample. Forty-four stars have photometric abundances [Fe/H]$\le -1.0$, although twelve 
are identified as confirmed or probable binary stars. Thirty-one metal-poor stars (including six binaries)
have accurate UBVRI Johnson/Cousins photometry, the most extensive such dataset compiled to date for
subdwarfs with well-determined trigonometric parallaxes. We shall discuss these stars in more detail in 
section 7.

\section {Spectroscopic Observations}

In addition to photometric observations, we have obtained spectroscopy of over
170 stars from Table 1. Those spectra cover a number of prominent metal lines
and molecular bands,
and therefore provide additional abundance information. 

\subsection {Las Campanas Observations}

Our observations were obtained in January and November, 1998, using the modular spectrograph
at Las Campanas Observatory. The initial observations were made from January 1 to 6 (UT),
with the spectrograph mounted on the 100-inch Du Pont telescope; the later observations
were made on November 28, 29 and 30 (UT), with the same
spectrograph mounted on the Swope 40-inch telescope. 
In both cases, we employed a 600 l/mm grating, blazed at 5000\AA\, with the spectrum centred
at 4700\AA\ on the detector, a SITE CCD chip. The 100-inch observations provide
wavelength coverage from 3780 to 6000\AA\ at a dispersion of 1.3 \AA pix$^{-1}$; the 40-inch
data span 3790 to 5900 \AA\ at a dispersion of 1.1 \AA pix$^{-1}$. The former observations have
a resolution of 2.6\AA; the latter, a resolution of 2.9\AA. 

The data were reduced using standard techniques incorporated in the {\sl IRAF} software analysis
package. The spectra were flat-field corrected using observations of tungsten lamps, and the
individual spectra extracted using the {\sl apextract} task. The wavelength calibration for
both datasets was set  using observations of hollow-cathode arc lamps obtained
at the start of each night. Given the relatively low resolution of these spectra,  we have
not attempted to determine radial velocities for the target stars. 

Once wavelength calibration was established, the spectra were set  on a flux scale using
observations of the standard stars LTT 1020, LTT 2415 and Hiltner 600 (Baldwin \& Stone, 1984). 
Inter-comparison of standards shows that the calibration is generally accurate to better than 5\%
on large scales ($>150\AA$). Since our main priority is measuring equivalent widths of individual atomic
features, these data are adequate for our present purposes. 

\subsection {Metallicities of programme stars}

The wavelength r\'egime spanned by our observations includes a number of strong atomic lines
and molecular bands in common use  as abundance indicators for F-, G- and K-type
stars.  Those include the Ca II H \& K lines, Ca I 4227\AA, the G-band at 4300\AA, 
the Mgb triplet and numerous FeI lines (notably at 5270 and 5331 \AA), besides the 
hydrogen Balmer lines H$\beta$ to H9. The latter, particularly H$\beta$, are important
in offering temperature calibration, since many of the Hipparcos stars lack reliable
photometric colours. 

Tables 9 and 10 list our equivalent width measurements of several
key features: the Ca K line, the G-band, the strongest Mgb feature (5170\AA), H$\beta$, and the Fe I
5270  and 5331 \AA\ lines (Fe1 and Fe2, respectively). 
The 100-inch spectra were measured by SM, and the 40-inch
spectra by INR. Although both sets of  measurements are
internally consistent, there are systematic differences in approach. There are
only a handful of stars in common between the two sets of observations, but each encompasses
a substantial number of stars of known abundance. Since our goals are qualitative, rather
than quantitative, we analyse the two datasets separately.

All of the measured features are strong lines, since our primary aim remains
identifying metal-poor stars likely to have [Fe/H]$\le -1$. The chosen
indicators are insensitive to metallicity variation at near-solar
abundance, but provide good discrimination at halo abundances. Repeat observations of
a number of stars show that the random uncertainties associated with the measured 
equivalent widths
are typically $\pm0.15$\AA. Those uncertainties are likely to be characteristic of the
errors in our observations of metal-poor stars. However, systematic errors, notably
regarding continuum placement, are undoubtedly present at a comparable level
in line-rich, disk-abundance dwarfs. 

We have calibrated our chosen abundance indicators using observations of stars with
spectroscopically-determined abundances, either from Table 2 or from the extensive catalogue
of Carney {\sl et al.} (1994). In both cases, we supplement those
calibrators with programme stars whose metallicity has been determined using Str\"omgren photometry
(Table 3).  Figure 14 illustrates the methods employed. Since our
chosen indices lack sensitivity at near-solar abundances, we have not attempted to separate 
metal-rich and intermediate abundance stars, and identify only the candidate halo-abundance
subdwarfs. Where necessary, we have also used our our spectroscopic data to estimate spectral types,
and determine whether those values are compatible with the (B-V)$_T$ colours. 

We have spectroscopy of 175 stars from Table 1. Forty-six  have prior metallicity
determinations, either from spectroscopic observations or Str\"omgren photometry.
The majority of the remainder have linestrengths consistent with disk-like 
abundances. Only thirteen are candidate subdwarfs, showing potential for
abundances [Fe/H]$ < -1.0$. The latter stars are identified in Tables 8 and 9
(`H' in column 8).
Twelve of the thirteen have UBV data and $\delta_{0.6}$ measurements: seven have [Fe/H]$_{0.6} \le -0.7$;
HIP 10353, 24935 and 97463 have [Fe/H]$_{0.6} \sim -0.6$; and HIP 5097 and 35232 have near-solar
metallicity estimates. HIP 17241 and 54834, in contrast, have line-strengths comparable 
with those of
extreme halo subdwarfs, such as HD 19445 and HD 64090. 
More detailed spectroscopic observations of these stars will provide more accurate abundance
estimates.

\begin{figure*}
\psfig{figure=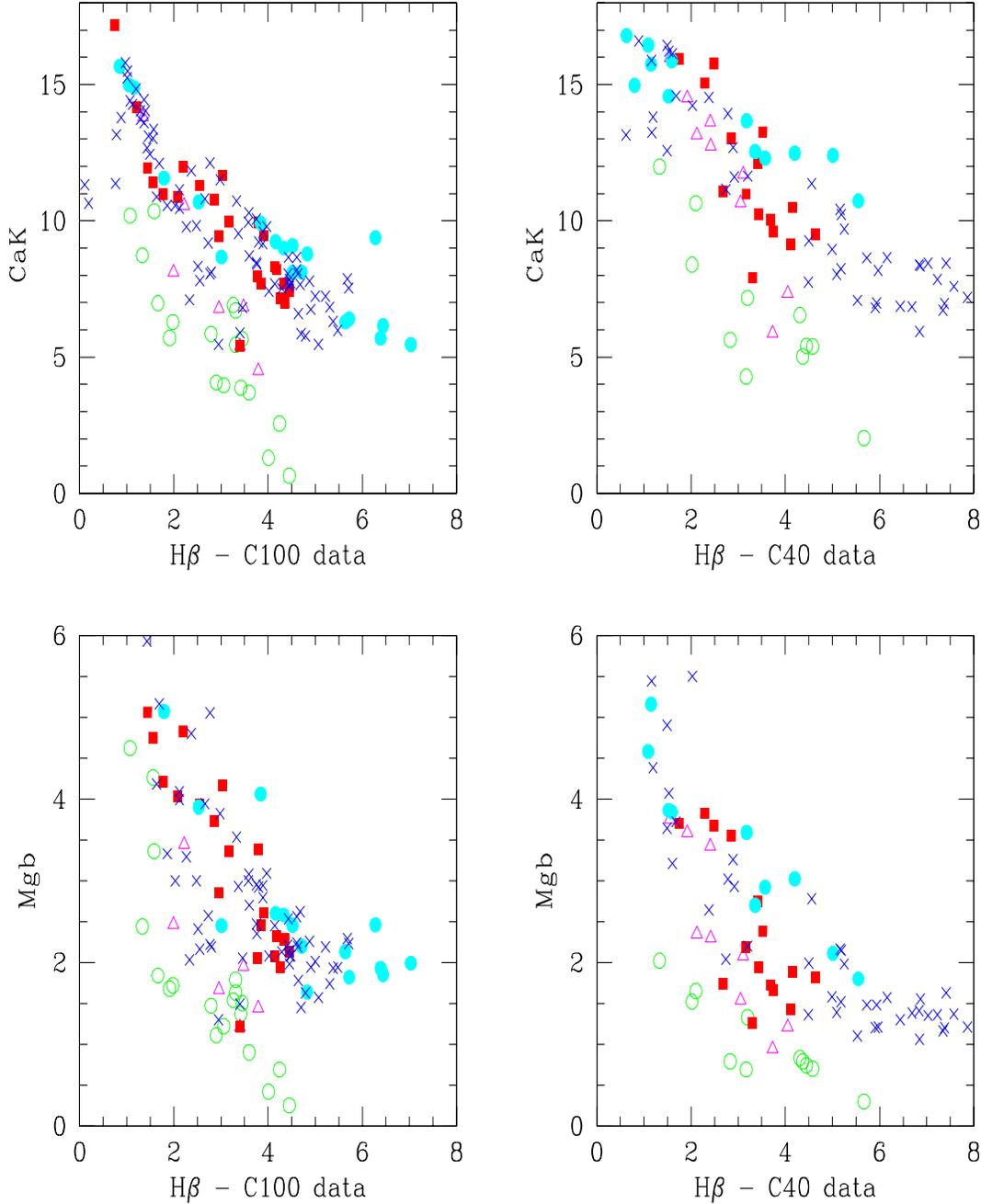,height=18cm,width=15cm
,bbllx=8mm,bblly=57mm,bburx=205mm,bbury=245mm,rheight=19cm}
\caption{Ca II K-line and Mgb equivalent width measurements from our Las Campanas observations
plotted against H$\beta$ equivalent widths. Stars with spectroscopic or Str\"omgren metallicities
[Fe/H]$> -0.3$ are plotted as solid points; open squares mark stars with $-0.3 \le [Fe/H] < -1.0$; 
solid triangles are halo subdwarfs with $-1.0 \le [Fe/H] < -1.5$; and open circles identify 
extreme subdwarfs. Hipparcos stars lacking such abundance measurements are plotted as crosses; 
candidate subdwarfs are marked as encircled crosses.  }
\end{figure*}

As Figure 14 shows, our C40 observations include a higher proportion of earlier-type
dwarfs, with H$\beta$ equivalent widths exceeding 6\AA. G271-11 is the only calibrator
with Balmer lines of comparable strength; both Ca II K and Mgb are significantly
stronger in that star than the Hipparcos stars. Carney {\sl et al.} (1994) measure a spectroscopic
abundance of [Fe/H]=-0.57; however, the UBV colours ( (U-B)=0.06, (B-V)=0.59 ) imply
$\delta_{0.6} = 0.06$, and [Fe/H]$\approx -0.2$. 
Fortunately, the C40 Hipparcos sample includes both Pleiads members, HIP 17481 and 17494,
and their location in the H$\beta$/Mg and H$\beta$/Ca K diagrams confirms that none
of the earlier type stars is a candidate halo subdwarf.

There are only nine stars where our abundance estimates are based solely on the
LCO spectroscopic data: HIP 895, 19637, 21478, 29510, 34145, 68452, 68870, 103287 and 110621.
As already noted, the line indices measured for HIP 34145 suggest a halo metallicity.
HIP 895, 68452 and 110621 have Ca II, Mgb and Fe equivalent widths of intermediate
strength, and are classed as mildly metal-poor (`I' in Table 1); the remaining stars
have near-solar abundances.

Several stars require individual mention:
\begin{description}
\item [HIP 3531AB:] also known as BD -8 133, our observations indicate that this is a close double 
star, separation $\sim 5$ arcseconds. The primary is a late-type K dwarf; the secondary, an
M2/M3 dwarf, $\sim 2.5$ magnitudes  fainter. Both are near-solar abundance disk dwarfs.
\item [HIP 4750:] Ca, Mg and Fe equivalent widths are consistent with [Fe/H]$< -1$; the G-band
is significantly stronger than average. We identify this as a likely CH-strong subdwarf.
\item [HIP 10360:] Our spectroscopy confirms that this star is of spectral type A. The
 strong Ca II K line suggests a relatively high metallicity.
\item [HIP 28121/28122:] As noted above, these stars form the binary BD +10 936A/B. The
fainter star, HIC 28121, was not observed by Hipparcos. Our measured linestrengths indicate
that both stars are at most mildly metal poor, [Fe/H]$> -0.5$. 
\item [HIP 43490:] also known as CD -79 347A, this subdwarf has a common proper-motion companion, 
$\sim1$ arcminute South and $\sim2$ magnitudes fainter. Photometry and spectroscopy of
CD -79 347B will therefore provide additional calibration of the [Fe/H]$\sim -1$  main sequence.
\item [HIP 44436:] As noted above, our spectroscopy shows that this is an early-type star.
In addition to strong Balmer lines, we detect Mg II 4481\AA\ with an equivalent width of 0.5 \AA.
\item [HIP 49574:] Another candidate CH-strong subdwarf
\item [HIP 52285:] also known as BD +4 2370, spectral type K2. The companion, 
BD +4 2370B, is $\sim1$ magnitude fainter. 
\item [HIP 65940:] Upgren (1972) classifies this star as spectral type K2V.
\item [HIP 106204:] Upgren (1972) classifies this star as spectral type K7V.
\item [HIP 106904:] Another subdwarf with CH bandstrengths apparently stronger than normal
for metal-poor stars.
\item [HIP 117882:] As noted in \S4.1, the linestrength measured for this star indicate
a near-solar abundance, inconsistent with the absolute magnitude implied by the Hipparcos
parallax.
\end{description}

 \begin {table}
 \begin{center}
\caption{Spectroscopic indices: LCO 100-inch data}
 \begin{tabular}{rcccccccl}
  \hline \hline 
Name & Ca K & CH & Mg & Fe1 & Fe2 & H$\beta$ & [Fe/H] \\
\hline 
    HIP 435&   7.0&   4.4&   2.3&   1.6&   1.7&   4.3& -0.97 \\
       1719&  14.4&  10.9&   8.2&   3.9&   3.0&   1.4&       \\
       3139&   7.4&   3.5&   2.1&   1.6&   1.5&   4.4& -0.49 \\
       3531$^1$&  13.2&  10.8&  11.4&   6.3&   7.1&   0.8&       \\
       3531B$^{1,2}2$&  11.3&   4.4&  11.3&   3.4&   4.9&   0.1&     \\
       4450&   9.8&   6.0&   3.0&   2.2&   2.2&   2.5&       \\
       4576&   8.5&   4.1&   2.5&   2.0&   2.0&   3.8&       \\
       4750&   7.1&   8.5&   2.0&   1.1&   1.1&   2.3&  H      \\
       4981&  10.9&   7.9&   4.0&   2.2&   2.3&   2.1& -0.58 \\
       5004&  13.8&   9.4&   6.4&   2.2&   2.4&   1.4& -1.02 \\
       5097&   8.1&   4.3&   2.2&   1.7&   1.8&   2.8&   H  \\
       6758&  11.1&   8.3&   4.1&   2.6&   2.3&   2.1&       \\
       6758&   9.8&   6.5&   3.3&   2.3&   2.3&   2.3&       \\
       7303&  10.8&   7.7&   3.9&   1.9&   1.7&   2.7&       \\
       7459&   8.1&   4.3&   2.5&   1.2&   1.4&   2.0& -1.17 \\
       7687&  13.7&   9.8&   7.2&   3.7&   3.5&   1.3&       \\
       7772&   7.2&   3.0&   2.0&   1.9&   1.8&   5.0&       \\
       8389&  15.5&  10.8&  10.1&   4.2&   4.0&   1.0&       \\
       8558&   4.5&   1.3&   1.4&   1.0&   1.1&   3.8& -1.10 \\
       9634&   7.7&   3.4&   2.1&   1.5&   1.6&   4.5&       \\
      10208&   7.5&   3.5&   2.1&   1.6&   1.6&   4.3&       \\
      10360$^3$&   4.7&   1.2&   1.6&   1.5&   1.4&   9.5&      \\
      10385&  14.3&  10.8&  10.5&   5.4&   5.5&   1.1&       \\
      10637&   7.8&   3.8&   2.1&   1.6&   1.4&   4.4&       \\
      11435&   8.7&   4.1&   2.5&   2.0&   2.0&   4.4&       \\
      13849&  15.2&  11.1&  10.2&   5.1&   5.2&   1.0&       \\
      14192&   9.2&   4.6&   2.6&   2.1&   2.0&   2.7&       \\
      15756&  14.0&  11.4&   9.4&   5.1&   5.0&   1.3&       \\
      16479&   8.2&   3.7&   2.2&   1.9&   2.0&   4.6&       \\
      16089&   7.5&   3.0&   2.2&   2.1&   1.7&   5.7&       \\
      17085&   8.1&   3.4&   2.2&   1.8&   1.8&   4.7& -0.22 \\
      17241&   5.5&   2.0&   1.3&   0.7&   0.8&   3.0&    H   \\
      19007&  14.2&  10.0&   7.6&   3.5&   3.4&   1.2& -0.62 \\
      21000A&   9.2&   4.6&   2.6&   2.1&   2.0&   4.2& -0.16 \\
      21000B$^2$&   9.0&   5.0&   2.6&   2.1&   2.0&   4.3& -0.16 \\
      21125&  12.1&   9.3&   5.2&   2.5&   2.2&   1.7&       \\
      21609&  10.4&   6.5&   3.4&   1.4&   1.3&   1.6& -1.76 \\
      22177$^1$&  11.4&   8.4&  10.2&   5.8&   6.6&   0.8&      \\
      22632&   6.7&   2.9&   1.8&   0.8&   0.7&   3.3& -1.59 \\
      23573&  13.1&  10.2&   6.5&   2.7&   2.6&   1.5&       \\
      24296&  10.5&   7.1&   4.0&   2.2&   2.3&   2.1&       \\
      24316&   6.3&   3.1&   1.7&   0.7&   0.6&   2.0& -1.52 \\
      24935&   8.3&   4.4&   2.4&   1.5&   1.5&   2.5&   H  \\
      24421&  10.1&   5.8&   3.0&   2.1&   2.1&   3.8&       \\
      26688&   5.9&   3.4&   1.5&   1.0&   1.0&   4.7&       \\
      29322$^1$&  10.6&   5.7&   8.8&   4.0&   4.4&   0.2&     \\
      31639&  12.7&   9.3&   5.9&   2.7&   2.5&   1.4&       \\
      33283&  13.6&  10.1&   8.2&   3.2&   2.8&   1.4&       \\
      34146&   8.3&   4.0&   2.1&   1.1&   1.3&   4.1& -0.40 \\
      35232&   7.8&   4.2&   2.2&   1.5&   1.5&   2.5&    H  \\
      36818&  10.6&   6.1&   3.3&   1.6&   1.9&   1.9& -0.75 \\
      39911&  11.0&   7.5&   4.2&   1.8&   2.0&   1.8& -0.92 \\
      41563&   6.8&   2.6&   1.7&   1.8&   1.8&   5.3&       \\
\hline
 \end{tabular}
 \end{center}
 \end{table}

\setcounter{table}{8}
 \begin {table}
 \begin{center}
\caption{Spectroscopic indices: LCO 100-inch data (contd.)}
 \begin{tabular}{rcccccccl}
  \hline \hline 
Name & Ca K & CH & Mg & Fe1 & Fe2 & H$\beta$ & [Fe/H]  \\
\hline 
   43445&   8.4&   4.5&   2.3&   1.6&   1.5&   3.8&       \\
      43490&  10.6&   5.8&   3.0&   1.5&   1.1&   2.0&   H  \\
      43973&  11.8&   8.3&   4.8&   2.2&   2.0&   2.4&       \\
      44116&   7.2&   3.5&   1.9&   1.4&   1.5&   4.2& -0.58 \\
      44124&   5.4&   2.2&   1.6&   0.7&   0.8&   3.3& -2.00 \\
      44435&   6.3&   2.5&   2.1&   1.9&   1.3&   5.6& -0.06 \\
      44436&   1.2&   0.0&   0.6&   0.4&   0.4&  15.6&      \\
      46120&   6.4&   3.7&   1.7&   0.6&   0.6&   1.9& -2.10 \\
      47171&  10.7&   7.2&   3.5&   2.1&   2.2&   3.3&       \\
      47948&  12.1&   8.9&   5.1&   2.7&   2.7&   2.8&       \\
      49574&   6.8&   4.9&   2.0&   1.3&   1.3&   3.5&   H  \\
      49785&   9.2&   4.9&   2.9&   1.8&   1.7&   3.8&       \\
      50532&  10.3&   6.2&   3.1&   2.2&   2.1&   3.6&       \\
      50965&  10.8&   7.2&   3.7&   1.8&   1.9&   2.8& -0.39 \\
      51156&   7.7&   3.6&   2.1&   1.6&   1.7&   4.5&       \\
      51298&   8.1&   3.6&   2.2&   1.9&   2.0&   4.6&       \\
      51300&   7.2&   3.2&   2.2&   1.8&   1.8&   5.2&       \\
      51769&  11.9&   9.1&   5.1&   2.2&   1.8&   1.4& -0.65 \\
      54641&   6.8&   3.9&   2.0&   1.0&   0.9&   3.5& -1.04 \\
      54834&   5.9&   2.4&   1.5&   0.8&   0.8&   3.4&   H  \\
      55790&   5.7&   2.4&   1.5&   0.7&   0.7&   3.1& -1.56 \\
      55978&   6.4&   2.5&   1.8&   1.8&   1.2&   5.7&  0.07 \\
      57360&   6.8&   3.0&   1.7&   0.9&   1.0&   3.0& -1.22 \\ 
      59750&   7.9&   4.0&   2.0&   0.4&   0.4&   3.8& -0.78 \\
      60251&   9.9&   6.1&   3.0&   2.1&   2.2&   3.6&       \\
      60632&   5.4&   2.0&   1.2&   0.6&   0.6&   3.4& -1.45 \\
      60852&   8.2&   4.6&   2.3&   1.7&   1.8&   4.2& -0.54 \\
      63781&   6.3&   2.2&   1.9&   1.8&   1.5&   5.4&       \\
      64765&   5.7&   2.3&   1.9&   1.8&   1.6&   6.4& -0.15 \\
      65040&  12.0&   8.4&   4.8&   2.2&   1.9&   2.2& -0.71 \\
      65201&   3.9&   1.2&   1.4&   0.6&   0.6&   3.4& -1.96 \\
      65940$^3$&  14.8&  11.2&   9.7&   4.9&   4.7&   1.2&       \\
      66169&  14.0&  10.5&   7.3&   3.6&   3.4&   1.4&       \\
      66815&   9.4&   6.4&   3.0&   1.7&   1.7&   2.7&       \\
      67189&   7.7&   4.1&   2.6&   2.1&   2.0&   4.7&       \\
      67655&  11.4&   8.5&   4.8&   2.1&   1.8&   1.6& -0.92 \\
      66815&   9.5&   6.3&   2.8&   1.7&   1.5&   3.0& -0.64 \\
      68165&  15.7&   9.8&   9.9&   4.1&   3.9&   0.9& -0.23 \\
      68452&   7.4&   3.7&   2.1&   1.6&   1.6&   4.0&       \\
      68870&   7.8&   3.3&   2.3&   1.9&   2.0&   4.9&       \\
      70152&  15.0&  11.4&  10.1&   5.4&   5.4&   1.0& -0.03 \\
      70681&  10.6&   6.3&   3.4&   1.5&   1.5&   2.2& -1.25 \\
      70689&   6.0&   2.5&   1.9&   1.7&   1.3&   5.5&       \\
      71886&   7.6&   3.5&   2.0&   1.6&   1.5&   4.4&       \\
      73798&  10.9&   8.1&   4.2&   1.7&   1.7&   1.6&    H   \\
      74078&   8.7&   4.5&   2.6&   2.0&   1.9&   4.6&       \\
      74994&   5.5&   2.6&   2.0&   2.0&   1.9&   7.0& -0.06 \\
     107873&   9.2&   5.4&   2.8&   2.1&   2.0&   3.9&       \\
     108095&  10.0&   6.3&   3.4&   2.1&   1.9&   3.2&       \\
     109869&  11.5&   7.5&   3.8&   2.8&   2.6&   3.0&       \\
     111426&   8.1&   4.0&   2.2&   1.8&   1.5&   4.4&       \\
     112384&   5.8&   2.6&   1.6&   1.2&   0.9&   4.8&       \\
     113430&   8.1&   4.0&   2.5&   1.8&   1.8&   4.5& -0.22 \\
\hline
 \end{tabular}
 \end{center}
 \end{table}

\setcounter{table}{8}
 \begin {table}
 \begin{center}
\caption{Spectroscopic indices: LCO 100-inch data (contd.)}
 \begin{tabular}{rcccccccl}
  \hline \hline 
Name & Ca K & CH & Mg & Fe1 & Fe2 & H$\beta$ & [Fe/H]  \\
\hline 
  113868&  13.4&   9.9&   6.4&   3.5&   3.3&   1.6&       \\
     114271&   3.7&   1.3&   0.9&   0.4&   0.6&   3.6& -1.80 \\
     114487&   5.5&   2.4&   1.6&   1.2&   1.0&   5.1&       \\
     114735&   6.6&   2.7&   1.8&   1.5&   1.5&   4.6&       \\
     114837&   9.9&   6.1&   3.4&   2.3&   2.3&   3.8& -0.44 \\
     115031&   7.7&   4.2&   2.5&   2.0&   1.1&   4.1&       \\
     115361&   7.7&   3.8&   2.3&   1.8&   1.9&   4.3& -0.39 \\
     116437&   9.5&   5.5&   2.9&   1.8&   1.8&   3.4&       \\
     117121$^1$&  13.8&  10.0&  10.9&   5.7&   5.7&   0.9&    \\
     117823&   8.7&   5.3&   2.7&   1.6&   1.6&   3.6&       \\
     117882&   9.1&   4.2&   2.5&   2.1&   2.0&   4.5&  0.20 \\
     118165&   7.9&   3.4&   2.3&   2.0&   2.0&   5.7&       \\
  & \\
  vB 100    &   9.4&   2.5&   2.5&   2.3&   2.1&   6.3&  0.15 \\
  vB 101    &   8.8&   4.2&   1.6&   2.2&   2.5&   4.8&  0.15 \\
  vB 105    &   9.9&   7.5&   4.1&   3.0&   2.9&   3.8&  0.15 \\
  vB 142    &  11.6&   8.2&   5.1&   3.5&   3.2&   1.8&  0.15 \\
  HD 3567   &   6.9&   2.6&   1.5&   0.9&   1.1&   3.3& -1.50 \\
  HD 19445  &   4.0&   2.0&   1.2&   0.5&   0.6&   3.1& -1.81 \\
  HD 84937  &   2.6&   0.6&   0.7&   0.3&   0.5&   4.2& -2.04 \\
  G1-9     &  11.7&   8.5&   4.2&   2.6&   2.6&   3.0& -0.51 \\
  G1-28    &  14.9&  10.3&   8.2&   3.5&   3.2&   1.2& -0.37 \\
  G8-15    &  30.0&   5.5&   9.9&   4.5&   3.7&   1.2& -0.85 \\
  G30-52   &  10.2&   8.9&   4.6&   3.9&   3.9&   1.1& -2.10 \\
  G32-26   &  20.7&   6.9&   4.3&   3.0&   3.5&   1.6& -1.53 \\
  G64-12   &   0.9&   0.0&   0.3&   0.2&   0.5&   4.2& -3.52 \\
  G158-100 &   8.7&   4.7&   2.4&   0.9&   1.4&   1.3& -2.64 \\
  G270-71  &   9.4&   5.7&   2.6&   1.9&   2.2&   3.9& -0.49 \\
  G271-11  &  11.3&   8.1&   3.9&   2.0&   2.0&   2.5& -0.57 \\
  G271-34  &  10.7&   7.8&   3.9&   1.9&   1.8&   2.5& -0.03 \\
\hline
 \end{tabular}
 \end{center}
Notes: \\
1. M dwarf;  2. Possible companion; 3. K dwarf \\
vB 100, 101, 105 \& 142 are all members of the Hyades cluster
 \end{table}

 \begin {table}
 \begin{center}
\caption{Spectroscopic indices: LCO 40-inch data }
 \begin{tabular}{rcccccccl}
  \hline \hline 
Name & Ca K & CH & Mg & Fe1 & Fe2 & H$\beta$ & [Fe/H]  \\
\hline 

 HIP 895   &   7.0&   2.9&   1.5&   1.6&   0.6&   5.9&        \\
    1051   &   7.1&   3.2&   1.1&   1.2&   0.4&   5.5&        \\
    3855$^1$   &  16.6&  18.2&  11.0&   4.6&   2.4&   0.9&        \\
    6251   &   9.7&   7.4&   2.0&   1.5&   0.5&   5.2&        \\
    7935   &   8.6&   5.4&   1.5&   1.5&   0.5&   5.7&        \\
    8298   &  14.6&  10.8&   3.7&   2.9&   1.4&   1.7&        \\
   10353   &   7.8&   3.5&   1.4&   1.3&   0.5&   4.5&   H    \\
   10375   &   6.8&   3.5&   1.2&   1.2&   0.5&   5.9&        \\
   14594   &   4.3&   1.7&   0.7&   0.3&   0.2&   3.2&  -1.81 \\
   17481$^2$   &   5.3&   1.9&   1.0&   1.5&   0.4&  10.0&  -0.1  \\
   17497$^2$   &   6.7&   3.1&   1.3&   1.6&   0.5&   9.1&  -0.1  \\
   19637   &   8.2&   3.7&   1.2&   1.2&   0.5&   6.0&        \\
   18700   &  10.2&   9.3&   2.2&   1.6&   0.7&   5.2&        \\
   19797   &   5.0&   1.7&   0.8&   0.4&   0.3&   4.4&  -1.57 \\
   21478   &  11.4&   7.9&   2.8&   2.0&   0.8&   4.6&        \\
   25717   &   6.2&   3.7&   1.3&   2.1&   0.6&   8.7&        \\
   26676   &  12.7&  10.9&   2.3&   1.6&   0.9&   2.4&        \\
   28122A$^3$  &   7.2&   3.5&   1.2&   1.5&   0.6&   7.9&        \\
   28122B  &   8.9&   4.3&   1.6&   1.6&   0.6&   5.0&        \\
   28188   &   8.2&   4.8&   1.5&   1.2&   0.6&   5.2&        \\
   29510   &   6.8&   2.4&   1.1&   1.9&   0.4&  10.5&        \\
   32009   &  13.9&  15.7&   3.0&   2.7&   1.4&   2.8&        \\
   32308$^1$   &  15.9&  12.3&   5.4&   4.7&   2.4&   1.2&        \\
   34145   &  11.1&   8.5&   2.0&   1.8&   0.6&   2.7&   H    \\
   36491   &   9.6&   5.7&   1.7&   1.1&   0.5&   3.7&  -0.90 \\
   36818   &  11.1&   6.8&   1.7&   1.4&   0.6&   2.7&  -0.75 \\
   38541   &   8.4&   7.9&   1.5&   0.9&   0.5&   2.0&  -1.70 \\
   42278   &   5.9&   1.8&   1.0&   1.1&   0.4&   9.7&        \\
   47161   &   8.4&   4.4&   1.4&   1.8&   0.5&   6.8&        \\
   48146   &  10.7&   9.4&   1.8&   1.8&   0.6&   5.6&  -0.05 \\
   52285A$^{1,3}$  &  16.4&  12.2&   3.6&   3.5&   1.7&   1.5&        \\
   52285B$^{1,3}$  &  13.1&  15.4&   7.6&   5.0&   2.5&   0.6&        \\
   54519$^1$   &  12.6&  12.9&   4.9&   4.2&   2.2&   1.5&        \\
   54993   &   7.6&   3.0&   1.4&   1.8&   0.5&   7.6&        \\
   95800   &   6.7&   2.5&   1.2&   1.3&   0.4&   7.3&        \\
   97127   &   6.8&   2.7&   1.4&   1.8&   0.5&   6.7&        \\
   97463   &  11.6&   8.6&   2.2&   1.7&   0.6&   3.2&   H    \\
  100207   &   6.9&   3.3&   1.3&   1.4&   0.5&   6.4&        \\
  101103   &   5.9&   2.4&   1.1&   1.1&   0.4&   6.8&        \\
  101650   &   8.4&   4.2&   1.4&   1.8&   0.6&   7.0&        \\
  101892   &   8.4&   4.3&   1.6&   1.7&   0.6&   7.4&        \\
  101987   &   8.4&   5.1&   1.5&   1.7&   0.7&   6.9&        \\
  103287   &  16.1&  15.7&   3.2&   3.4&   1.5&   1.6&        \\
  103714   &  10.4&   8.6&   2.2&   2.7&   0.9&   5.2&        \\
  104289   &  12.7&  14.4&   3.3&   1.9&   1.1&   2.9&        \\
  106204$^1$   &  13.2&  18.8&  11.9&   5.4&   2.9&   1.2&        \\
  106904   &  11.6&  11.0&   2.9&   1.7&   0.9&   2.9&   H    \\
  108006   &   6.7&   3.3&   1.4&   1.4&   0.5&   8.4&        \\
  108598   &  13.8&  11.9&   4.4&   2.9&   1.3&   1.2&        \\
  108655   &   7.0&   2.1&   1.2&   1.5&   0.4&   7.4&        \\
  109067   &  14.5&  12.3&   2.6&   1.8&   1.0&   2.4& -0.97  \\
  110621   &   8.0&   3.5&   1.4&   1.4&   0.4&   5.1&        \\
  110776   &  14.2&  17.5&   5.5&   4.1&   1.7&   2.0&        \\
  111374   &   7.8&   3.8&   1.4&   1.5&   0.4&   7.2&        \\
  113542   &   6.1&   1.0&   1.2&   1.4&   0.4&   9.2&        \\
\hline
 \end{tabular}
 \end{center}
 \end{table}

\setcounter{table}{9}
 \begin {table}
 \begin{center}
\caption{Spectroscopic indices: LCO 40-inch data (contd.)}
 \begin{tabular}{rccccccl}
  \hline \hline 
Name & Ca K & CH & Mg & Fe1 & Fe2 & H$\beta$ & [Fe/H] \\
\hline 

  114627   &   8.6&   5.2&   1.6&   1.5&   0.5&   6.2&        \\
  114837   &   9.3&   6.2&   2.0&   1.9&   0.7&   4.5&        \\
  115194   &  16.2&  11.4&   4.1&   3.1&   1.4&   1.5&        \\
 & \\
-33 3337  &   5.9&   3.3&   0.9&   0.5&   0.2&   3.7&  -1.33 \\
-3 740    &   3.0&      &   0.8&   1.0&   0.3&  18.4&  -2.73 \\
   HD 16031 &   5.4&   1.9&   0.7&   0.4&   0.2&   4.6&  -1.82 \\
 HD 59984   &   7.9&   6.7&   1.3&   1.2&   0.3&   3.3&  -0.64 \\ 
 Gl 105A   &  16.5&  15.9&   4.6&   4.2&   2.3&   1.1&        \\
  Gl 135   &  12.5&  13.1&   2.7&   2.1&   0.9&   3.4&        \\
 Gl 764.1A &  15.8&  12.3&   5.2&   4.3&   2.0&   1.1&        \\
 Gl 764.1B &  16.8&      &   7.3&   4.7&   2.5&   0.6&        \\
 Gl 775    &  15.0&  15.9&   6.4&   5.2&   2.7&   0.8&        \\  
  G2-47    &  13.6&  14.8&   3.4&   2.5&   1.2&   2.4&  -1.00 \\
  G5-42    &  12.3&   8.8&   2.9&   2.6&   1.2&   3.6&   0.04 \\
  G5-46    &  10.2&   8.1&   1.9&   1.6&   0.6&   3.4&  -0.34 \\
  G41-34   &  13.7&  21.8&   3.6&   3.3&   1.5&   3.2&  -0.20 \\
  G41-41   &   2.0&   0.5&   0.3&   0.1&   0.1&   5.7&  -2.80 \\
  G43-7    &  15.1&  15.8&   3.8&   3.3&   1.2&   2.3&  -0.68 \\
  G71-27   &   9.5&   7.8&   1.8&   1.7&   0.5&   4.6&  -0.63 \\
  G73-67   &  13.0&  11.9&   3.5&   2.8&   1.3&   2.8&  -0.43 \\
  G82-5    &  13.3&  11.0&   2.4&   2.0&   1.1&   3.5&  -0.70 \\
  G82-18   &  14.5&  21.0&   3.6&   4.0&   2.2&   1.9&  -1.41 \\
  G82-44   &   7.2&   4.8&   1.3&   0.6&   0.4&   3.2&  -2.10 \\
  G82-47   &   9.7&   6.4&   1.6&   2.0&   0.6&   6.8&  -0.55 \\
  G83-42   &  16.0&  12.4&   3.8&   2.3&   1.2&   1.5&  -1.47 \\
  G83-45   &  12.1&   6.7&   1.8&   2.0&   0.9&   7.4&  -2.49 \\
  G83-51   &  10.5&  15.3&   2.1&   1.5&   1.2&   5.2&  -1.51 \\
  G84-9    &  10.0&   6.2&   1.7&   1.4&   0.5&   3.7&  -0.94 \\
  G84-16   &  12.0&   9.9&   2.0&   1.3&   0.8&   1.3&  -1.95 \\
  G84-37   &   9.1&   5.4&   1.4&   1.0&   0.3&   4.1&  -0.92 \\
  G89-14   &   5.4&   1.9&   0.7&   0.5&   0.2&   4.4&  -1.90 \\
  G89-21   &  11.7&   9.8&   2.1&   1.4&   0.7&   3.1&  -1.15 \\
  G94-3    &  12.4&   8.2&   1.9&   0.9&   0.7&   1.8&        \\
  G98-42   &  15.9&  14.6&   3.8&   4.0&   1.9&   1.6&  -0.18 \\
  G99-48   &  10.6&   8.5&   1.6&   0.9&   0.5&   2.1&  -2.14 \\
  G110-34  &   5.6&   2.2&   0.8&   0.6&   0.3&   2.8&  -1.73 \\
  G112-43  &   6.5&   2.7&   0.8&   0.7&   0.2&   4.3&  -1.51 \\
  G112-44  &   7.3&   4.4&   1.2&   0.9&   0.3&   4.1&  -1.27 \\
  G112-56  &  14.0&  11.5&   2.8&   2.2&   0.9&   3.3&        \\
  G113-22  &  10.7&   7.6&   1.5&   1.2&   0.5&   3.0&  -1.30 \\
  G113-24  &  10.5&  10.6&   1.9&   1.4&   0.6&   4.2&  -0.51 \\
  G160-1   &  12.4&   8.9&   2.1&   2.1&   0.8&   5.0&   0.07 \\
  G160-3   &  12.5&  14.6&   3.0&   2.5&   1.0&   4.2&  -0.23 \\
  G160-13  &  15.8&  16.5&   3.7&   2.8&   1.3&   2.5&  -0.45 \\
  G271-11  &   8.8&   5.2&   1.7&   1.9&   0.7&   8.3&  -0.57 \\
  G271-34  &  11.0&   7.9&   2.2&   1.7&   0.7&   3.2&  -0.78 \\
  G271-57  &  15.9&  16.9&   3.7&   2.8&   1.4&   1.8&  -0.62 \\
  G271-62  &   2.8&   0.8&   0.6&   0.2&   0.2&   4.3&        \\
  G271-69  &  12.1&  11.4&   2.8&   2.3&   1.1&   3.4&  -0.56 \\
  G271-75  &  14.6&  15.2&   3.9&   3.7&   1.6&   1.5&  -0.29 \\
\hline 
 \end{tabular}
 \end{center}
Notes: 1. late-type K/early-type M dwarf; 2. Pleiades member; 3. known
binary.
 \end{table}

\section {Discussion}

\subsection {The subdwarf sample}

Combining all of the photometric and spectroscopic observations discussed in the 
previous sections, we have ancillary data for 270 of the 317 stars identified originally
as candidate subdwarfs with parallaxes measured to an accuracy of 15\%.  
SIMBAD lists BV photometry and spectral type for most of the remaining stars, although with 
no attributed source for either parameter.
Several stars have discrepant colours and spectral types: 
\begin{itemize}
\item HIP 88231 and HIP 95924 are classified as A0 and B9 respectively. Both
lie close to the Galactic Plane, and are likely to be distant, reddened early-type
stars, like HIP 81617 and 90724. 
\item HIP 70622, 76670 and 78952 are all classed as G5, but have (B-V)$_l = 0.46$
\item HIP 83070 has a fainter companion at a separation of 10 arcseconds which may
affect the Hipparcos astrometry
\item HIP  80781 is also noted as double in the Hipparcos catalogue.
\end{itemize}
More accurate photometry and spectroscopy of all forty-seven stars currently lacking
population classification is clearly desirable.

 \begin {table*}
 \begin{center}
\caption { Confirmed subdwarfs: photometry and abundances}
 \begin{tabular}{rrrrrrrrcccl}
  \hline \hline 
 HIP & M$_V$ & $\sigma_V$ & U-B & B-V  & V-R & V-I & (b-y) &  [Fe/H]$_{sp}$ &[Fe/H]$_S$ & [Fe/H]$_{0.6}$ & \\
  \hline 
4750 & 5.27 & 0.35 & -0.107&   0.566&  0.315&   0.641&     &     &     &  -1.1 & n \\
5004 & 6.33 & 0.25 &  0.182&   0.758&  0.442&   0.877& 0.472 & -1.02 & -1.13 & -1.0 & * \\
7459 & 5.43 & 0.23 &  -0.147&   0.514&  0.329&   0.679& 0.365 & -1.17 & -1.09 & -1.2 & * \\
8558 & 3.89 & 0.18 &  -0.147&   0.369&   0.234&   0.486&      &  -1.10 &  &  -1.2 & * \\
14594 & 5.11 & 0.10 &       &   0.49 &      &     & 0.351 & -1.95 & -1.71 & & * \\
16404 & 6.17 & 0.20 &     & 0.65 &      &     & 0.451 & -1.92 & -1.91 & & *, bin \\
17241 & 4.81 & 0.29 &   -0.184&   0.447&  0.290&   0.612&    &    &    & -1.3 & n \\
19797 & 4.77 & 0.24 &    & 0.36 &    &    & 0.322 & -1.55 & -1.52 & & * \\
21609 & 6.01 & 0.13 &  -0.076&   0.640&  0.387&   0.798&0.452 & -1.76 & -1.81 & -1.6 & *, bin \\
22632 & 5.10 & 0.17 &   -0.197&   0.489&  0.309&   0.634& 0.358 & -1.59  & -1.47 & -1.6 & * \\
24316 & 5.25 & 0.16 &  -0.199&   0.496&   0.312&   0.647&  0.371 & -1.55 & -1.60 & -1.6 & * \\
26676 & 5.99 & 0.32 &  0.024&   0.658&    0.414&   0.801& 0.435 & -1.10 & -1.29 & -1.0 & * \\
34145 & 5.35 & 0.26 &    & 0.55 &    &    &    & H &    &    & n \\
38541 & 6.02 & 0.07 &    & 0.62 &    &    & 0.430 & -1.7 & -1.75 & & *, bin \\
40778 & 4.80 & 0.33 &    & 0.48 &    &    & 0.339 & -1.7 & -1.45 & & * \\
43490 & 5.30 & 0.14 &   -0.113&   0.550&  0.321&   0.664&    & H &    & -1.1 & n \\
44124 & 5.12 & 0.32 &  -0.185&   0.486&  0.317&   0.647& 0.349 & -1.96 & -1.34 & -1.4 & * \\
46120 & 6.20 & 0.13 &   -0.182&   0.577&  0.352&   0.728& 0.399 & -2.09 & -2.1 & -1.9 & * \\
48152 & 3.80 & 0.19 &  -0.203&   0.398& 0.263&   0.559& 0.302 & -2.07 & -2.24 & -1.6 & * \\
53070 & 4.63 & 0.13 &  -0.190&   0.482&   0.306&   0.636&0.344 & -1.4 & -1.29 & -1.5 & *, bin \\
54641 & 4.41 &0.10 &  -0.153&   0.481&  0.295&   0.607& 0.335 & -1.04 & -1.22 & -1.1 & * \\
54834 & 4.69 & 0.16 &  -0.177&   0.464&  0.296&   0.609&     &    &    & -1.3 &  n \\
55790 & 4.29 & 0.31 &  -0.192&   0.470&   0.300&   0.621& 0.343 & -1.56 & -1.62 & -1.4 & *, bin \\
57360 & 4.22 & 0.22 &  -0.164&   0.466&   0.293&   0.602& 0.33 & -1.21 & -1.31 & -1.1 & *, bin \\
57450 & 5.59 & 0.6 &   -0.15&   0.55&    &    & 0.398 & -1.26 & -1.45 & -1.4 & * \\
57939 & 6.61 & 0.02 &    & 0.75 &    &     & 0.484 & -1.32 & -1.41 &    & * \\
60632 & 4.88 & 0.27 &  -0.206&   0.447&  0.286&   0.595&   0.330& -1.55 & -1.79 & -1.5 & *, bin \\
65201 & 4.75 & 0.21 &   -0.204&   0.458&   0.312&   0.641& 0.349 & -1.86 & -1.96 & -1.5  & *, bin \\
70681 & 5.71 & 0.17 &  -0.082&   0.601&   0.359&   0.727& 0.400 & -1.25 & -1.16 & -1.2 & * \\
72461 & 4.79 & 0.32 &    & 0.44 &    &    & 0.332 & -2.15 & -2.0 &  & * \\
73798 & 5.90 & 0.26 &  -0.036&   0.642&   0.375&   0.755&    &    &    & -1.2 & n \\
81170 & 6.19 & 0.16 & 0120 & 0.746 & 0.447 & 0.900  & 0.474 & -1.20 & -1.39 &-1.30 & *, bin \\
88648 & 6.14 & 0.31 & -0.144&   0.611& 0.387&   0.782&  0.430 &    & -1.8 &-1.8 & *, bin \\
89215 & 6.50 & 0.26 & 0.188 & 0.773 & 0.435  & 0.860  & 0.474 & -1.36 & -1.34 & -1.2 & * \\
89554 & 4.25 & 0.15 &  -0.181&   0.449&  0.290&   0.602& 0.327 & -1.44 & -1.43 & -1.3 & *\\
94704 & 6.75 & 0.35 &   0.02&   0.66& 0.40&   0.84&    &    &    & -1.0 &  n \\
95190 & 4.79 & 0.35 &  -0.081&   0.570&    0.341&   0.696&    &    &    & -1.0 & n, bin \\
95996 & 5.15 & 0.36 &    &  0.49&    0.33&   0.65& 0.355 &    & -1.38 &    & *, bin\\
98020 & 5.85 & 0.10 &    & 0.60 &    &    &  0.416 & -1.6 & -1.57 &    & *, bin\\
99267 & 5.51 & 0.21 &    & 0.51 &    &    &    & -2.01 &    &    & * \\
100568 & 5.46 & 0.12 &  -0.126&   0.546& 0.332&   0.678&0.381 & -1.10 &  -1.27 & -1.3 & * \\
103269 & 6.03 & 0.23 &  -0.11&   0.62&    &    &    & -1.70 &    & -1.6 & * \\
106924 & 6.25 & 0.18 &  -0.07&   0.63&    &  &  & -1.62 &    & -1.4 & * \\
114271 & 4.03 & 0.19 &  -0.188&   0.420&   0.280&   0.580&    & -1.80 &    & -1.4 & n  \\
3026 & 4.15 & 0.34 & -0.131 & 0.469 & 0.291 & 0.609 &    & -1.30 &    & -0.9 & * \\
\hline
 \end{tabular}
 \end{center}
Notes: No Lutz-Kelker corrections have been applied to the absolute
magnitude listed in column 2; column 9 lists spectroscopic abundance estimates (see Table 2 - high-resolution
analyses are given preference for stars with multiple estimates); column 10 lists
Str\"omgren-based metallicities; column 11 gives $\delta(0.6)$ values; 
column 12 indicates whether the subdwarf was known previously (*) or 
is an addition (n), and
identifies known or suspected binaries (bin).
 \end{table*}

\begin{figure*}
\psfig{figure=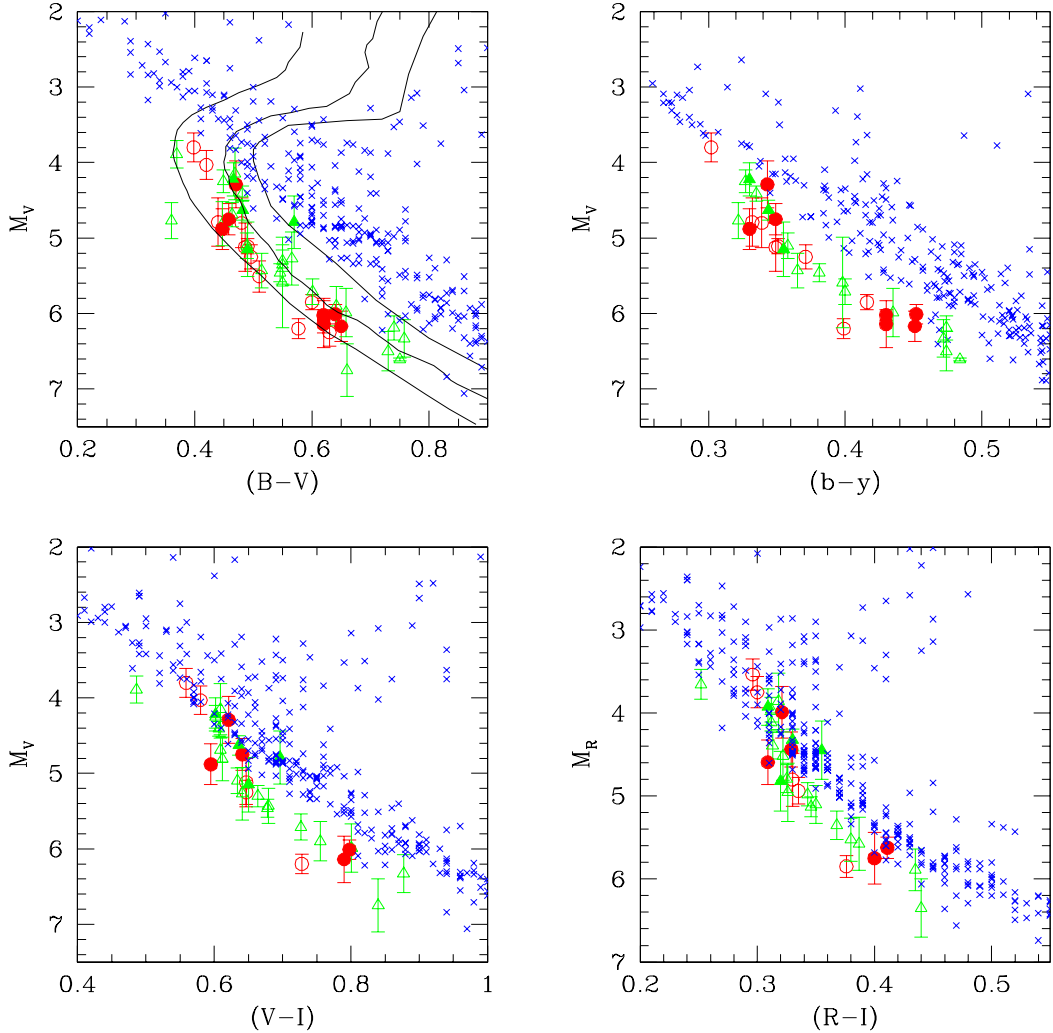,height=14cm,width=15cm
,bbllx=8mm,bblly=57mm,bburx=205mm,bbury=245mm,rheight=15cm}
\caption{Colour magnitude diagrams for the stars listed in Tables 11 and 12. In each case, disk main
sequence stars are plotted as crosses. Subdwarfs from Table 11 with [Fe/H]$\le -1.6$ are plotted as dots;
triangles mark higher abundance subdwarfs; known or suspected binaries are plotted as solid
symbols; data for stars from Table 12 are plotted as five-point stars. The
cluster (M$_V$, (B-V)) sequences are identical to those plotted in figure 2. }
\end{figure*}

Table 11 summarises the available photometric data and metallicity 
estimates for forty-five stars we consider 
confirmed as halo subdwarfs. We have omitted
stars with abundance determinations [Fe/H]$ > -1$, including those with
ambiguous results (e.g. HIP 35232: [Fe/H]$_{0.6} = -0.2$, spectral class  `H').
We add HIP 3026 from Table 8, which has a formal
parallax uncertainty of ${\sigma_\pi \over \pi} = 0.144$,
UBVRI photometry and an abundance [Fe/H]$\sim -1$.
The extensive observations described in this paper succeed in
making only nine additions (including HIP 114721)
to the canon of fiducial subdwarfs.  However, the data presented in Table 11 mark
the most reliable and homogeneous compilation of broadband photometry currently
available. These values should be used in preference to those derived
by averaging results from inhomogeneous sources (e.g. as in Carretta {\sl et al.}, 2000).

\begin{figure}
\psfig{figure=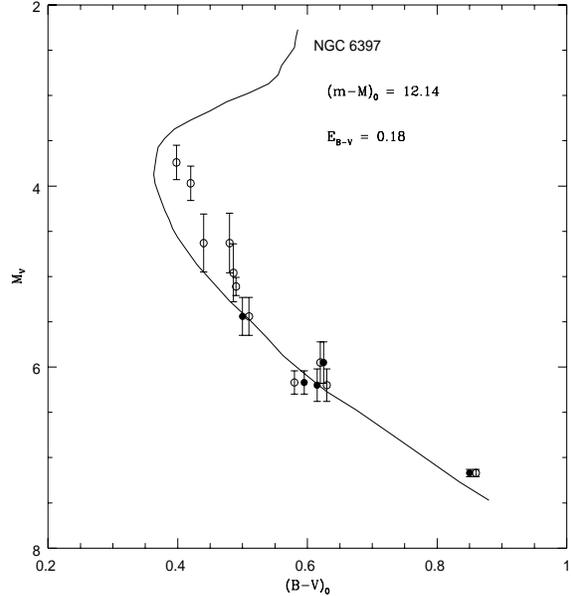,height=8cm,width=8cm
,bbllx=8mm,bblly=57mm,bburx=205mm,bbury=245mm,rheight=9cm}
\caption{The (M$_V$, (B-V)) diagram for extreme subdwarfs, [Fe/H]$< -1.6$.
The absolute magnitudes
have been adjusted for Lutz-Kelker bias using Hanson's n=3 relation. Open
circles plot the stars at their measured colours; solid points mark the 
location of the redder stars after correction to [Fe/H]=-1.82. 
The NGC 6397 sequence, at the specified distance modulus and
reddening,  is superimposed directly on the diagram. }
\end{figure}

Amongst the stars listed in Table 11, at least twelve
were previously known or suspected to be binary stars.
We add HIP 95190 to that list, based on its location in the HR diagram.
Thirty-one stars
(19 currently classed as single) have Str\"omgren data; 34 stars (25 single) have UBV
measurements; and 32 stars (22 single) have VRI (Cousins) photometry. 
Carretta {\sl et al.} (2000) have recently compiled their own sample of high-weight
metal-poor stars, with reliable abundance estimates and Hipparcos parallaxes accurate
to ${\sigma_\pi \over \pi} < 0.12$. Their dataset spans a wider range in
both colour and abundance than our sample, and includes eleven stars with 
[Fe/H]$< -1$ which are not in Table 11. For completeness, we have searched the 
literature for UBVR$_C$I$_C$ and Str\"omgren observations of those stars, and
those data are given in Table 12. Six of the twelve stars (HIP 7869, 33221, 49616, 
73385, 76976 and HD 211998) are subgiants.

 \begin {table*}
 \begin{center}
\caption { Subdwarfs from Carretta {\sl et al.}: photometry and abundances}
 \begin{tabular}{rrrrrrrrcccl}
  \hline \hline
 HIP & M$_V$ & $\sigma_V$ & U-B & B-V  & V-R & V-I & (b-y) &  [Fe/H]$_{sp}$ &[Fe/H]$_S$ & [Fe/H]$_{0.6}$ & \\
  \hline
7869 & 4.07 & 0.11 & -0.07 & 0.56 & 0.34 & 0.69 & 0.372 & -1.13 & -0.82  & -0.8 & R1, HD 10607 \\
    & 8.18 & 0.11 & 1.01 & 1.21 & 0.75 & 1.45 &   &   &  &  & R1, LHS 1279 \\
18915 & 7.17 & 0.04 & 0.37 & 0.87 &  &  &  0.533 & -1.69 & -1.64 & -1.8 & R1, HD 25329 \\
31332 & 6.24 & 0.23 & 0.52 & 0.94 & 0.57 & 1.08 &  & -2.11 &  & -2.0 & R2, HD 46663 \\
33221 & 3.87 & 0.26 & -0.15 & 0.48 &  & & 0.334 & -1.33 &  & -1.1 & C1, CD -33 3337 \\
49616 & 3.42 & 0.19 & -0.02 & 0.72 & 0.42 & 0.87 & 0.502 & -1.91 & -2.00 & -2.2 & R2, HD 89499 \\
73385 & 3.75 & 0.24 & -0.15 & 0.56 &  & & 0.401 & -1.73 & -1.98 & -1.5 & R1, HD 132475 \\
74234 & 7.07 & 0.11 & 0.34 & 0.84 & 0.52 & 1.01 & 0.524 & -1.28 & -1.24 & -1.1 & R1, HD 134440 \\
74235 & 6.71 & 0.09 & 0.16 & 0.78 & 0.45 & 0.91 & 0.484 & -1.30 & -1.33 & -1.6 & R1, HD 134439 \\
76976 & 3.43 & 0.09 & -0.19 & 0.49 & & & 0.380 & -2.40 & -0.6 & -1.5 & CL, HD 140283 \\
79537 & 6.84 & 0.03 & 0.28 & 0.82 & 0.48 & 0.94 & 0.509 & -1.64 & -1.22 & -1.2 & B, C2, Gl 615 \\
      & 2.98 & 0.14 & -0.06 & 0.65 & & & 0.450 & -1.43 & -1.51& -1.5 &  S, HD 211998 \\
\hline
 \end{tabular}
 \end{center}
Notes: LHS 1279 is a cpm companion of HD 10607;
HD 89499 is a spectroscopic binary;  \\
CD -33 3337 is identified as an astrometric binary by Carretta {\sl et al.}, 2000; \\
 $[Fe/H]_{sp}$ from Carretta {\sl et al.}, 2000; \\
All of the Str\"omgren data are from Schuster \& Nissen, 1988;  \\
References for UBVRI:  B - Bessell, 1990 (BVRI); C1 - Cousins, 1972; C2 - Carney, 1980 (UBV); 
CL - Carney \& Latham, 1987;   R1 - Ryan, 1992; R2 - Ryan, 1989;
S - SIMBAD database
 \end{table*}

Figure 15 plots the 
resultant colour-magnitude diagrams, combining data from both tables 11 and 12.
The disk main sequence is defined in the Johnson/Cousins system
by nearby stars with ${\sigma_\pi \over \pi} < 0.15$ and BVRI photometry from
Bessell (1990); the Str\"omgren sequence is defined by the same stars plotted in Figure 3.
We have included the (M$_R$, (R-I)) diagram to illustrate the
insensitivity to abundance of the location of the main sequence in that plane.
With few strong absorption features in either band, decreasing
abundance produces a relatively small change in the differential blanketing. Such
is not the case at lower luminosities, where variations in TiO bandstrength 
lead to more substantial offsets between disk and halo (Gizis, 1997). 

\subsection {Metal-poor subdwarfs and globular cluster distances}

Our main aim in undertaking this program was the identification of previously-unrecognised
unevolved extreme subdwarfs. We have, not unexpectedly, had
little success. Table 11 includes nine single stars with  metallicities
below [Fe/H]=-1.6, of which only four (HIP 46120, 99267, 103269 \& 106924)
are redder than (B-V)=0.50. However,  all of those stars save HIP 40778
have abundances derived from high-resolution observations, and, crucially,
HIP 46120 has [Fe/H]$< -2$. The last-mentioned star is the only lower main-sequence
star with both an accurate parallax and a reliable abundance measurement\footnote{
Carretta {\sl et al.} (2000) include HIP 46120 = CD-80:328 in their recent re-analysis
of globular cluster distances, but
they adopt [Fe/H]=-1.75, rather than -2.07, as derived by
Ryan \& Deliyanis (1998). They comment that, at the
adopted metallicity, the star occupies an anomalous
position in the colour-magnitude diagram.
Those anomalies are not present at the lower abundance adopted in our
analysis.}.

It is not our intention to re-examine globular cluster distances in this
paper. Nonetheless, Figure 16 offers a single comparison, matching the nine 
extreme subdwarfs discussed above against NGC 6397 in the (M$_V$, (B-V)) plane.
Table 12 includes three subdwarfs from Carretta {\sl et al.} (2000) with [Fe/H]$<-1.6$:
HIP 18915, 31332 and 79537. However, Figure 15 shows that HIP 31332
lies above the disk main sequence, suggesting either that the star is
a binary or that there is an error in either the abundance measurement
or the photometry; moreover, HIP 79537 has discordant metallicity estimates, 
with Ryan \& Deliyannis (1998) measuring [Fe/H]=-1.38,
rather than -1.64. We therefore include only
HIP 18915 in Figure 16.

Figure 15 plots the subdwarf absolute magnitudes derived directly from the
Hipparcos trigonometric parallax measurements; in Figure 16, 
the absolute magnitudes have been corrected for Lutz-Kelker bias using
the n=3 approximation from Hanson (1979). The corrections are small:
$\Delta_{LK} < 0.2$ mag. for all stars, and $\Delta_{LK} < 0.1$ for
(B-V)$> 0.5$. We have also used theoretical models from Straniero
\& Chieffi (1991) to determine the appropriate colour corrections
to adjust the lower main-sequence stars to match [Fe/H]=-1.82, 
the abundance derived by Gratton {\sl et al.} (1997) for NGC 6397. The NGC
6397 fiducial sequence, adjusted to a true distance modulus of 12.14
mag. and a reddening of E$_{B-V}$=0.18 mag., is simply superimposed
on the diagram. The subdwarf data are not inconsistent with the
adopted cluster distance modulus. Following the arguments outlined by Reid (1998),
the corresponding distance modulus
inferred for the Large Magellanic Cloud is (m-M)$_0$=18.62. 

\section {Conclusions}

We have searched the Hipparcos catalogue for previously unrecognised halo
subdwarfs which might be added to the current meagre sample of
stars suitable for globular cluster main-sequence fitting. Our survey, covering
263 of 317 candidates with (B-V)$_T < 0.8$ and ${\sigma_\pi \over \pi} \le 0.15$,
shows that few such stars remain hidden in the
current database. However, in the course of this exercise we have compiled the
first catalogue of reliable UBVR$_C$I$_C$ photometry of spectroscopically-confirmed
halo subdwarfs with accurate trigonometric parallax measurements. Our sample 
includes a bare handful of extreme subdwarfs, notably HIP 46120, suitable for matching against
metal-poor clusters such as M15, M68 and M92. 

Globular cluster main-sequence fitting distance determinations have been undertaken
almost exclusively in the (M$_V$, (B-V)) plane. This concentration reflects
necessity rather than choice: the (M$_R$, (R-I)) plane, for example, may
offer significant advantages in greater tolerance to abundance uncertainties, but
even with the observations contributed in this paper, many key subdwarfs still
lack reliable photometry. Filling the gaps in Tables 11 and 12 should be a
high priority.

As has been pointed out elsewhere, the dominant
uncertainty in cluster distance analysis lies with the available photometry, rather
than parallax measurements.  Indeed, at a more general level, it remains the case
that many stars in the Hipparcos catalogue, which have astrometric measurements
at sub-milliarcsecond precision, have received no attention from ground-based
observatories since the completion of the Henry Draper or Bonner Durchmusterung
catalogues. We are in the paradoxical situation is knowing their distances
with higher accuracy than their apparent magnitudes or spectral types. 
This is a matter which should be borne in mind when considering future 
large-scale astrometric projects.

\subsection*{Acknowledgements}

INR would like to thank the director and the Time Assignment 
Committee of the Carnegie Observatories
for the allocation of telescope time on both the 100-inch DuPont and 40-inch Swope telescopes
at Las Campanas Observatory. Partial support for SM was provided by a grant from
the Space Telescope Science Institute,  GO-08146.01-97A.
Both the SIMBAD database and the ADS bibliographic service were invaluable in this
research project.


\begin{thebibliography}{}

\bibitem [Anthony-Twarog \& Twarog] {att} Anthony-Twarog, B.J., Twarog, B.A. 2000, AJ, 120, 1311

\bibitem [Axer {\sl et al.}, 1994] {ax94} Axer, M. Fuhrmann, K., Gehren, T. 1994, A\&A, 291, 895

\bibitem [Baldwin \& Stone] {bs} Baldwin, J.A., Stone, R.P.S., 1984, MNRAS, 206, 241

\bibitem [Bessell, 1979] {b79} Bessell, M.S. 1979, PASP, 91, 589

\bibitem [Bessell, 1983] {b83} Bessell, M.S. 1983, PASP, 95, 480 

\bibitem [Bessell, 1990] {b90} Bessell, M.S. 1990, A\&AS, 83, 357

\bibitem [Bessell, 2000] {b20} Bessell, M.S. 2000, PASP, 112, 961

\bibitem [Bessell \& Weis, 1983] {b83} Bessell, M.S., Weis, E.W.  1987, PASP, 99, 642

\bibitem [Bi\'emont {\sl et al.}, 1991] {b91 }Bi\'emont, E., Baudoux, M., Kurucz, R.L., Ansbacher, W., 
Pinnington, E.H. 1991, A\&A, 249, 538

\bibitem [Bond \& Mac Connell, 1975] {bm75} Bond, H.E., MacConnell, D.J. 1971, ApJ, 165, 51

\bibitem [Carney] {c78} Carney, B.W. 1978, AJ, 83, 1087

\bibitem [Carney, 1979] {ca79} Carney, B.W. 1979, ApJ, 233, 211

\bibitem [Carney, 1983] {c83} Carney, B.W. 1983, AJ, 88, 610

\bibitem [Carney \& Latham, 1987] {cl} Carney, B.W., Latham, D.W. 1987, AJ, 93, 116

\bibitem [Carney {\sl et al.}, 1994] {clla} Carney, B.W., Latham, D.W., Laird, J.B., Aguilar, L.A. 1994, 
AJ, 107, 2240

\bibitem [Carretta \& Gratton, 1997] {cg97} Carretta, E., Gratton, R.G. 1997, A\&AS, 121, 95

\bibitem [Carretta et al, 2000] {car20} Carretta, E., Gratton, R.G., Clementini, G.,
Fusi Pecci, F., 2000, ApJ, 533, 215

\bibitem [Carretta et al, 2000] {car20b} Carretta, E., Gratton, R.G., Sneden, C. 2000b,
A\&A, 356, 238

\bibitem [Celis, 1975] {c75} Celis, L. 1975, A\&AS, 22, 9

\bibitem [Clementini et al, 1999] {cl99} Clementini, G., Gratton, R.G., Carretta, E., Sneden, C. 1999, 
MNRAS, 302, 22

\bibitem [Cousins, 1972] {c72} Cousins, A.W.J. 1972, MNASSA, 31, 7

\bibitem [Cousins, 1973] {c73} Cousins, A.W.J. 1973, Mem. RAS, 77, 223

\bibitem [Cousins \& Stoy, 1962] {cs62} Cousins, A.W.J., Stoy, R.H. 1962 Royal Obs. Bull. 64, 103

\bibitem [Cutis] {cut} Cutispoto, G., Tagliaferri, G., Giommi, P., Gouiffes, C., 
Pallavicini, R., Pasquini, L., Rodono, M. 1991, A\&AS, 87, 233

\bibitem [Dean, 1981] {dnn} Dean, J. 1981, MNASSA, 40, 14

\bibitem [de Geus {\sl et al.}, 1991] {dg91} de Geus, E.J., Lub, J., van der Grifte, E., 1991, A\&AS, 85, 915

\bibitem[Eggen, 1990] {eg1} Eggen, O.J. 1990, AJ, 100, 1159

\bibitem[ESA, 1997]{E97} ESA, 1997, The Hipparcos Catalogue

\bibitem [Fabricius] {fm20} Fabricius, C., Makarov, V.V., 2000a, A\&AS, 144, 45

\bibitem [Fabricius] {fm20} Fabricius, C., Makarov, V.V., 2000b, A\&A, 356, 141

\bibitem [Falin \& Mignard] {fm} Falin, J.L., Mignard. F. 1999, A\&AS, 185, 281

\bibitem [Ferro {\sl et al.}] {fer} Ferro, A.A., Parrao, L., Schuster, W.,
Gonz\'alez-Bedolla, S., Peniche, R., Pena, J.H.  1991, A\&AS, 83, 225

\bibitem [Figueras et al, 1990] {F1} Figueras, F., Jordi, C., Rossello, G., Torra, J., 1990, A\&AS, 82, 57

\bibitem [Franco] {fran} Franco, G.A.P. 1994, A\&AS, 104, 9

\bibitem [Gizis, 1997] {g97} Gizis, J.E. 1997, AJ, 113, 806

\bibitem [Giclas et al, 1963] {g63} Giclas, H.L., Burnham, R., Thomas, N.G. 1963, Lowell Obs. Bull., 6, 1

\bibitem [Gratton {\sl et al.}, 1997a] {gcc} Gratton, R.G., Carretta, E., Castelli, F., 1997a, A\&A, 314, 191

\bibitem [Gratton {\sl et al.}, 1997b] {g97} Gratton, R.G., Fusi-Pecci, F., Carretta, E., Clementini, G., Corsi, C.E.,
Lattanzi, M. 1997b, ApJ, 491, 749

\bibitem [Guetter, 1980] {g1980} Guetter, H.H. 1980, PASP, 92, 215

\bibitem [Hanson, 1979] Hanson, R.B. 1979, MNRAS, 186, 875

\bibitem [Harris, 1996] {ha96} Harris, W.E. 1996, AJ, 112, 1487

\bibitem [Hauck \& Mermilliod, 1998] {hm98} Hauck, B., Mermilliod, M. 1998, A\&AS, 129, 431

\bibitem [Kenyon et al] {k94} Kenyon, S.J., Dobrzycka, D., Hartmann, L. 1994, AJ, 108, 1872

\bibitem[Kilkenny {\sl et al.}, 1998] {k98} Kilkenny, D., van Wyk, F., Roberts, G., Marang, F., Cooper, D. 1998,
MNRAS, 294, 93

\bibitem [Knude, 1979] {kn} Knude, 1981 A\&AS, 44, 225

\bibitem [Lee, 1984] {l84} Lee, S.G. 1984, AJ, 89, 702

\bibitem [Lub \& Pel, 1977] {lp77} Lub, J., Pel, J.W. 1977, AA, 54, 137

\bibitem [Manfroid {\sl et al.}] {man} Manfroid, J., Oblak, E., Pernier, B.  1987, A\&AS, 69, 505

\bibitem [Menzies {\sl et al.}, 1989] {men89} Menzies, J.W., Cousins, A.W.J., Banfield, R.M., Laing, J.D. 1989,
 SAAO Circ. 13, 1

\bibitem [Milone {\sl et al.}, 19nn] {mil} Milone, E.F., Wilson, W.J.F., Fry, D.J.I., 
Schiller, S.J. 1994, PASP, 106, 1120

\bibitem [Oblak] {ob} Oblak, E. 1991, A\&AS, 83, 467

\bibitem [Oja, 1985] {oj85} Oja, T. 1985, A\&AS, 59, 461

\bibitem [Oja, 1986] {oj86} Oja, T. 1986, A\&AS, 65, 405

\bibitem [Oja, 1991] {oj91} Oja, T. 1991, A\&AS, 89, 415

\bibitem [Olsen, 1994a] {ol1} Olsen, E.H. 1994, A\&AS, 104, 429

\bibitem [Olsen, 1994b] {ol2} Olsen, E.H. 1994, A\&AS, 106, 257

\bibitem [Perlmutter, S.] {p98} Perlmutter, S. {\sl et al.} 1998, Nature, 391, 51

\bibitem [Perryman et al, 1998] {p98} Perryman, M.A.C. et al., 1998, A\&A, 331, 81

\bibitem [Reid, 1993] {r93} Reid, I.N. 1993, MNRAS, 265, 785

\bibitem [Reid, 1998] {r98} Reid, I.N. 1998, AJ, 115, 204

\bibitem[Reid, 1999]{R99} Reid, I.N., 1999, Ann. Rev. Astr. Ap., 37, 191

\bibitem [Riess, 2000] {r20} Riess, A.G. {\sl et al.} 2000, ApJ, 536, 62

\bibitem [Roman, 1955] {r55} Roman, N.G. 1955, ApJS, 2, 195

\bibitem [Rossello {\sl et al.}] {ro} Rossello, G., Figueras, F., Jordi, C., Nunez, J., 
Paredes, J.M., Sala, F., Torra, J. 1988, A\&AS, 75, 21

\bibitem [Ryan, 1989] {r1} Ryan, S.G. 1989, AJ, 98, 1693

\bibitem [Ryan, 1992] {r2} Ryan, S.G. 1992, AJ, 104, 1144

\bibitem [Ryan \& Deliyannis, 1998] {rd98} Ryan, S.G., Deliyannis, C.P. 1998, ApJ, 500, 398

\bibitem [Ryan \& Norris] {rn1} Ryan, S.G., Norris, J.E. 1991, AJ, 101, 1835

\bibitem [Sandage, 1969] {sa69} Sandage, A. 1969, ApJ. 158, 1115

\bibitem [Sandage \& Kowal, 1986] {sk} Sandage, A, Kowal, C. 1986, AJ, 91, 1140

\bibitem [Schmidt et al, 1998] Schmidt, B.P. {\sl et al.}, 1998, ApJ, 507, 46

\bibitem [Schuster \& Nissen, 1988] {sc88} Schuster, W.J., Nissen, P.F. 1988, A\&AS, 73, 225

\bibitem [Schuster \& Nissen, 1989] {sc89} Schuster, W.J., Nissen, P.F. 1989, A\&A, 221, 65

\bibitem [Schuster {\sl et al.}, 1993] {sc93} Schuster, W.J., Parrao, L., Contreras Mart\'inez, M. E. 
1993, A\&AS, 97, 951

\bibitem [Str\"omgren, 1966] {str66} Str\"omgren, B. 1966,  Ann. Rev. Astr. Ap., 4, 433

\bibitem [Twarog, 1980] {tw80} Twarog, B. 1980, ApJS, 44, 1

\bibitem [Upgren et al] {u1} Upgren, A.R., 1972, AJ, 77, 486

\bibitem [Wallerstein \& Carlson, 1960] {wa60} Wallerstein, G., Carlson, M. 1960, ApJ, 132, 276

\bibitem [Weis, 1986] {we1} Weis, E., 1986, AJ, 91, 626

\bibitem [Weis, 1993] {we2} Weis, E., 1993, AJ, 105, 1962

\bibitem [Weis, 1996] {we3} Weis, E., 1996, AJ, 112, 2300

\bibitem [Wildey {\sl et al.}, 1962] {wi62} Wildey, R.L., Burbidge, E.M., Sandage, A.R., Burbidge, G.R.
1962, ApJ, 135, 94

\end{thebibliography}
\end{document}